\documentclass[UTF8,a4paper,twocolumn]{aastex61}

\usepackage{CJK}
\usepackage{amsmath}

\newcommand{\oh}{12\,+\,log(O/H)}

\newcommand{\vs}{vs.}

\newcommand{\ha}{\hbox{H$\alpha$}}
\newcommand{\hb}{\hbox{H$\beta$}}

\newcommand{\hii}{\hbox{H\,{\sc ii}}}

\newcommand{\gsim}{\lower.5ex\hbox{$\; \buildrel > \over \sim \;$}}
\newcommand{\lsim}{\lower.5ex\hbox{$\; \buildrel < \over \sim \;$}}
\newcommand{\oi}{\hbox{[O\,{\sc i}]}}
\newcommand{\oii}{\hbox{[O\,{\sc ii}]}}
\newcommand{\oiii}{\hbox{[O\,{\sc iii}]}}
\newcommand{\nii}{\hbox{[N\,{\sc ii}]}}
\newcommand{\sii}{\hbox{[S\,{\sc ii}]}}

\newcommand{\avgas}{$A_{V,\mathrm{gas}}$}
\newcommand{\avstar}{$A_{V,\mathrm{star}}$}
\newcommand{\rav}{$A_{V,\mathrm{star}}/A_{V,\mathrm{gas}}$}
\newcommand{\re}{$R_{\mathrm{e}}$}
\newcommand{\sigha}{$\mathrm{\Sigma_{H\alpha}}$}
\newcommand{\ewha}{$\mathrm{EW_{H\alpha}}$}

\submitjournal{AAS}

\shorttitle{Variations in Dust Attenuation Ratio}
\shortauthors{Lin et al.}

\graphicspath{{/Users/jayson/data/work/try_diff_av/}}

\begin{document}

\title{A Variant Stellar-to-nebular Dust Attenuation Ratio on Subgalactic and Galactic Scales}

\begin{CJK*}{UTF8}{gbsn}

\email{zesenlin@mail.ustc.edu.cn, xkong@ustc.edu.cn}
\author[0000-0001-8078-3428]{Zesen Lin (林泽森)}
\affil{Key Laboratory for Research in Galaxies and Cosmology, Department of Astronomy, University of Science and Technology of China, Hefei 230026, China}
\affil{School of Astronomy and Space Sciences, University of Science and Technology of China, Hefei 230026, China}

\author{Xu Kong (孔旭)}
\affil{Key Laboratory for Research in Galaxies and Cosmology, Department of Astronomy, University of Science and Technology of China, Hefei 230026, China}
\affil{School of Astronomy and Space Sciences, University of Science and Technology of China, Hefei 230026, China}




\begin{abstract}

The state-of-the-art geometry models of stars/dust suggest that dust attenuation toward nebular regions (\avgas) is always larger than that of stellar regions (\avstar). Utilizing the newly released integral field spectroscopic data from the Mapping Nearby Galaxies at Apache Point Observatory survey, we investigate whether and how the \rav\ ratio varies from subgalactic to galactic scales. On a subgalactic scale, we report a stronger correlation between \avstar\ and \avgas\ for more active \hii\ regions. The local \rav\ is found to have moderate nonlinear correlations with three tracers of diffuse ionized gas (DIG), as well as indicators of gas-phase metallicity and ionization. The DIG regions tend to have larger \rav\ compared to classic \hii\ regions excited by young OB stars. Metal-poor regions with a higher ionized level suffer much less nebular attenuation and thus have larger \rav\ ratios. A low-\avgas\ and high-\rav\ sequence, which can be resolved into DIG-dominated and metal-poor regions, on the three BPT diagrams is found. Based on these observations, we suggest that besides the geometry of stars/dust, local physical conditions such as metallicity and ionized level also play an important role in determining the \rav. On a galactic scale, the global \rav\ ratio has strong correlations with stellar mass ($M_*$), moderate correlations with star formation rate (SFR) and metallicity, and weak correlations with inclination and specific SFR. Galaxies with larger $M_*$ and higher SFR that are more metal-rich tend to have smaller \rav\ ratios. Such correlations form a decreasing trend of \rav\ along the star-forming main sequence and mass--metallicity relation. The dust growth process accompanied by galaxy growth might be one plausible explanation for our observations.

\end{abstract}




\section{Introduction} \label{sec:intro}

Almost half of optical emission is shielded by dust and reaches us in the form of infrared (IR) radiation \citep{Dole2006}, especially for stellar light from star-forming galaxies (SFGs) that contribute to the cosmic IR background up to about 95\% \citep{Viero2013}. This so-called dust extinction/attenuation effect prevents us from accurately measuring some important galactic properties, such as star formation rate (SFR), stellar mass ($M_*$), and so on. Dust is produced from the late stage of stellar evolution---namely supernovae (SNe) and stellar winds of asymptotic giant branch stars (e.g., \citealt{Dwek1998,Aoyama2017,Gjergo2018})---grows in the interstellar medium (ISM), especially in dense molecular clouds (e.g., \citealt{Hirashita2014}); and is destroyed by SN shocks (e.g., \citealt{Aoyama2017}). Such dust processes substantially alter the dust abundance and the size distribution of dust grains (e.g., \citealt{Asano2013,Aoyama2017,Hirashita2019}) and hence affect the absorption and scattering of light (see \citealt{Galliano2018} for a review). Therefore, to better understand the formation and evolution of galaxies, we should know how and how much stellar light is reprocessed by the dust extinction/attenuation effect.

Besides the dust properties, the geometry between dust and stars is also believed to be an important factor for dust attenuation in terms of variations in the shape of the dust attenuation law \citep{Salmon2016,Narayanan2018} and the IRX-$\beta$ relation \citep{Meurer1999,Kong2004,Ye2016,Popping2017,Narayanan2018a}. Particularly, it is thought of as the main origin of the different levels of dust attenuation between stellar light and emission lines (e.g., \citealt{Price2014,Reddy2015,Koyama2019}). Plenty of observations on both local (e.g., \citealt{Calzetti1997,Kreckel2013,Zahid2017,Koyama2019}) and high-redshift (e.g., \citealt{Garn2010,Wuyts2011,Wuyts2013,Price2014,Pannella2015,Theios2019}) SFGs reveal that nebular emission lines tend to suffer more dust attenuation compared to stellar continuum. Such extra attenuation can be explained by a widely used two-component dust model \citep{Charlot2000,Wild2011,Chevallard2013} in which dust inside a galaxy consists of a diffuse, optically thin component and a denser, optically thick one (the birth cloud) that relates to star-forming regions. Thus, emission from ionized gas travels through the outer envelope of the birth clouds and then the ambient ISM, while the majority of stellar light propagates only the ambient ISM and suffers less dust reddening compared to emission lines \citep{Charlot2000}.

By utilizing the observation of local starburst galaxies and bright star-forming regions, \citet{Calzetti1997} presented the relation between the color excess $E(B-V)$ of stars and gas as $E(B-V)_{\mathrm{star}}=0.44E(B-V)_{\mathrm{gas}}$; i.e., the reddening of stellar light is only nearly half of the one toward nebular regions. A similar relation based on $V$-band attenuation ($A_{V}$) was reported by the spatially resolved study of \citet{Kreckel2013} in which they found $A_{V,\mathrm{star}}=0.47\pm0.006A_{V,\mathrm{gas}}$, assuming $R_V=3.1$ for several relatively face-on local galaxies. More recent studies also suggest that \rav\ may be not a constant for all types of galaxies \citep{Wild2011,Koyama2015,Zahid2017,Koyama2019,Qin2019a}. It is found to be correlated with $M_*$ \citep{Koyama2015,Zahid2017,Koyama2019}, specific SFR (sSFR; \citealt{Wild2011,Koyama2015,Koyama2019,Qin2019a}), and inclination \citep{Wild2011}.

High-redshift studies also revealed a wide diversity of the stellar-to-nebular reddening ratio, varying from 0.44 \citep{Wuyts2011} to $\sim 1$ \citep{Pannella2015,Puglisi2016}, as well as the dependences of \rav\ on the physical properties of galaxies \citep{Price2014,Reddy2015,Puglisi2016}.

However, most the aforementioned studies focused on dust attenuation on a global scale, deriving the \rav\ using the integrated fluxes of galaxies. With the development of the integral field unit (IFU) technique, spatially resolved spectra were observed for thousands of galaxies in several IFU surveys, such as SAMI (using the Sydney-Australian Astronomical Observatory Multi-object Integral Field Spectrograph; \citealt{Croom2012}), Calar Alto Large Integral Field Area (CALIFA; \citealt{Sanchez2012}), and Mapping Nearby Galaxies at Apache Point Observatory (MaNGA; \citealt{Bundy2015}). Taking advantage of these data sets, we are able to study the dust attenuation on subgalactic scales that would help us obtain a more comprehensive understanding of dust effects. However, only a few works paid attention to the subgalactic dust attenuation until now.

On the basis of spatially resolved spectra of several local galaxies, \cite{Kreckel2013} showed that the dust attenuation ratios within individual galaxies exhibit a dependence on the $\sii\lambda6717$--to--$\mathrm{H\alpha}$ line ratio, which is believed to be a tracer of the transition between diffuse ionized gas (DIG) and \hii\ regions (e.g., \citealt{Hoopes2003}). Adopting a critical $\sii\lambda6717/\mathrm{H\alpha}$ ratio of 0.2, they found an \rav\ ratio of 0.7 and 0.5 for DIG- and \hii-dominated regions, respectively. Recently, using the high-resolution IFU observations of NGC 5626, a lenticular galaxy with a redshift of 0.023, \citet{Viaene2017} found that the continuum attenuation is independent from the ionized gas attenuation in this early-type galaxy. This observation raises the question of whether the dust attenuations of the stellar continuum and ionized gas indeed correlate tightly with each other and whether the correlation could be broken under certain conditions.

On the other hand, adopting the ultraviolet (UV) slope $\beta$ as an indicator of stellar attenuation, \cite{Calzetti1994} observed a linear correlation between $\beta$ and the nebular attenuation, indicating a tight and monotonic relation between the stellar and nebular dust attenuation. Based on this property, the widely used dust attenuation law of \cite{Calzetti2000} was constructed. More recent works following the same methodology to derive the attenuation law either in the local universe (e.g., \citealt{Battisti2016,Battisti2017a}) or at high redshift (e.g., \citealt{Reddy2015}) also rely on this relation. Therefore, it is important to verify the applicability of this finding to a broader range of galaxies than those used in previous studies.

Making use of IFU data, this work aims at investigating whether the dust attenuation of the stellar continuum and ionized gas correlate with each other and, if so, whether their ratio correlates with local and global properties for \hii\ regions. This paper is organized as follows. In Section \ref{sec:data}, we give a brief overview of the data, as well as sample selection. We present results of correlations between the reddening ratio and local physical properties in Section \ref{sec:resolved}. We further analyze how \rav\ varies with the global properties of galaxies in Section \ref{sec:global}. Finally, we summarize in Section \ref{sec:summary}. Throughout this paper, we adopt a flat $\Lambda$CDM cosmology with $H_0=71~\mathrm{km~s^{-1}~Mpc^{-1}}$, $\Omega_{\Lambda}=0.73$, $\Omega_{\mathrm{m}}=0.27$, and a \cite{Salpeter1955} initial mass function (IMF).

\section{Data and Sample Selection} \label{sec:data}

\subsection{MaNGA Survey and Pipe3D VAC}
\label{subsec:survey_VAC}

MaNGA, as one of the major spectroscopic programs of the Sloan Digital Sky Survey IV (SDSS-IV; \citealt{Blanton2017}), plans to observe $\sim10,000$ nearby galaxies with a redshift range of 0.01--0.15, spanning a wide range of $M_*$, SFR, and environment \citep{Bundy2015,Yan2016}. Spectral observations of the MaNGA survey cover a wavelength range of 3600--10300 \AA\ with a spectral resolution of $R\sim2000$ and a spatial resolution, i.e., the FWHM of the reconstructed point spread function (PSF), of 2\farcs5 \citep{Law2016}. In this work, we make use of MaNGA data from the SDSS DR15 \citep{Aguado2019}, which includes 4824 data cubes.

Our main data are from the MaNGA value-added catalog (VAC) of the Pipe3D pipeline \citep{Sanchez2018}, which is released as one part of the SDSS DR15\footnote{\url{https://www.sdss.org/dr15/manga/manga-data/manga-pipe3d-value-added-catalog/}}. This pipeline is designed to implement full spectral fitting with stellar population models for IFU spectra. Details and algorithms can be found in \citet{Sanchez2016a,Sanchez2016}. Here we provide a brief summary of the procedure of stellar continuum fitting. Spatial binning is performed for each data cube to achieve a nominal continuum signal-to-noise ratio (S/N) of 50 before fitting. A set of 156 simple stellar population (SSP) templates from \cite{CidFernandes2013} accounting for 39 stellar ages (from 1 Myr to 13 Gyr) and four metallicities ($Z/Z_{\odot}=0.2$, 0.4, 1, and 1.5) is used to model the coadded spectra of each spatial bin. The fitting procedure includes two steps: first, a fitting is performed to extract the nonlinear parameters, such as velocity, velocity dispersion, and dust attenuation, assuming an extinction curve of \citet{Cardelli1989} and a selective extinction of $R_V=3.1$; then, a robust modeling is applied to the corrected spectra after accounting for the nonlinear effects derived from the previous step, via a linear combination of the 156 SSPs with the Monte Carlo method to obtain the coefficients of the best-fit model and the corresponding uncertainties.

After the continuum modeling, the pipeline rescales the best-fit spectral energy distribution of one spatial bin by the $V$-band flux of each spaxel within that bin and takes it as the best-fit model of each spaxel. This resulting ``fake'' spectrum is then subtracted from the observed one to generate a pure emission line spectrum. Individual emission lines are fitted spaxel by spaxel with a single Gaussian profile. It is noteworthy that the continuum modeling is implemented for spatial bins, whereas the ionized gas analysis is performed for individual spaxels. Finally, the Pipe3D VAC provides data products for 4815 galaxies, including the results of the analyses of both stellar continua and emission lines \citep{Aguado2019}.

\subsection{Reanalysis of Emission Lines}
\label{subsec:reanalysis}

Given the aim of this study, we need dust attenuation of both stellar continuum and emission lines, where the former one is provided for spatial bins and the latter one is given for individual spaxels in the Pipe3D VAC. To reconcile the spatial scale of these two parameters, we determine to use spatial bins to study and reanalyze the emission lines on the same spatial scale. This choice can also prevent the oversampling of low-S/N spectra for which the stellar properties should be the same if they belong to the same spatial bin, as well as avoid any potential uncertainties from the $V$-band flux-based assignment of the underlying stellar contribution. For simplicity, in the following, the term ``spaxel'' refers to a given spatial element regardless of the number of spaxels within it.

We reconstruct the best-fit spectra for each spaxel using the weights of 156 SSP templates from the Pipe3D VAC. The resulting spectra are in good agreement with the coadded observed spectra and subtracted to create pure emission line spectra for further emission line analysis. We fit each emission line with a single Gaussian function using the Python version of the MPFIT\footnote{\url{https://code.google.com/archive/p/astrolibpy/}} software \citep{Markwardt2009}. The S/Ns of emission lines are calculated via the method introduced in \citet{Ly2014}. We use the Balmer decrement to correct the galactic internal attenuation. Under the assumption of Case B recombination, we employ an intrinsic flux ratio of $\mathrm{H\alpha/H\beta}=2.86$ \citep{Storey1995}, an extinction curve of \citet{Cardelli1989}, and $R_V=3.1$ to do the dereddening. In short, we redo the analysis of ionized gas but on the spatial bin scale.

\subsection{Sample Selection}
\label{subsec:sample}

\begin{figure}[htb]
    \centering
    \includegraphics[width=0.475\textwidth]{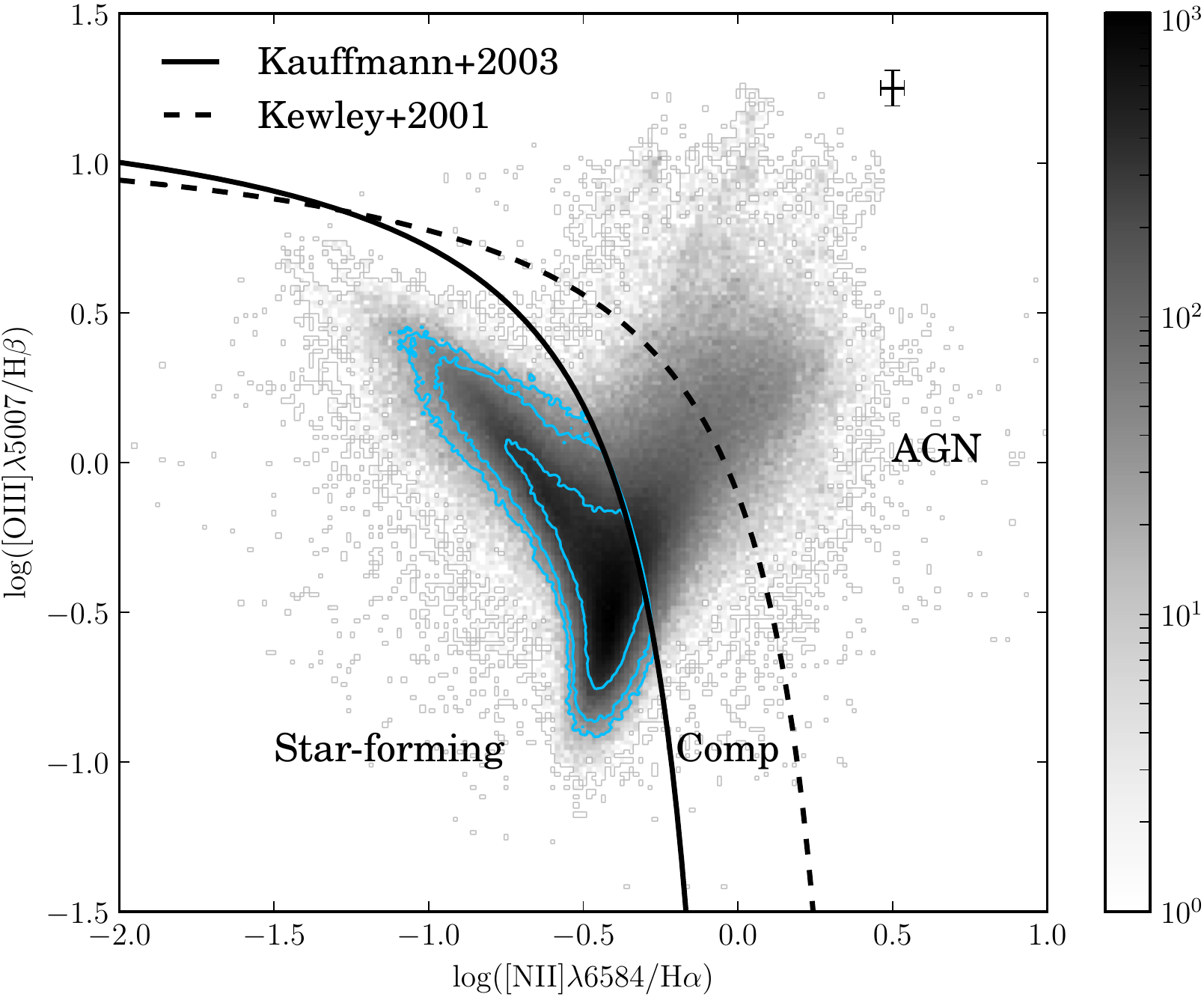}
    \caption{The BPT diagram of the selected spaxels. The dashed black curve represents the theoretical maximum starburst line of \citet{Kewley2001}, and the solid black curve shows the demarcation between the pure star-forming region and the AGN region from \citet{Kauffmann2003a}. The gray region shows the density distribution of all of the selected spaxels, while the blue contours are drawn at 68\%, 95\%, and 99\% of the star-forming spaxels only, respectively. The black error bar denotes the median uncertainties of all star-forming spaxels.
    \label{fig:bpt}}
    \end{figure}

To ensure a reliable estimate of the dust attenuation of gas (\avgas), we limit our analysis to spaxels with S/Ns of \ha\ and \hb\ greater than 5 and S/Ns of \oiii$\lambda$5007, \nii$\lambda$6584, and \sii$\lambda$6717 greater than 3. Setting lower S/N cuts should not change our results (see Appendix \ref{appen:snr_cut} for a detailed discussion). Given that the spatial binning scheme performed by Pipe3D requires not only an S/N goal but also a continuity criterion, many of the spatial bins cannot reach the set S/N of 50 due to the balance between these two requirements \citep{Sanchez2016}. Therefore, an additional continuum S/N cut, which we set to 10, is also needed to ensure an unbiased estimate of stellar attenuation (\avstar; \citealt{Sanchez2016a}). The distribution of these selected spaxels on the standard Baldwin--Phillips--Terlevich (BPT) diagram \citep{Baldwin1981}, which is widely used to remove contamination of active galactic nuclei (AGNs) from star-forming regions, is shown in Figure \ref{fig:bpt}. We further select star-forming spaxels based on the \citet{Kauffmann2003a} demarcation. The quality control\footnote{\url{https://data.sdss.org/datamodel/files/MANGA_PIPE3D/MANGADRP_VER/PIPE3D_VER/QC_MaNGA.html}} provided by the Pipe3D VAC is also considered; only galaxies flagged as zero (i.e., do not find any problem) are included in our sample. The above criteria result in more than 450,000 spaxels from 2975 galaxies, which is taken as our final sample. The median and 68\% range of the physical size of these spaxels is $0.20_{-0.12}^{+0.54}$ kpc$^2$.

Recent IFU studies show that there is a subgalactic main-sequence (SGMS) relation between the stellar mass surface density ($\Sigma_{*}$) and the SFR surface density ($\Sigma_{\mathrm{SFR}}$) held down to subkiloparsec scale (e.g., \citealt{Wuyts2013,Cano-Diaz2016,Hsieh2017,Liu2018}). The SFR is computed from the dust-corrected \ha\ luminosity based on the conversion of \cite{Kennicutt1998}. We find that our spaxels form an evident SGMS that shows remarkable agreement with another two SGMSs \citep{Hsieh2017,Liu2018} derived from the MaNGA data of SDSS DR14, which suggests that our sample well represents the population of subgalactic star-forming regions.

The stellar dust attenuation \avstar, which is a product of the continuum modeling process described in Section \ref{subsec:survey_VAC}, is taken from the Pipe3D VAC. Its statistical uncertainty propagated from the uncertainties of the observed spectrum is derived by the Monte Carlo approach during the continuum modeling and is estimated to have a typical value of 0.007 mag. Therefore, the total uncertainty should be dominated by the systematic uncertainty. According to our S/N criterion, we expect a nearly unbiased \avstar\ estimation with a dispersion better than 0.26 mag \citep{Sanchez2016a}. The nebular dust attenuation \avgas\ is a byproduct of the emission line dereddening in which the uncertainty propagation is carefully performed. Under the aforementioned assumptions, the median and 1$\sigma$ dispersion of the statistical uncertainty of \avgas\ are estimated be $0.08_{-0.04}^{+0.07}$ mag.

The stellar and nebular attenuation are both derived on the basis of the \cite{Cardelli1989} extinction curve. The choice of extinction/attenuation curve might affect the resulting attenuation values. However, for \avstar, \cite{Sanchez2016a} claimed that no major differences were found when the \cite{Cardelli1989} extinction curve was replaced by the \cite{Calzetti2000} attenuation law or a $\lambda^{-1.3}$ extinction curve due to the small differences between these extinction/attenuation laws in the optical wavelength range. The replacement of the extinction/attenuation curve in the calculation of \avgas\ only introduces a constant scaled factor. For these reasons, the choice of extinction/attenuation curve could alter the absolute values of \rav, but the relative values, as well as most of our following results, should remain almost unchanged.

\section{Relation between \rav\ and local properties}
\label{sec:resolved}

\subsection{Correlate or Not?}
\label{subsec:if_correlation}

We first examine whether \avgas\ correlates with \avstar. For the whole spaxel population, the Pearson linear and Spearman rank correlation coefficients are 0.40 and 0.34, respectively. However, if the spaxels are divided into bins of \sigha, we find that the correlation coefficients of the binned subsamples strongly change with \sigha. 

    \begin{figure}[htb]
    \centering
    \includegraphics[width=0.5\textwidth]{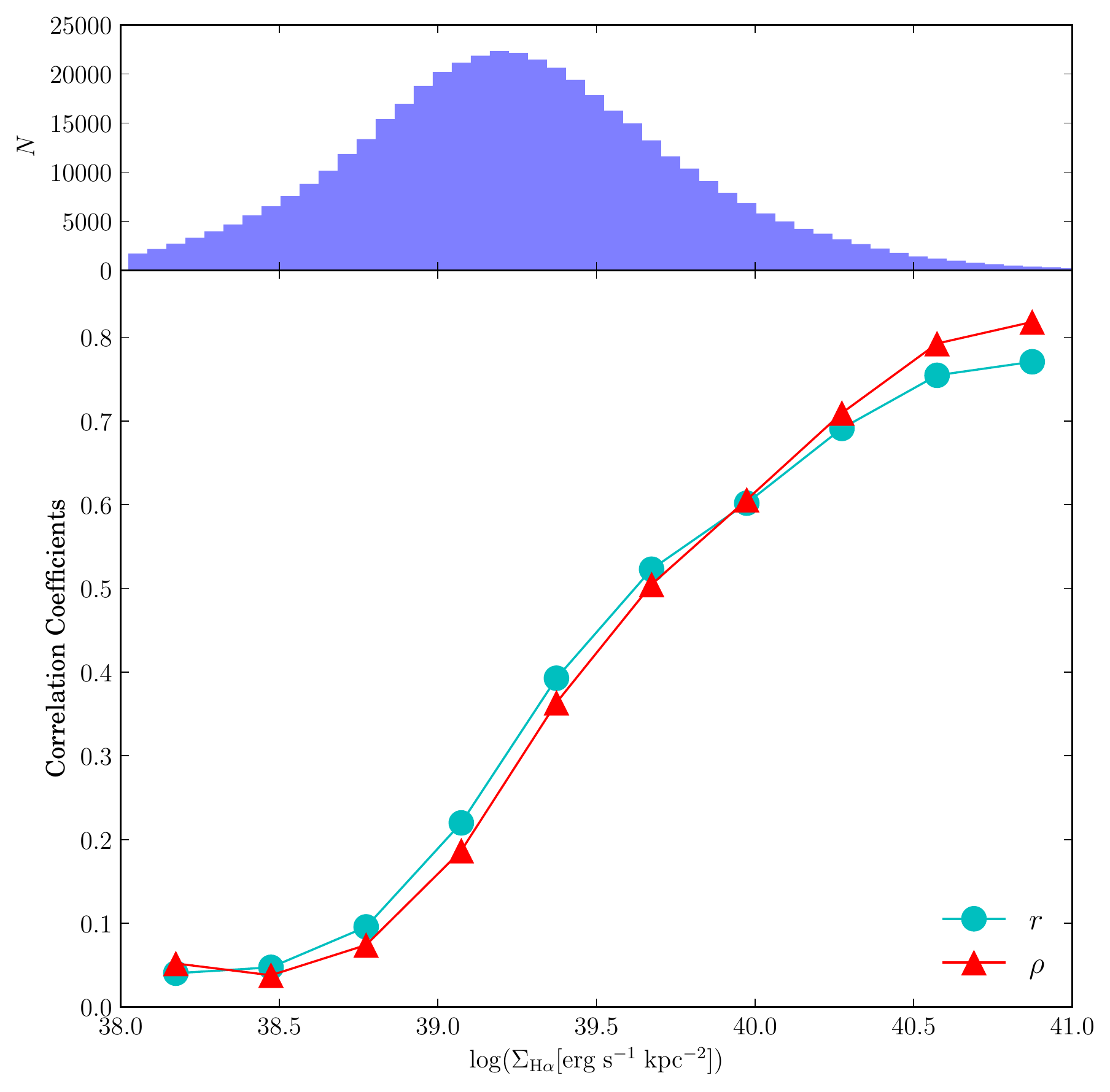}
    \caption{Pearson and Spearman rank correlation coefficients between \avstar\ and \avgas\ as a function of \sigha. The cyan circles denote the Pearson linear correlation coefficients, while the red triangles indicate the Spearman rank correlation coefficients. A histogram of \sigha\ is also given in the top panel.
    \label{fig:cc_sigha}}
    \end{figure}

We show how the Pearson ($r$) and Spearman rank ($\rho$) correlation coefficients between \avstar\ and \avgas\ vary with \sigha\ in Figure \ref{fig:cc_sigha}. Obviously, both correlation coefficients display a substantial increase from $\lesssim0.1$ at the low-\sigha\ end to $\sim0.8$ at the high-\sigha\ end. Thus, the nebular attenuation indeed correlates with the stellar attenuation, although the strength of this correlation strongly correlates with the level of local star formation.

\subsection{Correlation Analysis}
\label{subsec:cc}

In this subsection, we consider how the \rav\ ratio correlates with local physical properties, such as the \ha\ luminosity (i.e., local SFR) and its surface brightness ($\Sigma_{\mathrm{H\alpha}}$, i.e., $\Sigma_{\mathrm{SFR}}$); stellar mass within each spaxel ($M_{*,\mathrm{spaxel}}$) and its surface density ($\Sigma_{*}$); local sSFR and its observational indicator, equivalent width of \ha\ (\ewha); stellar age tracer $D4000$ \citep{BruzualA.1983}; and several emission line indices.

The O3N2 and N2 indices, which are defined as
\begin{equation}
    \mathrm{O3N2}\equiv\log\left(\frac{\oiii\lambda5007}{\mathrm{H\beta}}\times\frac{\mathrm{H\alpha}}{\nii\lambda6584}\right)
\end{equation}
and $\mathrm{N2}\equiv\log(\nii\lambda6584/\mathrm{H\alpha})$, respectively, are widely used as diagnostics of gas-phase metallicity (e.g., \citealt{Pettini2004,Kewley2008,Marino2013}). Moreover, N2O2, defined as $\log(\nii\lambda6574/\oii\lambda3727)$,\footnote{Due to the limited spectral resolution of MaNGA, all $\oii\lambda3727$ mentioned in this work represents the sum of the $\oii\lambda3726,3729$ doublet.} is also considered as an excellent metallicity tracer (e.g., \citealt{Dopita2000,Dopita2013}), especially when the DIG may have a nonnegligible contribution to emission lines \citep{Zhang2017}. However, the empirical calibrations between these indices and oxygen abundance are different among previous works, leading to discrepancies up to $\Delta[\log(\mathrm{O/H})]=0.7$ dex \citep{Kewley2008}. \citet{LinZ2017} also reported an inconsistency between metallicities derived from O3N2 and N2; even the calibrations were derived based on the same data set. For these reasons, here we directly use O3N2, N2, and N2N2 as indicators of metallicity but do not apply any empirical calibrations.

\begin{figure}[htb]
    \centering
    \includegraphics[width=0.5\textwidth]{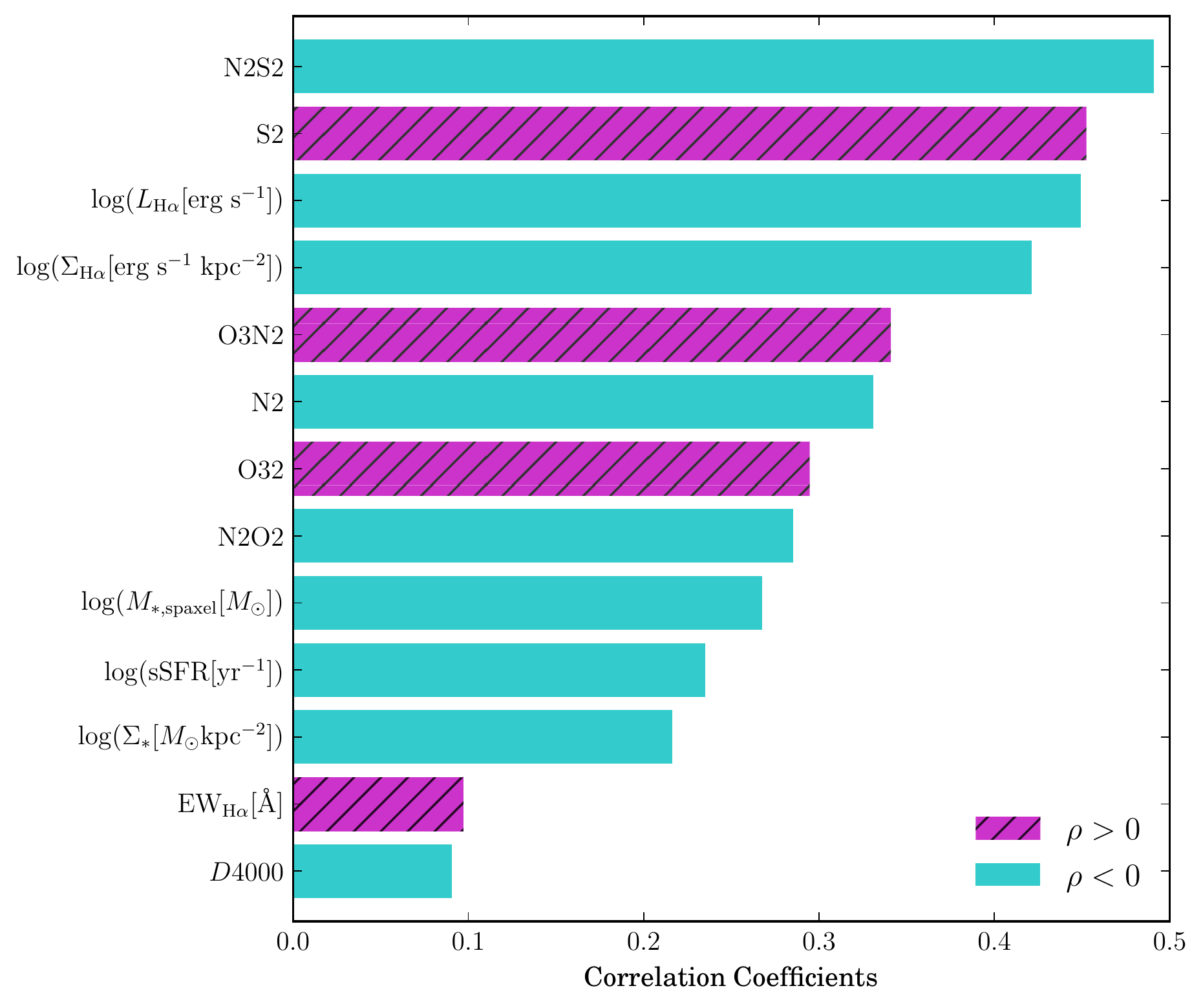}
    \caption{Spearman rank correlation coefficients between the \rav\ ratio and several local properties. The hatched magenta bars present positive correlation coefficients, while the cyan bars indicate the absolute value of negative correlation coefficients. The order of these parameters is sorted by the absolute value of their correlation coefficients.
    \label{fig:cc}}
    \end{figure}

Given that the status of ionized gas may play an important role in determining the \rav\ ratio \citep{Kreckel2013}, we also include DIG indicators in the analysis. Besides the $\sii\lambda6717/\mathrm{H\alpha}$ ratio, $\Sigma_{\mathrm{H\alpha}}$ \citep{Zhang2017} and \ewha\ \citep{Lacerda2018} are also used to separate \hii- and DIG-dominated regions. Furthermore, as suggested by M. Lin et al. (2020, in preparation), the \nii/\ha\ \vs\ \sii/\ha\ diagram can be a useful tool to infer the ionization source of DIGs. Thus, we define $\mathrm{S2}\equiv\log(\sii\lambda6717/\mathrm{H\alpha})$ and $\mathrm{N2S2}\equiv\log(\nii\lambda6584/\sii\lambda6717)$ and add these two indices in our correlation analysis, together with $\mathrm{O32}\equiv\log(\oiii\lambda5007/\oii\lambda3727)$, to distinguish differences in ionization parameter \citep{Baldwin1981,Dopita2013}. Figure \ref{fig:cc} shows the Spearman rank correlation coefficients ($\rho$) between the \rav\ ratio and all of the above parameters.

The DIG indicator of N2S2 gives the strongest correlation with the \rav\ ratio. However, its physical meaning is not clear at present; we defer discussion of this index to Section \ref{subsec:n2s2}. The other two DIG tracers of S2 and $\Sigma_{\mathrm{H\alpha}}$ also present moderate ($|\rho|>0.4$) correlations with the attenuation ratio, which suggest that emission line regions with smaller $\Sigma_{\mathrm{H\alpha}}$ (more diffuse) and higher $\sii\lambda6717$-to-$\mathrm{H\alpha}$ ratio (smaller ionization parameter; e.g., \citealt{Haffner2009}) would have larger \rav\ values. In other words, at fixed \avstar, DIG regions tend to suffer less nebular attenuation with respect to classical \hii\ regions. Such correlations are in good agreement with the result of \cite{Kreckel2013} and can be explained by the two-component dust model of \citet{Charlot2000}. In this model, the \hii-dominated regions are embedded in dense and dustier clouds in which massive stars are born, surrounded by the diffuse ISM, where most of the DIG-dominated regions are located. Irrespective of what provides the ionizing photons of the DIG, leaky \hii\ regions (e.g., \citealt{Haffner2009}), or hot low-mass evolved stars (e.g., \citealt{Flores-Fajardo2011,Zhang2017}), the possible excitation mechanisms of the DIG imply that these regions should be more uniformly mixed with the diffuse ISM.

However, \ewha\ (and sSFR), which is also considered an alternative DIG indicator by \cite{Lacerda2018}, contrary to the other three DIG tracers, shows a much weaker correlation with the attenuation ratio. As elucidated in \cite{Zhang2017}, the dependence of \ewha\ on metallicity makes it a poor tracer of DIG in some cases, resulting in an almost flat distribution together with a very large scatter on the \ewha\ (sSFR) \vs\ \rav\ plane for our sample. By comparison, we present the contour of spaxel density on the $\Sigma_{\mathrm{H\alpha}}$--\rav\ plane in Figure \ref{fig:sigma_rav}. Obviously, the dispersion of spaxels significantly decreases with increasing \sigha, suggesting the \rav\ ratio converges to a constant that close to the value of \citet{Calzetti1997} at the high-\sigha\ end. The different behavior between \sigha\ (or $L_{\mathrm{H\alpha}}$) and \ewha\ implies that the dust attenuation of local star-forming regions is more sensitive to the current star formation activity rather than the past one, which is also implied by the weaker correlation between \rav\ and $M_{*,\mathrm{spaxel}}$ (or $\Sigma_*$).

Moderate correlations are also found for emission line indices of both metallicity and ionization. Given the positive (negative) relation between oxygen abundance and the N2, N2O2 (O3N2) index (e.g., \citealt{Kewley2008,Dopita2013,Marino2013}), the correlations presented in Figure \ref{fig:cc} imply that the ratio between \avstar\ and \avgas\ tends to be smaller in more metal-rich regions. Moreover, regions with a higher ionized level (larger O32) tend to have a larger \rav\ ratio. We will return to these two parameters with a more detailed discussion in Section \ref{subsec:phycon}. The weakest correlation is shown by $D4000$, suggesting that \rav\ is stellar age--insensitive on the physical scale of our spaxels.

The Pearson correlation coefficients between the \rav\ ratio and these properties are also calculated. However, all of them are found to be smaller than the corresponding Spearman rank correlation coefficients, indicating weaker linear correlations between the \rav\ ratio and the explored parameters.

\subsection{Correlation with Physical Condition of Ionized Gas}
\label{subsec:phycon}

As mentioned in Section \ref{subsec:cc}, the physical condition of ionized gas, such as metallicity and ionized level, may help us determine the dust attenuation ratio. However, it is hard to drawn reliable conclusions from Figure \ref{fig:cc} alone due to the small correlation coefficients. On the one hand, although we find moderate correlations between the \rav\ ratio and metallicity indicators, the relation with metallicity still has a large uncertainty because of the nonnegligible role of ionization in emission line ratios \citep{Dopita2013}. For example, \cite{Mao2018} suggested that the radial gradients of O3N2 and N2 indices across \hii\ regions can be attributed to the variations of ionized level rather than the metallicity. On the other hand, there is a contradiction that a larger \rav\ ratio is preferred for highly ionized gas (i.e., large O32 index), as well as DIG, for which a smaller ionization parameter compared to classic \hii\ regions is always suggested (e.g., \citealt{Haffner2009}). To better understand the role of metallicity and ionization parameter in determining the \rav\ ratio and reconciling the seemingly conflicting observations, the connection between these two parameters should be disentangled.

    \subsubsection{N2O2 \vs\ O32 Diagram}
    \label{subsubec:n2o2_o32}

\begin{figure}[htb]
        \centering
        \includegraphics[width=0.475\textwidth]{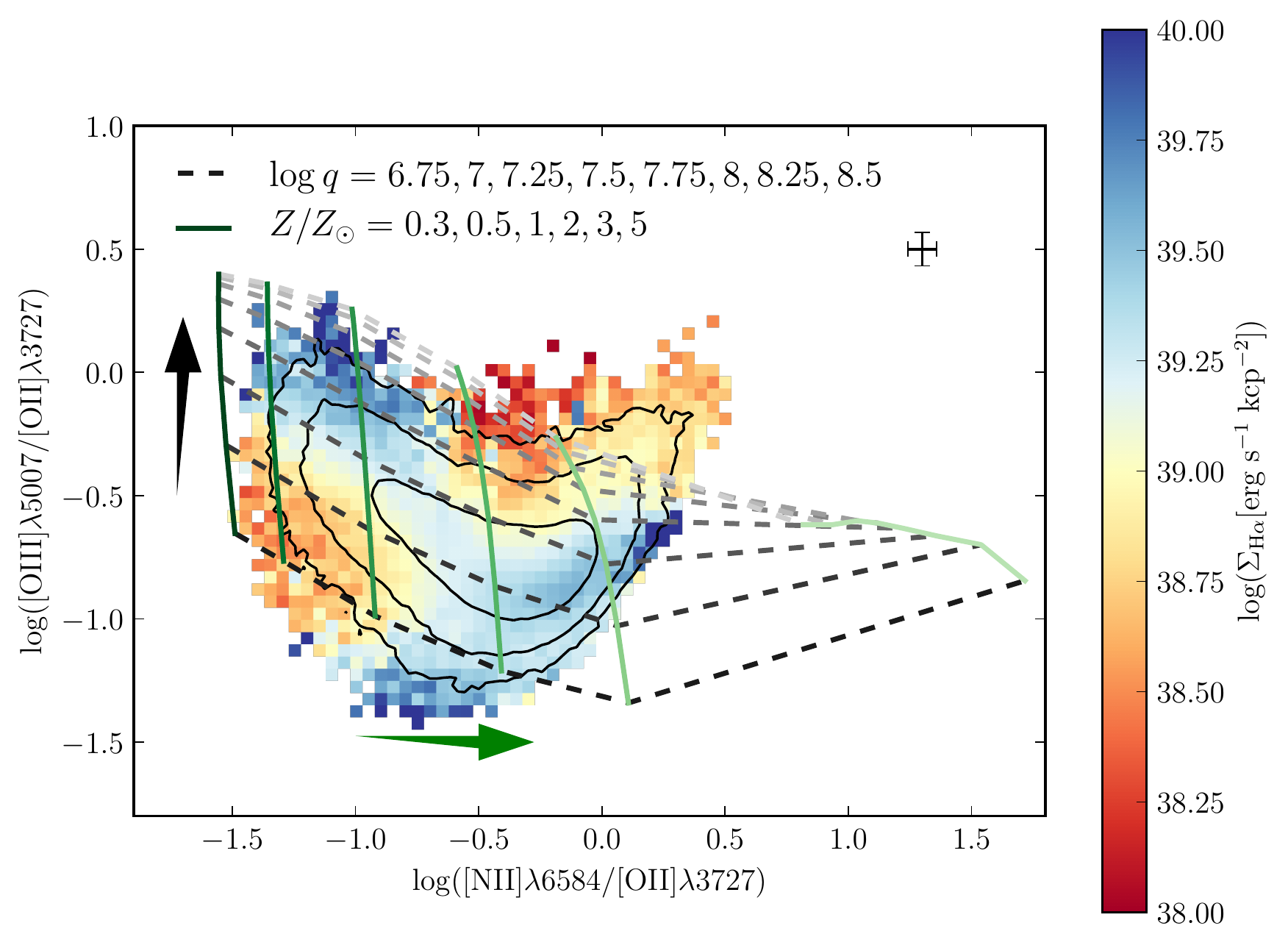}
        \caption{Median map of \sigha\ on the N2O2 \vs\ O32 diagram. The black solid lines indicate contours encompassing 68\%, 95\%, and 99\% of all spaxels. The grid consisting of green solid lines and black dashed lines is an ionization model of the spherical \hii\ region with $\kappa=\infty$ from \cite{Dopita2013}, where the green solid lines represent constant metallicities and the black dashed lines represent constant ionization parameters. The colors of these lines become transparent as the values increase. The horizontal green arrow denotes the direction of increasing metallicity $Z$, and the vertical black arrow denotes the direction of increasing ionization parameter $q$. The values of $Z$ and $\log q$ plotted in this diagram are also listed. The typical uncertainties of N2O2 and O32 are shown in the upper right corner.
        \label{fig:zq_ha}}
        \end{figure}

    We first explore how the DIG tracer (\sigha) distributes in the metallicity--ionization parameter space. The median map of \sigha\ on the N2O2 \vs\ O32 diagram for all spaxels is shown in Figure \ref{fig:zq_ha}, overlapping with density contour (black solid lines). In addition, we also plot the grid of the ionization model of the spherical \hii\ region with $\kappa=\infty$\footnote{Here the $\kappa$ describes how the energy distribution of electrons in \hii\ regions departs from a standard Maxwell--Boltzmann equilibrium energy distribution, while $\kappa=\infty$ represents the Maxwell--Boltzmann distribution \citep{Nicholls2012,Dopita2013}.} from \cite{Dopita2013}, in which emission line fluxes at different metallicities $Z$ and different ionization parameters $\log(q)$ are computed. This diagram clearly separates the metallicity from the ionization parameter and will be suitable for examining the respective effect of abundance or ionization when the other is fixed. As we expected, most of the selected spaxels are covered by the theoretical grid of the \hii\ region, except for a branch perpendicular to the main star-forming cloud at the upper right corner, which is identified as an AGN sequence \citep{Dopita2013}. At the metal-poor end, there is a trend that the median \sigha\ increases with increasing ionization. However, the trend apparently inverts at the metal-rich end, possibly due to the contamination from AGN-dominated spaxels.

    We further separate all spaxels into \hii- and DIG-dominated\footnote{The distribution of all DIG-dominated spaxels on the BPT diagram might extend across the \cite{Kauffmann2003a} boundary \citep{Zhang2017} and covers the composite or even the AGN regime of Figure \ref{fig:bpt}, where it is already excluded by our selection of star-forming spaxels. Thus, the DIG-dominated population in this work is incomplete. The following analysis of DIG-dominated spaxels is restricted to this subsample that resides in the star-forming regime.} populations adopting the criterion of \cite{Zhang2017}, i.e., \hii-dominated spaxels with \sigha\ $>10^{39}~\mathrm{erg~s^{-1}~kpc^{-2}}$ and DIG-dominated spaxels with \sigha\ $<10^{39}~\mathrm{erg~s^{-1}~kpc^{-2}}$. \cite{Kreckel2013} adopted $\sii\lambda6717/\ha=0.2$ to separate the spaxels into these two populations, which is based on observations in the Milky Way \citep{Madsen2006}. However, due to the beam-smearing effect resulting from the low spatial resolution of MaNGA \citep{Zhang2017}, as well as the metallicity dependence of the $\sii\lambda6717/\ha$ ratio, this simple boundary cannot be directly applied to the MaNGA data.

    According to the \sigha\ criterion, the fractions of spaxels that are classified as \hii- and DIG-dominated regions are 68.2\% and 31.8\%, respectively. Two DIG-dominated loci are found in Figure \ref{fig:zq_ha}; one is at the low-$Z$, low-$q$ regime, and the other is the AGN sequence. The distributions of deprojected galactocentric distances of spaxels in both loci span a wide range and peak at $\sim 1.5$ and $\sim 2.5$ kpc for the AGN sequence and the low-$Z$, low-$q$ locus, respectively. In combination with the properties of the underlying low-S/N and low-\sigha\ population shown in Appendix \ref{appen:snr_cut}, we argue that these loci should be predominantly attributed to LI(N)ER-like regions or DIG excited by hot, low-mass, evolved stars. The coverage of the star-forming cloud is in good agreement with the distribution of SDSS galaxies from \cite{Kewley2006} shown in \cite{Dopita2013} and extends from the low-$Z$, high-$q$ end to the high-$Z$, low-$q$ end.

    \subsubsection{Dust attenuations on the N2O2 \vs\ O32 Diagram}
    \label{subsubec:dust_n2o2_o32}

\begin{figure*}[htb]
        \centering
        \includegraphics[width=0.95\textwidth]{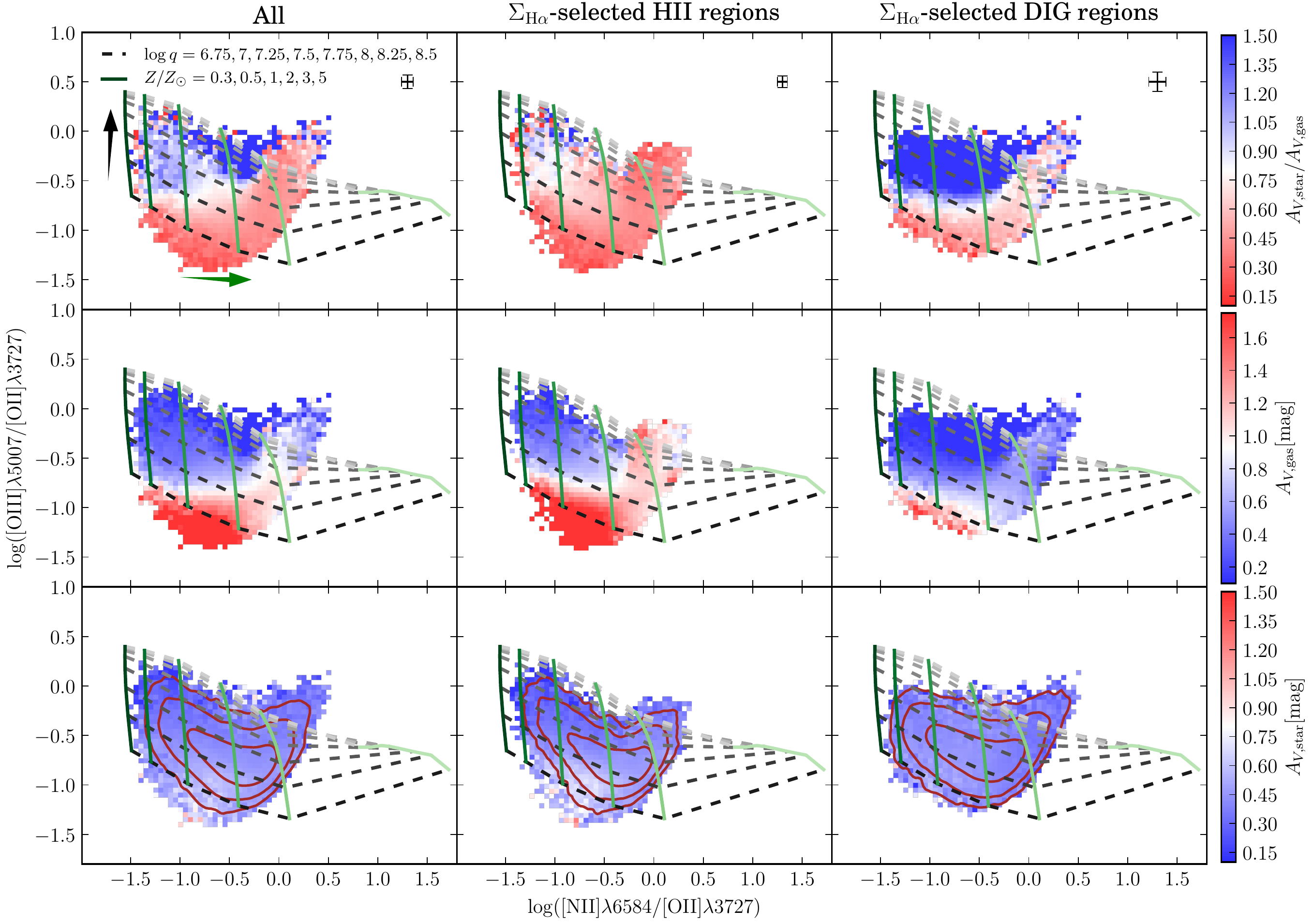}
        \caption{Median maps of \rav\ (top), \avgas\ (middle), and \avstar\ (bottom) on the N2O2 \vs\ O32 diagram. In each case (row), we plot results for all spaxels (left), \hii-dominated regions (middle), and DIG-dominated regions (right). The overlapping $Z$--$q$ grid is the same as Figure \ref{fig:zq_ha}. The contours overplotting in the bottom row enclose 68\%, 95\%, and 99\% of the spaxels on each diagram. The typical uncertainties of N2O2 and O32 for each population are shown at the upper right corner of the top row.
        \label{fig:zq_grid}}
        \end{figure*}

    To see how dust attenuation (and ratio) vary with $Z$ and $q$, we present the median map of \rav, \avgas, and \avstar\ for all of the spaxels and the \hii- and DIG-dominated spaxels in Figure \ref{fig:zq_grid}, overplotting with the theoretical grid of star-forming regions from \cite{Dopita2013}. The medians and 68\% limits of \rav\ are $0.55_{-0.26}^{+0.68}$, $0.48_{-0.22}^{+0.32}$, and $0.93_{-0.56}^{+7.89}$ for all spaxels, \hii-dominated, and DIG-dominated, respectively. As a comparison, \cite{Kreckel2013} found the \rav\ ratios of 0.5 and 0.7 for their selected \hii- and DIG-dominated regions, respectively, which are consistent with ours.

    Although the ionization level of DIG regions is found to be systematically lower compared to that of \hii\ regions, the large overlap of the contours of both subsamples with the theoretical grid in Figure \ref{fig:zq_grid} indicates that both of them cover a similar and wide range in parameter space of either metallicity or ionization. Comparing the subplots in Figure \ref{fig:zq_grid} column by column, the overall features are very similar for all three populations except for the systematic shift of absolute value between them. Comparing the subplots row by row, it is evident that the overall trend of the \rav\ ratio is dominated by that of \avgas, while the \avstar\ maps exhibit a relatively uniform distribution on the N2O2 \vs\ O32 diagram (i.e, the metallicity--ionization parameter grid), regardless of the population.

    The roles of gas-phase metallicity and ionization level in determining \rav\ are more obvious on this plot. For the \rav\ ratio, all three populations show a significantly higher value at the low-$Z$, high-$q$ corner, leading to the moderate correlations between \rav\ and metallicity--ionization parameter found in Section \ref{subsec:cc}. As for \avgas, the spaxels occupying the low-$Z$, high-$q$ corner suffer smaller dust attenuation, indicating that \avgas\ decreases with $Z$ ($\log q$) being lower (higher). This trend might be a geometric result if the low-\avgas\ spaxels are in the outer disks of their host galaxies. We adopt N2O2 $<0$ and O32 $>-0.5$ to select the low-\avgas\ regime and use the deprojected galactocentric distances normalized by the effective radii of host galaxies (i.e., $r/R_{\mathrm{e}}$) to measure their location on the disks. We find that the $r/R_{\mathrm{e}}$ distribution of the low-\avgas\ subsample is similar to that of the whole sample. However, spaxels in this low-\avgas\ regime tend to reside in less massive galaxies compared to the whole sample. Therefore, the trend of the low-\avgas\ subsample with lower metallicity and higher ionization parameter does not result from the geometric effect and is mainly due to the smaller $M_*$ of their host galaxies.

    Intriguingly, the low-$Z$, high-$q$ corner of the DIG-dominated spaxels has \rav\ larger than 1, indicating that the stellar light is more attenuated than the emission lines. Spaxels in this regime only account for nearly 16\% of the DIG-dominated population. Comparing with the \hii-dominated spaxels in the same regime, these DIG-dominated spaxels have comparable \avstar\ but smaller \avgas, suggesting that $A_{V,\mathrm{star}}/A_{V,\mathrm{gas}}>1$ mainly results from a smaller \avgas. The distributions of these spaxels on several emission line diagrams are consistent with the model predictions of leaky \hii\ regions presented in \cite{Zhang2017} (their Figure 18). Meanwhile, the $D4000$ and \ewha\ of these DIG-dominated spaxels are also comparable to normal star-forming regions. Taking into account the relatively higher ionization level of these spaxels, we argue that this DIG might be attributed to the leaky \hii\ region scenario in which ionized gas leaks from the dusty \hii\ regions, while the stellar continuum is dominated by the background stellar populations that are physically unassociated with the leaky \hii\ regions, leading to $A_{V,\mathrm{star}}>A_{V,\mathrm{gas}}$.

    In summary, the smaller value of the ionization parameter of DIG regions with respect to \hii\ regions is a systematic effect, within which they still have a large dispersion in this parameter, resulting in a large scatter of the local \rav\ ratio. Furthermore, the influences of metallicity and ionization level are independent of the classification of spaxels. Metal-poor regions with large $\log q$ tend to have smaller nebular attenuation and thus larger \rav\ values for either the \hii- or DIG-dominated subsample. These two effects may originate from different aspects of dust attenuation, where the former reflects the different stars/dust geometry for \hii- and DIG-dominated populations \citep{Kreckel2013}, and the latter implies the difference in dust properties (e.g., the size distribution of dust grains; \citealt{Relano2018}) or the dust-to-gas ratio ($\mathrm{D/G}\equiv m_{\mathrm{dust}}/m_{\mathrm{gas}}$; \citealt{Engelbracht2008,Galliano2011,Remy-Ruyer2014}) in different local environments.

    \subsubsection{Possible Reason for Correlations between \rav\ and Physical Conditions}
    \label{subsubec:exp_phyconds}

    Although most of the previous works attributed the integrated \rav\ to the geometry effect (e.g., \citealt{Price2014,Reddy2015,Koyama2019}), the conclusions drawn from Figure \ref{fig:zq_grid} imply that the local physical conditions of ionized gas may also have a significant effect on the \rav\ ratio. The dust attenuation at wavelength $\lambda$ ($A_{\lambda}$) along the line of sight $s$ can be written as
    \begin{equation}\label{eq:a_lambda}
        \begin{split}
        A_{\lambda} &= 1.086\tau_{\lambda}^{\mathrm{att}}=1.086\times\int_s\kappa_{\lambda}^{\mathrm{att}}\rho_{\mathrm{dust}}(s)\,\mathrm{d}s\\
                    &= 1.086\times\int_s\kappa_{\lambda}^{\mathrm{att}}\rho_{\mathrm{gas}}(s) \cdot \mathrm{D/G}\,\mathrm{d}s,
        \end{split}
    \end{equation}
    in which $\tau_{\lambda}^{\mathrm{att}}$ is the attenuation optical depth at wavelength $\lambda$, $\kappa_{\lambda}^{\mathrm{att}}$ is the corresponding total opacity (including the absorption and scattering effects), and $\rho_{\mathrm{dust}}(s)$ and $\rho_{\mathrm{gas}}(s)$ are the mass density of dust and gas along the line of sight, respectively. In fact, the term ``extinction'' should be distinguished from ``attenuation.'' The former always refers to dust effects for individual pointlike sources and includes absorption and scattering out of the line of sight, whereas the latter represents the net loss of light for extended emitters, including extinction and scattering into the line of sight in which the stars/dust geometry plays an important role (see \citealt{Calzetti2001} for a review). Therefore, the $\tau_{\lambda}^{\mathrm{att}}$ we used here is something like an effective optical depth rather than the center-to-edge optical depth of a cloud that describes the amount of dust, and the $\kappa_{\lambda}^{\mathrm{att}}$ depends on not only the dust properties but also the stars/dust geometry.

    Here we further express $A_{\lambda}$ as a function of $\rho_{\mathrm{gas}}$ and D/G. In the case of \avstar\ and \avgas, the differences between them that can be explained as geometry effects due to the two-component dust model \citep{Charlot2000,Wild2011,Chevallard2013} are in $\kappa_{\lambda}^{\mathrm{att}}$, $\rho_{\mathrm{gas}}$, and the integrated range $s$. However, variations in dust properties, including in $\kappa_{\lambda}^{\mathrm{att}}$, and D/G can also change $A_{\lambda}$ and thus the \rav\ ratio.

    The metallicity is found to be the main driver of the variation of D/G \citep{Engelbracht2008,Asano2013,Remy-Ruyer2014}, even within individual galaxies (e.g., \citealt{Chiang2018}). Given that the D/G increases with galaxies being more metal-rich, \avgas\ is expected to be higher toward the metal-rich end at fixed $\log q$, which is in good agreement with what we observe in Figure \ref{fig:zq_grid}. However, we should note that the gas-phase metallicity measured here does not directly reflect the metal enrichment of the diffuse ISM in most of the line of sight for classical \hii\ regions (except for the small portion surrounding \hii\ regions). Besides, due to the small $\rho_{\mathrm{gas}}$, the dust accretion is less efficient in the diffuse ISM \citep{Draine2009}. As a result, the impact of the emission line--based metal enrichment on the diffuse ISM is much weaker. Hence, \avstar\ is not significantly altered by observed gas-phase metallicity, and only a very weak trend along constant $\log q$ is exhibited for the \hii-dominated population in Figure \ref{fig:zq_grid}.

    Furthermore, \cite{Galliano2011} found that the D/G decreases with the interstellar radiation field (ISRF) being stronger in the Large Magellanic Cloud, mainly due to the enhanced destruction of dust grains by shock. A similar but much weaker trend is also hinted at by \cite{Relano2018} for M33. For this reason, the enhancement of the ISRF (elevation of $\log q$) would lead to a smaller \avgas, making the low-$Z$, high-$q$ regime more transparent.

    Therefore, the influences of the local physical conditions on either dust properties or D/G are one plausible explanation for the correlations between \rav\ and the metallicity--ionization parameter. The feature of Figure \ref{fig:zq_grid} is consistent with a picture in which less dust is surviving to shield the light in metal-poor ionized gas with a high ionized level; the resultant low nebular attenuation leads to a high \rav\ value.

\subsection{Correlation with Location in BPT Diagrams}
\label{subsec:bpt}

We have shown that the \rav\ value depends on the physical conditions of the ionized gas in a somewhat complex manner. Here we continue to study how this property correlates with the location of spaxels on the BPT diagrams that are always used to identify the excitation mechanism of the ionized gas (e.g., \citealt{Kewley2001,Kauffmann2003a,Kewley2006}).

    \begin{figure*}[htb]
    \centering
    \includegraphics[width=0.9\textwidth]{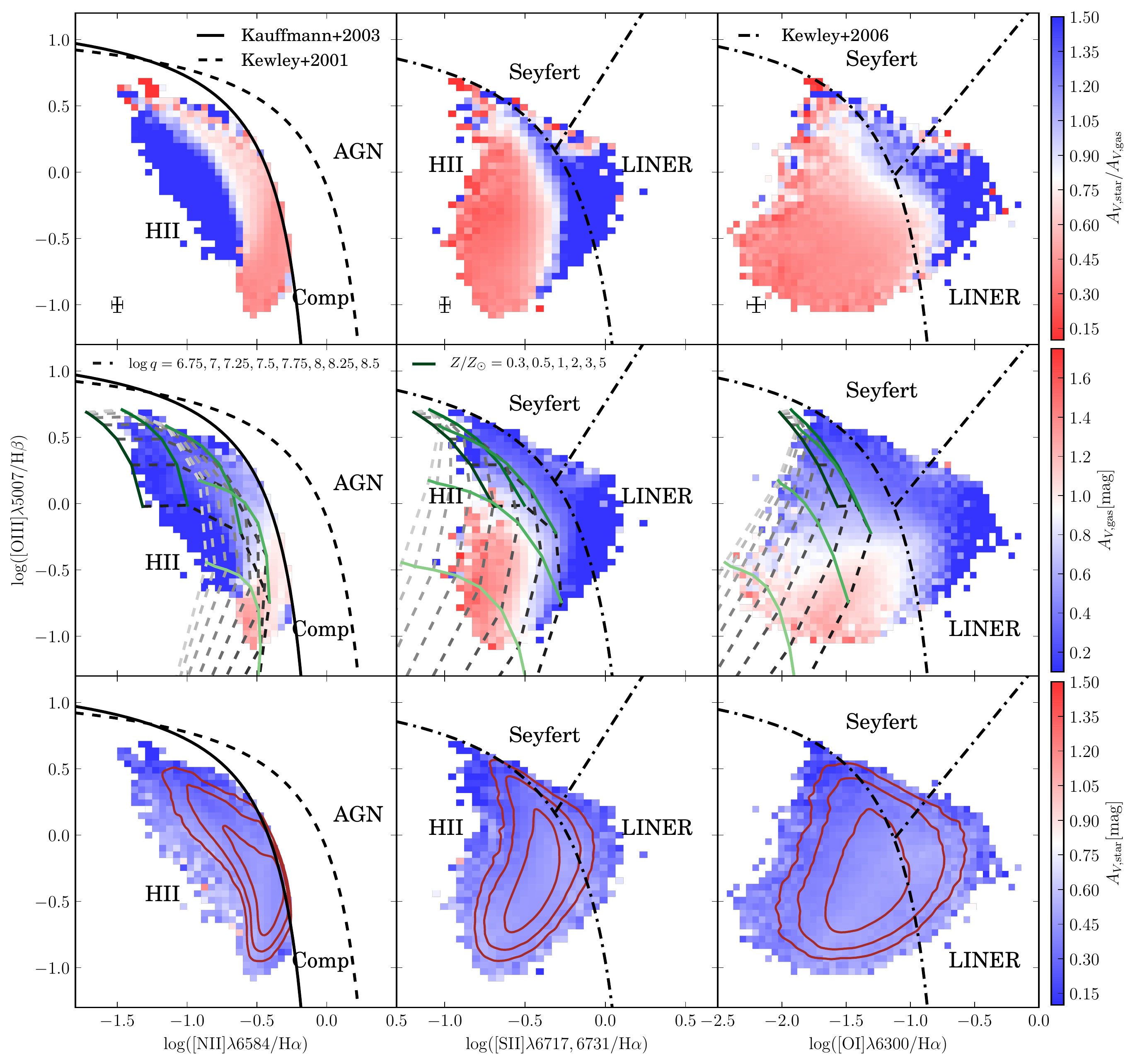}
    \caption{Median maps of \rav\ (top), \avgas\ (middle), and \avstar\ (bottom) on the BPT diagrams for all selected spaxels. On the \nii/\ha\ \vs\ \oiii/\hb\ diagnostic diagrams, the dashed curves indicate the extreme starburst classification line of \citet{Kewley2001}, while the solid curves are the division between pure star-forming and AGN-\hii\ composite regions from \cite{Kauffmann2003a}. On the \sii/\ha\ \vs\ \oiii/\hb\ and \oi/\ha\ \vs\ \oiii/\hb\ diagnostic diagrams, the dashed--dotted curves are demarcations between star-forming regions, LINERs, and Seyfert galaxies from \cite{Kewley2006}. The grid consisting of green solid lines (constant $Z$) and black dashed lines (constant $\log q$) denotes the predicted line ratios of the theoretical photoionization model of a spherical \hii\ region with $\kappa=\infty$ from \cite{Dopita2013}. The colors of these lines become faint as the value increases (see the directions of the arrows in Figure \ref{fig:zq_ha}). The values of $Z$ and $\log q$ plotted in this diagram are also given in the middle row. The contours overplotted in the bottom row encompass 68\%, 95\%, and 99\% of the spaxels on each diagram. The typical uncertainties are shown in the lower left corner of the top row.
    \label{fig:bpt_rav_all}}
    \end{figure*}

    \begin{figure*}[htb]
    \centering
    \includegraphics[width=0.9\textwidth]{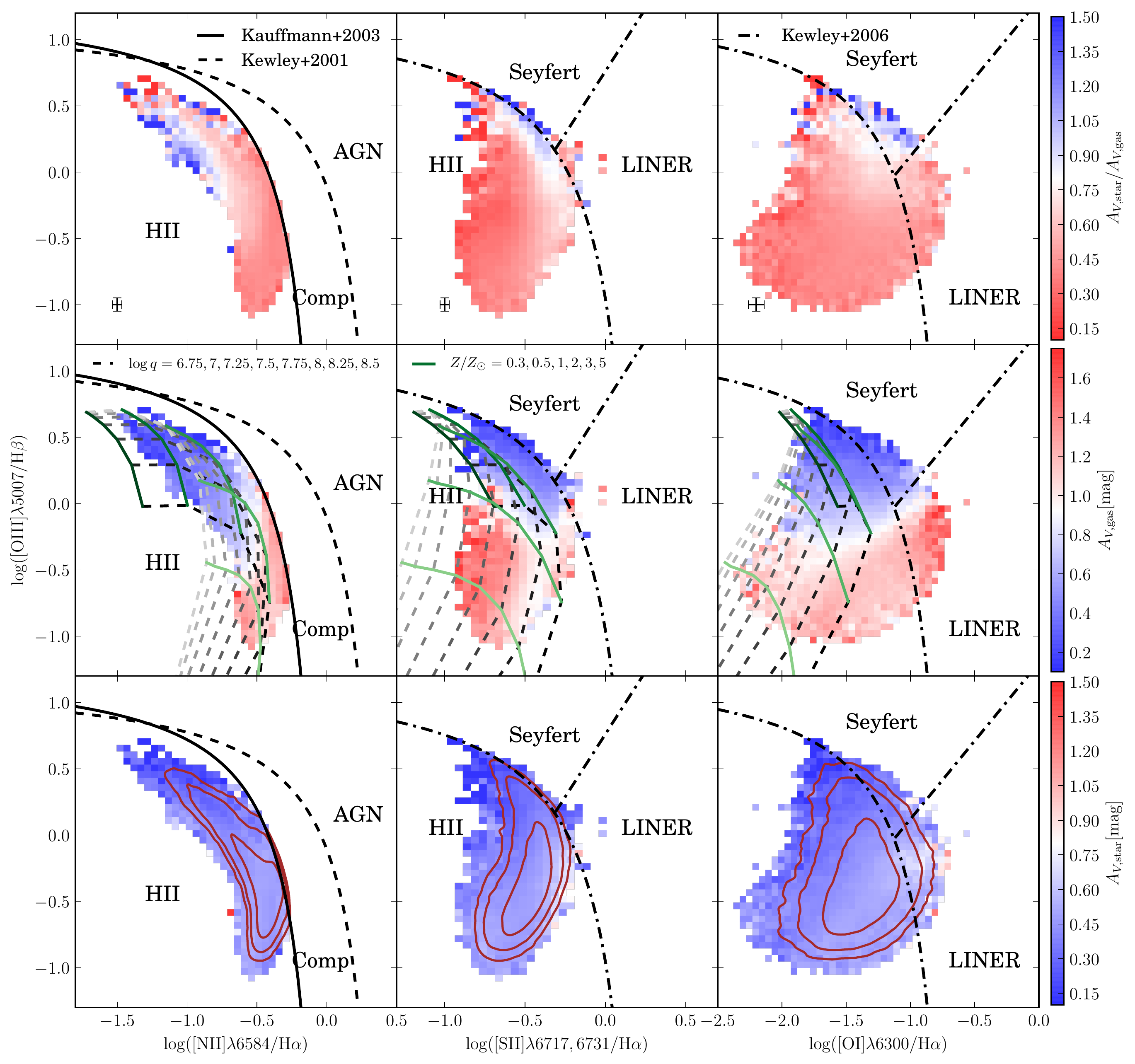}
    \caption{Median maps of \rav\ (top), \avgas\ (middle), and \avstar\ (bottom) on the BPT diagrams for \hii-dominated spaxels. The symbols are the same as in Figure \ref{fig:bpt_rav_all}.
    \label{fig:bpt_rav_hii}}
    \end{figure*}

We present the median maps of \rav, \avgas, and \avstar\ on the BPT diagrams for all selected spaxels in Figure \ref{fig:bpt_rav_all}. Predictions of the theoretical photoionization model of the spherical \hii\ region with $\kappa=\infty$ from \cite{Dopita2013} are also plotted for all three diagnostic diagrams. Although the \nii/\ha--\oiii/\hb\ BPT diagram (hereafter \nii--\oiii\ diagram) combined with the \cite{Kauffmann2003a} boundary is used to exclude spaxels with possible AGN contribution, we still find a small amount of spaxels lying beyond the \cite{Kewley2006} boundaries of star-forming regions on both the \sii/\ha\ \vs\ \oiii/\hb\ and \oi/\ha\ \vs\ \oiii/\hb\ diagrams (hereafter \sii--\oiii\ and \oi--\oiii\ diagram, respectively).

It is very interesting to find that the dust attenuation (and ratio) show some patterns on the BPT diagrams. Similar to the N2O2 \vs\ O32 diagram, the maps of \avstar\ on all three diagrams are fairly smooth, so that the overall trends of the \rav\ ratio are dominated by those of the \avgas, regardless of the diagrams. Notably, there is an intriguing low-\avgas\ regime and a consequent but more distorted high-\rav\ sequence on each diagnostic diagram. The low-\avgas\ and high-\rav\ regime occupies the low-\nii/\ha\ region on the \nii--\oiii\ diagram, the \cite{Kewley2006} boundary on the \sii--\oiii\ diagram, and the upper right region, which also extends across the \cite{Kewley2006} boundary, on the \oi--\oiii\ diagram. On the one hand, the closer locations of the low-\avgas\ sequences to the \cite{Kewley2006} boundaries on both the \sii--\oiii\ and \oi--\oiii\ diagrams imply that the excitation sources of these spaxels may be something other than massive young stars, so that they would be more DIG-like and mixed well with the diffuse ISM. On the other hand, observations from either the Milky Way \citep{Madsen2006} or nearby galaxies (e.g., \citealt{Hoopes2003,Kaplan2016,Zhang2017,Lacerda2018}) revealed enhancement of \sii/\ha\ and \nii/\ha\ ratios in DIG regions compared to classical \hii\ regions, indicating a lower ionization level of DIG. Although this low-\avgas\ population indeed extends to the regime with a high-\sii/\ha\ ratio, it seems to exhibit a smaller \nii/\ha\ ratio. The location of the sequence on the \nii--\oiii\ BPT diagram differs from that of the DIG-dominated regions selected by either low-\sigha\ \citep{Zhang2017} or low-\ewha\ \citep{Lacerda2018}, as well as the theoretical prediction of shocked gas from the photoionization model \citep{Alatalo2016}. We stress that such low \avgas\ at the small-\nii/\ha\ end does not arise from the relatively larger uncertainties of DIG-dominated spaxels, since the S/Ns of both \ha\ and \hb\ for the small-\nii/\ha\ spaxels are slightly higher than those of the large-\nii/\ha\ spaxels if we simply separate the sample by $\log(\nii\lambda6584/\ha)=-0.75$. Therefore, the low-\avgas\ and high-\rav\ sequences cannot be explained by DIG alone. We will show later that these sequences result from the combination of the DIG-dominated and metal-poor spaxels in which the latter ones lead to the low-\nii/\ha\ values on the \nii--\oiii\ diagram.

Moreover, the theoretical grids of both the \nii--\oiii\ and \sii--\oiii\ diagrams are able to cover most of the spaxels in our sample, whereas the grid on the \oi--\oiii\ diagram only occupies about half of the observed regime. As elucidated in \cite{Dopita2013}, their calculations of the \oi\ line are much less reliable than those of the \nii\ and \sii\ lines due to the fact that the \oi\ emission comes out of a very narrow zone near the ionization front in which shocks may be an important heating source \citep{Dopita1997}. Thus, the \oi--\oiii\ grid should be treated with caution. As can be seen in Figure \ref{fig:bpt_rav_all}, the low-\avgas\ and high-\rav\ sequences on these diagrams are all overlaid by the low-$Z$ ends of the corresponding grids. In Section \ref{subsec:phycon}, we already showed that metal-poor spaxels tend to have smaller \avgas\ and higher \rav\ ratios, irrespective of the populations we defined. Accordingly, metallicity may be another main factor to explain the location of the low-\avgas\ and high-\rav\ sequences in the BPT diagrams.

To further separate the effects of DIG and metallicity, we plot similar median maps for the \hii-dominated subsample in Figure \ref{fig:bpt_rav_hii}. The most significant difference between Figure \ref{fig:bpt_rav_all} and Figure \ref{fig:bpt_rav_hii} is the truncation of the low-\avgas\ and high-\rav\ sequences on the \sii--\oiii\ and \oi--\oiii\ diagrams. The removal of DIG contamination leads to the disappearance of the sequence at the low-\oiii/\hb, high-\sii/\ha\ (\oi/\ha) regime on the \sii--\oiii\ (\oi--\oiii) diagram, indicating the low ionization state of DIG (e.g., \citealt{Hoopes2003,Haffner2009,Kaplan2016,Zhang2017}). The remaining blue sequences in the \hii-dominated population can be well described by the low-$Z$ end of the theoretical grids, even on the ``unreliable'' \oi--\oiii\ diagram. Interestingly, the extended region of the \hii-dominated subsample is now well constrained by the \cite{Kewley2006} boundary on the \sii--\oiii\ diagram.

In short, DIG contamination alone cannot explain the low-\avgas\ and high-\rav\ sequences found on the \sii--\oiii\ and \oi--\oiii\ diagrams; metal-poor spaxels seem to be the origin of the remaining half of the sequences after removing DIG-dominated regions. This result reinforces the conclusions drawn from the analysis of the metallicity and ionization parameter in Section \ref{subsec:phycon}.

\subsection{N2S2--\rav\ Correlation}
\label{subsec:n2s2}

\begin{figure*}[htb]
    \centering
    \includegraphics[width=0.8\textwidth]{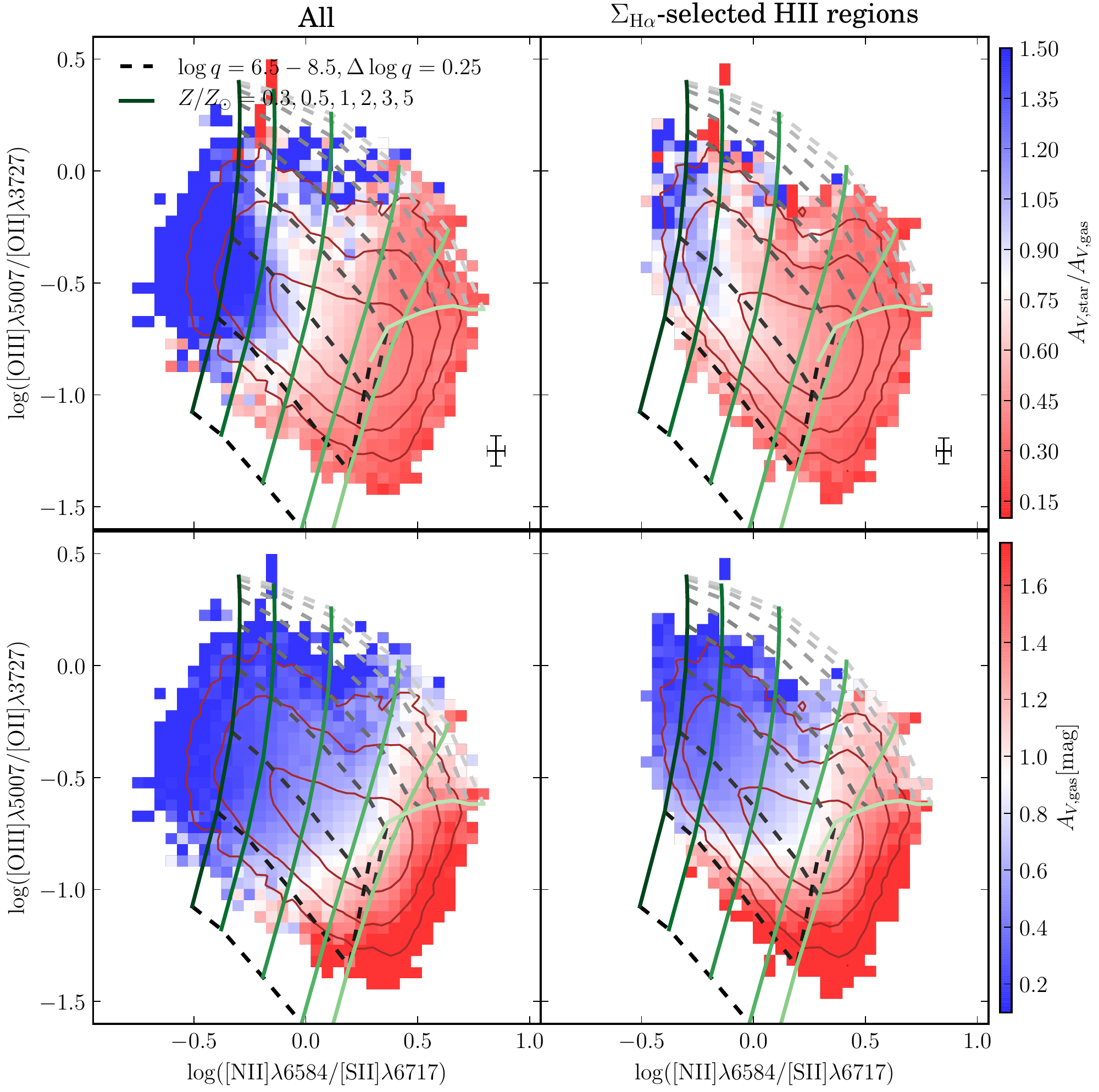}
    \caption{Median maps of \rav\ (top) and \avgas\ (bottom) on the N2S2--O32 diagram for all (left) and the \hii-dominated (right) spaxels. In each panel, the brown contours encompass 68\%, 95\%, and 99\% of the spaxels, respectively, while the grid denotes the predicted line ratios of the theoretical photoionization model of a spherical \hii\ region with $\kappa=\infty$ from \cite{Dopita2013}. The typical uncertainties are shown in the lower right corner.
    \label{fig:n2s2_rav}}
    \end{figure*}

In Section \ref{subsec:cc}, we find that the N2S2 index has the strongest correlation with the \rav\ ratio among the local properties we explored. We use the predictions of the theoretical model of \hii\ regions from \cite{Dopita2013} to examine how the N2S2 index varies with metallicity and ionization parameter and find that this index increases toward high-$Z$ (high-$q$) end when $q$ ($Z$) is fixed within the adopted parameter ranges (i.e., $Z=0.1Z_{\odot}-5Z_{\odot}$ and $\log q=6.5-8.5$). Observations of ionized gas also unveiled that DIG tends to have a smaller N2S2 ratio relative to the average value of traditional \hii\ regions excited by massive OB stars due to the softer ionized field \citep{Madsen2006,Haffner2009}. Namely, the N2S2 ratio is sensitive to both metallicity and ionization parameter; we thus use the N2S2--O32 diagrams to separate the effect of the ionization parameter. In Figure \ref{fig:n2s2_rav} we present the median maps of \rav\ and \avgas\ on the N2S2--O32 diagram for all spaxels and the \hii-dominated ones. Theoretical predictions of the ionization model of \hii\ regions from \cite{Dopita2013} are also plotted. The median map of \avstar\ is not shown here due to the fact that, similar to the maps in other emission line ratio diagrams (e.g., Figure \ref{fig:zq_grid} and Figure \ref{fig:bpt_rav_all}), it exhibits a uniform distribution across the diagram and does not reveal any more useful information.

The observed trends of \rav\ are dominated by those of \avgas. With the help of the O32 index, the role of the ionization parameter is well separated. We find that the N2S2 index decreases toward the low-metallicity end at fixed $\log q$, leading to a larger \rav\ (smaller \avgas). Although DIG tends to have smaller N2S2 values, the removal of the DIG-dominated population cannot totally flatten the trend between the N2S2 index and \rav, while the correlation between the N2S2 index and \avgas\ is still significant ($\rho=0.57$) for the \hii-dominated population. Therefore, we ascribe the emergence of the N2S2--\rav\ correlation to the combination of DIG contamination (i.e., geometry effect) and the small N2S2 values of metal-poor spaxels.

\subsection{Effect of Spatial Resolution}
\label{subsec:spatial_resolution}

The two-component dust model, which we use to describe the role of geometry in determining the expected relation between \avstar\ and \avgas, presumes that all line emission arises from \hii\ regions \citep{Charlot2000}. However, this presumption might not be true for all of our spaxels. Although the physical sizes of our spaxels have a median of 0.20 kpc$^2$, the reconstructed PSF of the MaNGA data cubes has an FWHM of $\sim2\farcs5$ \citep{Law2016}, corresponding to a physical size of $\sim1.5$ kpc at the mean redshift of the MaNGA targets ($\langle z\rangle=0.03$; \citealt{Bundy2015}). This scale is significantly larger than the typical size of the \hii\ regions \citep{Hunt2009,Lawton2010}, creating a problem regarding how the low spatial resolution of the MaNGA data cubes impacts the measured dust attenuations.

Based on a multiscale study of M33, \cite{Boquien2015} found that both nebular ($A_{\mathrm{H\alpha}}$) and stellar ($A_{\mathrm{FUV}}$) attenuation are scale-dependent. As the resolution downgrades (from 33 to 2048 pc), the dynamic ranges of both attenuations and the $A_{\mathrm{FUV}}/A_{\mathrm{H\alpha}}$ ratio become smaller, indicating a smooth effect at coarser resolutions where intense \hii\ regions and quiescent regions merge.

For our \hii-dominated spaxels, since \avstar\ is derived from the overall shape of the underlying stellar continuum and is not closely linked to the young stellar populations compared to $A_{\mathrm{FUV}}$ computed in \cite{Boquien2015}, the smooth effect on our stellar attenuation should be weaker than the one reported by these authors. However, the surrounding more diffuse, less attenuated regions might have a nonnegligible contribution to the emission line measurements of the spaxels, possibly leading to a smaller \avgas, on average, and thus a slightly larger \rav. Given that the systematic differences between \avstar\ and \avgas\ are still observed for the \hii-dominated population, we conclude that the smooth effect stemming from the coarse spatial resolution is too weak to eliminate the geometry effect of the two-component dust model. Therefore, the large PSF size of the MaNGA data has little impact on the comparison with the dust model, as well as most of the conclusions drawn from this section.

\subsection{The Smooth \avstar\ Maps}
\label{subsec:smooth_avstar}

Since the derivation of \avstar\ is independent of emission lines, the relatively uniform distribution of \avstar\ on the (gas-phase) metallicity--ionization parameter space (Figure \ref{fig:zq_grid}) and the BPT diagrams might be an effect of the methodology. However, it is noteworthy that the overall trends of \avgas\ observed in Figure \ref{fig:zq_grid} still exist in the maps of \avstar\ but are much weaker. Particularly, the median of \avstar\ at the low-$Z$, low-$q$ regime is slightly higher for the \hii-dominated population, which is similar to the case of \avgas.

Another effect that might contribute to the uniformity of \avstar\ is the smooth effect arisen from the large PSF size of the MaNGA data. A multiscale study of M33 suggested that either stellar or nebular attenuation suffers a smooth effect at some level \citep{Boquien2015}. As discussed in Section \ref{subsec:spatial_resolution}, the ranges of both attenuations become narrower as the resolution becomes coarser. However, the median maps of \avgas\ on either the N2O2 \vs\ O32 diagram or the BPT diagrams (Figure \ref{fig:bpt_rav_all}) are not as smooth as those of \avstar, implying that the coarse resolution should not be the main reason for the uniform \avstar\ distribution. Thus, we argue that the observed uniform distribution of \avstar\ indeed has a physical origin, indicating that the stellar attenuation is less relevant to the physical conditions of the ionized gas. In fact, as implied by the two-component dust model \citep{Charlot2000,Wild2011}, \avstar\ might reflect the local environment of the diffuse ISM along the line of sight rather than the birth clouds. Studies of young star clusters (YSCs) confirmed that YSCs with ages of $\sim10$ Myr are no longer embedded in their birth clouds \citep{Whitmore2014,Hollyhead2015,Grasha2019}. In other words, stellar populations older than this time scale might only suffer dust attenuation from the diffuse ISM.

Due to the lack of a second method to compute \avstar\ at present, we still cannot rule out the possibility that the derivation method has contributed to flattening the trends in the \avstar\ maps.

\section{Relation between \rav\ and Global Properties}
\label{sec:global}

\subsection{Determining the Global \rav}
\label{subsec:global_rav}

\begin{figure}[htb]
    \centering
    \includegraphics[width=0.475\textwidth]{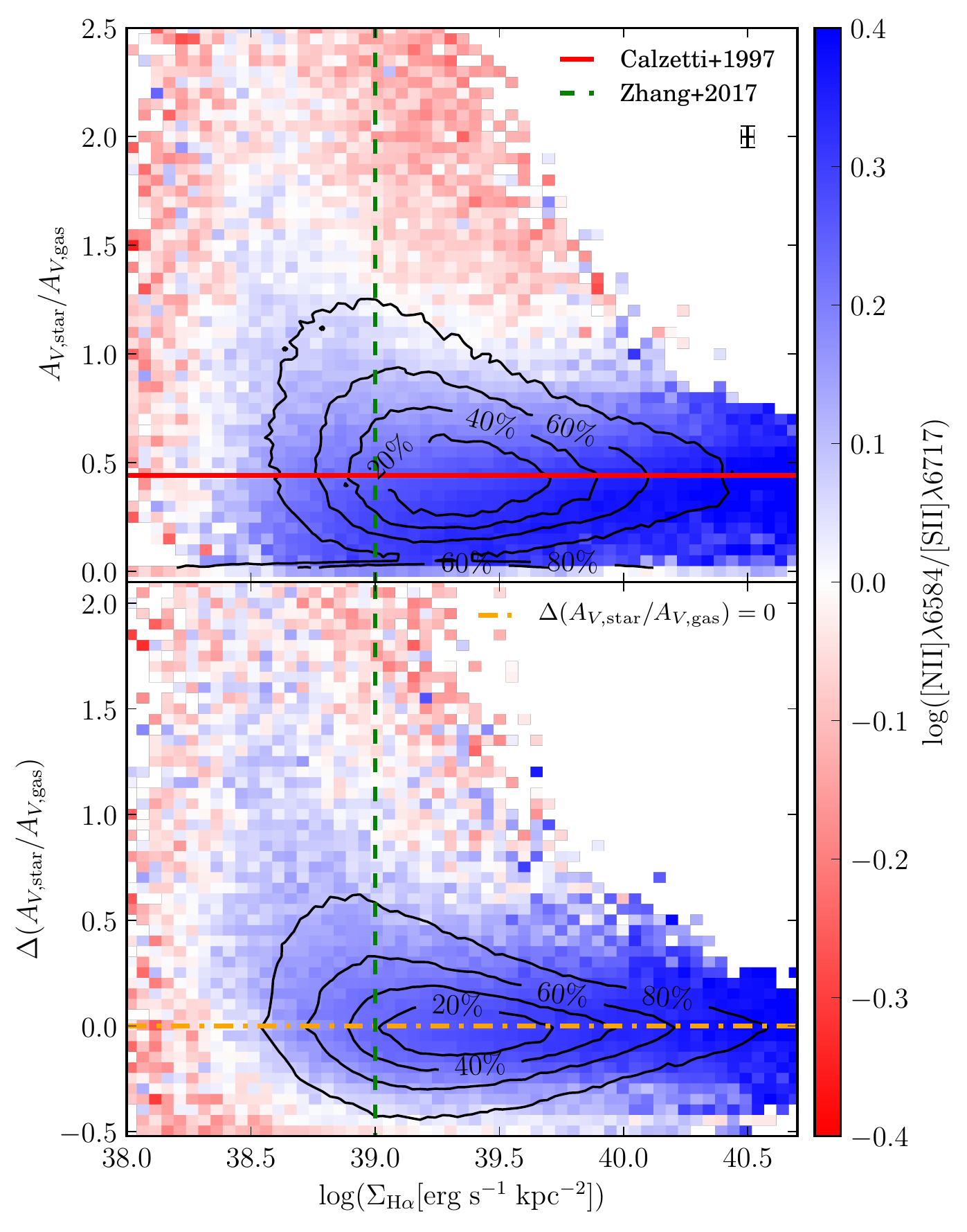}
    \caption{Density contours of all spaxels in the \rav--\sigha\ (upper panel) and $\Delta$(\rav)--\sigha\ (bottom panel) planes. The contours contain 20\%, 40\%, 60\%, and 80\% of the spaxels, respectively. The underlying maps are color-coded by the median value of $\log(\hii\lambda6584/\sii\lambda6717)$. The red solid line is the canonical value of $A_{V,\mathrm{star}}/A_{V,\mathrm{gas}}=0.44$ from \cite{Calzetti1997}, while the orange dashed--dotted line indicates $\Delta(A_{V,\mathrm{star}}/A_{V,\mathrm{gas}})=0$. The vertical green dashed lines show the criterion of \hii-dominated spaxels from \cite{Zhang2017}, i.e., $\Sigma_{\mathrm{H\alpha}}=10^{39}~\mathrm{erg~s^{-1}~kpc^{-2}}$. The typical uncertainties are shown in the upper right corner of the top panel. Only statistical uncertainties are plotted.
    \label{fig:sigma_rav}}
    \end{figure}

As shown in \cite{Kreckel2013}, the \rav\ ratio tends to converge to a constant at the high-\ha\ flux end within individual galaxies. After visual inspection of the \rav\ \vs\ \sigha\ plot of each galaxy, we find not only the similar trend presented in \cite{Kreckel2013} but also a galaxy-to-galaxy variation of the convergent value; i.e., the global \rav\ may vary between galaxies. To demonstrate this feature, we plot the density contours of all spaxels in the \rav--\sigha\ plane in the upper panel of Figure \ref{fig:sigma_rav}. Since the convergent value may be a variable between galaxies, we calculate a representative value of \rav\ via $\sigma$-clipped statistics. The median of the $3\sigma$ clipping after three iterations is taken as the global value of \rav\ for each galaxy. Inspection of the histogram of \rav\ within individual galaxies confirms that this parameter can be a good estimate of the peak of the distribution. Then the global \rav\ is subtracted from the individual ones within the corresponding galaxy to obtain $\Delta$(\rav). The density distribution of spaxels in the $\Delta$(\rav)--\sigha\ plane is given in the bottom panel of Figure \ref{fig:sigma_rav}.

Both \rav\ and $\Delta$(\rav) exhibit great reductions in scatters toward high \sigha, indicating a common geometry of dust/star and local physical conditions for high-\sigha\ spaxels. The convergent \rav\ of 0.425, resulting from a peak detection utilizing the PeakUtils package\footnote{\url{https://peakutils.readthedocs.io/en/latest/index.html}} based on the histogram of \rav, is slightly smaller than the one of \cite{Calzetti1997}. Comparison between the two sets of contours in Figure \ref{fig:sigma_rav} reveals that the scatter of $\Delta$(\rav) is further reduced after normalization by the global \rav\ of each galaxy. This result implies that the global values of \rav\ indeed vary between galaxies, which was already suggested by many previous works on either local (e.g., \citealt{Wild2011,Koyama2015,Zahid2017,Koyama2019}) or high-redshift (e.g., \citealt{Price2014,Puglisi2016}) galaxies.

\subsection{Observational Evidence of Correlations between \rav\ and Other Properties}
\label{subsec:observational_evidences}

Naturally, one would wonder whether the global \rav\ correlates with other physical properties of galaxies if it is thought of as a variable. Previous works demonstrated that this ratio may correlate with $M_*$ (e.g., \citealt{Koyama2015,Koyama2019,Puglisi2016,Zahid2017}), SFR (e.g., \citealt{Reddy2015}), sSFR (e.g., \citealt{Wild2011,Price2014,Koyama2019}), inclination \citep{Wild2011}, and so on. Making use of the Pipe3D VAC and the NASA-Sloan Atlas (NSA) catalog used by the MaNGA survey,\footnote{\url{https://www.sdss.org/dr15/manga/manga-target-selection/nsa/}} we calculate the Pearson correlation coefficients $r$ and the Spearman rank correlation coefficients $\rho$ between the global \rav\ ratio and several physical properties and list them in Table \ref{tab:cc}. Taking the medians of \avstar\ and \avgas\ of spaxels within individual galaxies as the corresponding global values, the correlation coefficients between these global attenuations and other physical properties are also given in Table \ref{tab:cc}. Note that these ``global'' dust attenuations (also the ``global'' \rav\ defined in Section \ref{subsec:global_rav}) are the typical values of local \hii\ regions within galaxies, which might be different from those derived from the integrated light. The global SFR is obtained by summing up the \ha\ luminosities over the field of view (FOV) and applying the \cite{Kennicutt1998} conversion. The metallicities are taken from the Pipe3D VAC, while $M_*$ and axis ratio $b/a$ are from the NSA catalog after correcting for the differences in the adopted cosmology and IMF. Although the Pipe3D VAC provides a global stellar mass of galaxies, it is found to be slightly smaller at the high-mass end due to the limited FOV of MaNGA. The sSFR is calculated by $\mathrm{SFR}/M_*$ using the adopted parameters. Here \oh\ is measured at the effective radius \re\ to represent the global oxygen abundance. Studies based on the CALIFA survey suggested that this parameter has a pretty good one-to-one relation with the average oxygen abundance over the entire FOV but with a smaller uncertainty, regardless of metallicity calibrations (\citealt{Sanchez2013,Sanchez2017}). To ensure a robust estimate of the global \rav, we only include galaxies with more than 50 selected spaxels in the following analysis. In total, 1949 galaxies remain after this selection. We stress that changing this critical number of spaxels cannot significantly alter the following results.

    \begin{deluxetable*}{cCCCCCC}[htb]
    \tablecaption{Correlation Coefficients between Global Dust Attenuation (Ratio) and Some Physical Properties\label{tab:cc}}
    \tablehead{
    \colhead{Parameters\tablenotemark{a}} & \colhead{$\log(M_*)$\tablenotemark{b}} & \colhead{$\log(\mathrm{SFR})$} & \colhead{$\log(\mathrm{sSFR})$} & \colhead{\oh\tablenotemark{c}} & \colhead{\oh\tablenotemark{c}} & \colhead{$b/a$\tablenotemark{d}} \\
    & \colhead{($M_{\odot}$)} & \colhead{($M_{\odot}~\mathrm{yr^{-1}}$)} & \colhead{($\mathrm{yr^{-1}}$)} & \mathrm{(O3N2)} & \mathrm{(N2)} &
    }
    \startdata
    \rav & -0.66 & -0.59 & 0.18 & -0.59 & -0.62 & -0.27\\
    & -0.72 & -0.62 & 0.24 & -0.60 & -0.63 & -0.32\\
    \avstar & -0.12 & -0.08 & 0.06 & 0.11 & 0.14 & -0.54\\
    & -0.13 & -0.09 & 0.07 & 0.05 & 0.05 & -0.55\\
    \avgas & 0.49 & 0.46 & -0.10 & 0.59 & 0.62 & -0.23\\
    & 0.54 & 0.50 & -0.15 & 0.62 & 0.65 & -0.21\\
    \enddata
    \tablenotetext{a}{For each parameter, the first row lists the Pearson correlation coefficients $r$, while the second row presents the Spearman rank correlation coefficients $\rho$.}
    \tablenotetext{b}{Stellar mass from $K$-correction fit for S\'{e}rsic fluxes, taken from the NSA catalog.}
    \tablenotetext{c}{Oxygen abundance \oh\ at the effective radius derived based on the O3N2 and N2 calibrations of \citet{Marino2013}.}
    \tablenotetext{d}{Axis ratio $b/a$ from the two-dimensional, single-component S\'{e}rsic fit in the $r$ band, taken from the NSA catalog.}
    \end{deluxetable*}

    \begin{figure}[htb]
    \centering
    \includegraphics[width=0.475\textwidth]{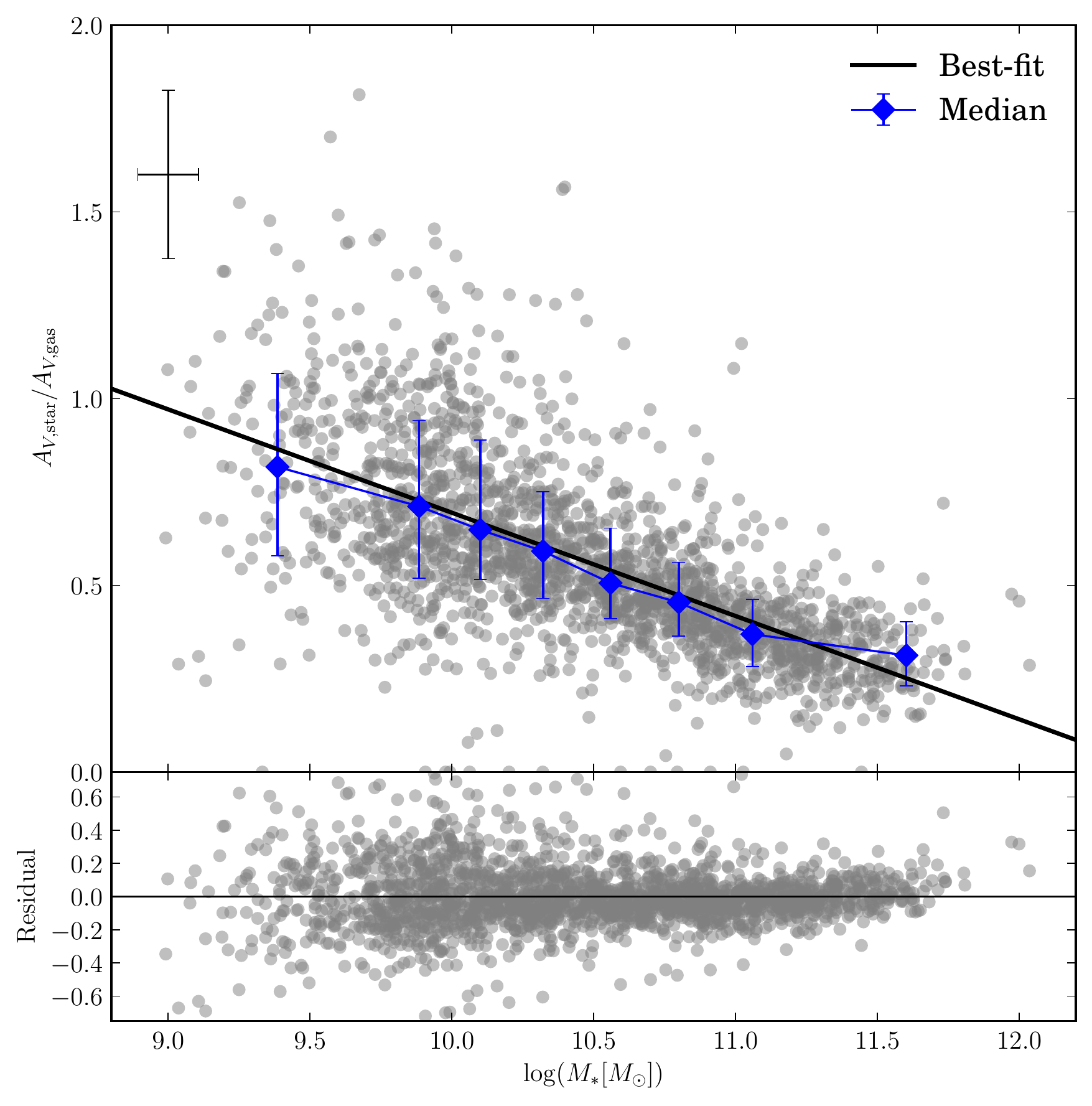}
    \caption{Global \rav\ ratio as a function of global stellar mass ($M_*$) of galaxies. The black solid line shows the best-fit result with a linear function for all galaxies. The upper panel shows the \rav\ \vs\ $M_*$ distribution for individual galaxies, while the bottom panel presents the residuals between the observed and the predicted \rav\ from the best-fit line. The blue diamonds with error bars indicate the medians and corresponding $1\sigma$ scatters (i.e., 16\%-84\% range) of the binned distributions for all galaxies. These bins are adjusted to equalize the number of galaxies in each bin for each set of bins. The typical uncertainties are denoted in the upper left corner.
    \label{fig:m_rav}}
    \end{figure}

From Table \ref{tab:cc}, it is clear that $M_*$ presents the strongest correlation with the global \rav\ in either $r$ or $\rho$, implying smaller \rav\ values for more massive galaxies. We show the global \rav\ as a function of $M_*$ for the selected galaxies in Figure \ref{fig:m_rav}, together with the best-fit result for all galaxies using a linear function in the form of
\begin{equation}
\label{eq:linear}
    A_{V,\mathrm{star}}/A_{V,\mathrm{gas}} = a_0 + a_1 x,
\end{equation}
in which $x=\log (M_*/M_{\odot})$. The typical uncertainty of \rav\ shown in the figure is the median of the standard deviation of \rav\ within each galaxy. Since the NSA catalog does not contain a formal uncertainty for $M_*$, we give a rough estimation based on the comparison of $M_*$ from \cite{Kauffmann2003} and the kcorrect code from which the NSA $M_*$ was calculated \citep{Blanton2007}.

The best-fit result (black solid line in Figure \ref{fig:m_rav}) is given by $a_0=3.460\pm0.075$, $a_1=-0.277\pm0.007$, and is in good agreement with the binned median curve. The scatter of residuals is $M_*$-dependent, and more massive galaxies have smaller scatters. Comparing the observations with the predictions from the best-fit curve, the standard deviation of the residuals is 0.19.

    \begin{figure}[ht]
    \centering
    \includegraphics[width=0.475\textwidth]{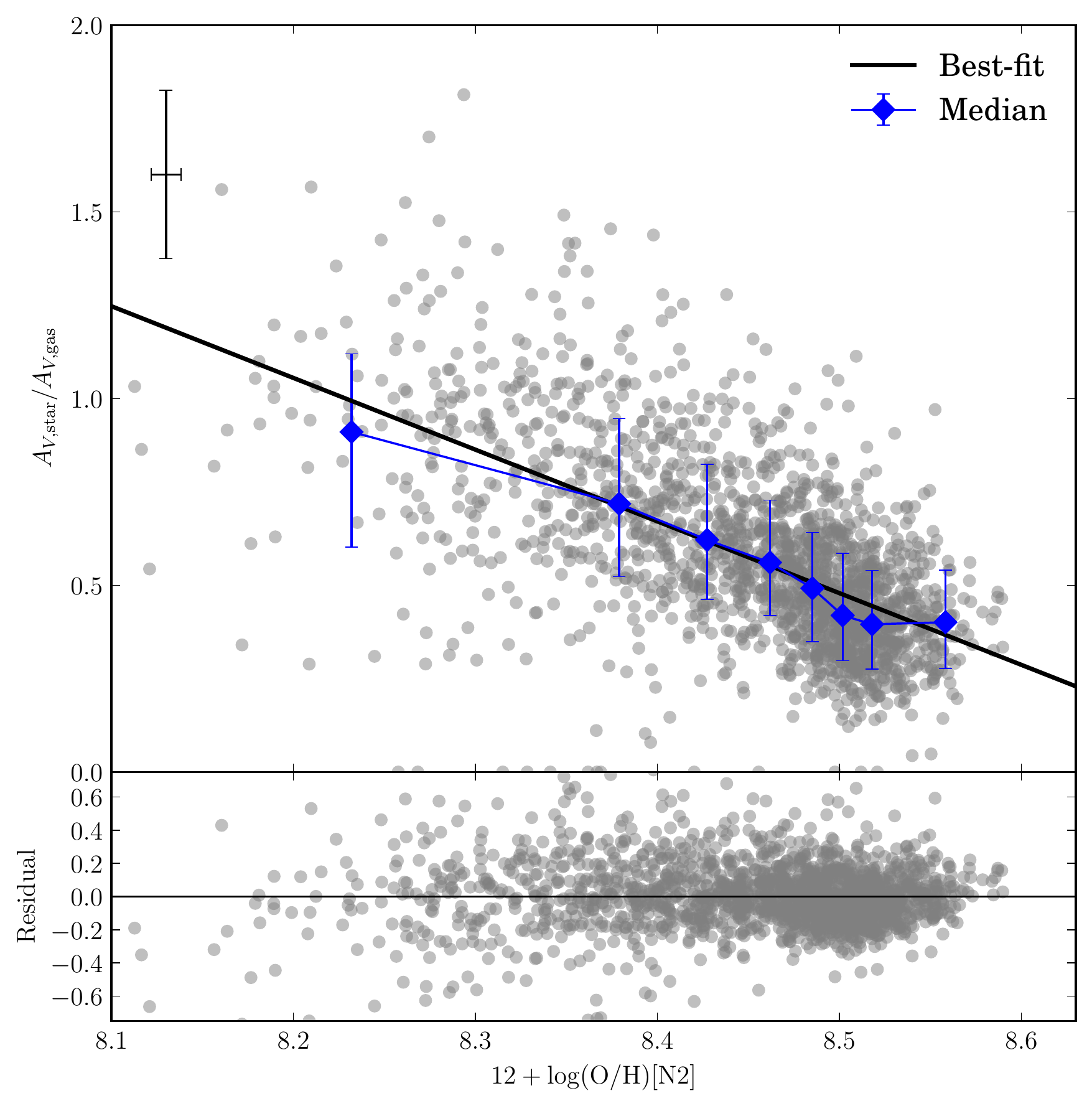}
    \caption{Global \rav\ ratio as a function of oxygen abundances derived from N2 indicators based on the \citet{Marino2013} calibrations. The symbols are the same as in Figure \ref{fig:m_rav}. The typical uncertainties are shown in the upper left corner.
    \label{fig:oh_rav}}
    \end{figure}

Table \ref{tab:cc} also suggests that the correlation between the attenuation ratio and oxygen abundance, regardless of indicators, is only slightly weaker than that of $M_*$. Taking the N2 index as the fiducial indicator, we show how the global \rav\ varies with gas-phase metallicity in Figure \ref{fig:oh_rav}. The oxygen abundance derived from the O3N2 diagnostic presents a very similar distribution as Figure \ref{fig:oh_rav}. We fit the data points with a linear function of Equation (\ref{eq:linear})
in which $x=12+\log(\mathrm{O/H})$. The best-fit result is $a_0=16.804\pm0.486$, $a_1=-1.921\pm0.058$, with substantial scatter for N2-based metallicities (black solid line in Figure \ref{fig:oh_rav}), and is also consistent with the binned medians denoted by blue diamonds. The standard deviation of the residuals is 0.20.

    \begin{figure*}[ht]
    \centering
    \includegraphics[width=\textwidth]{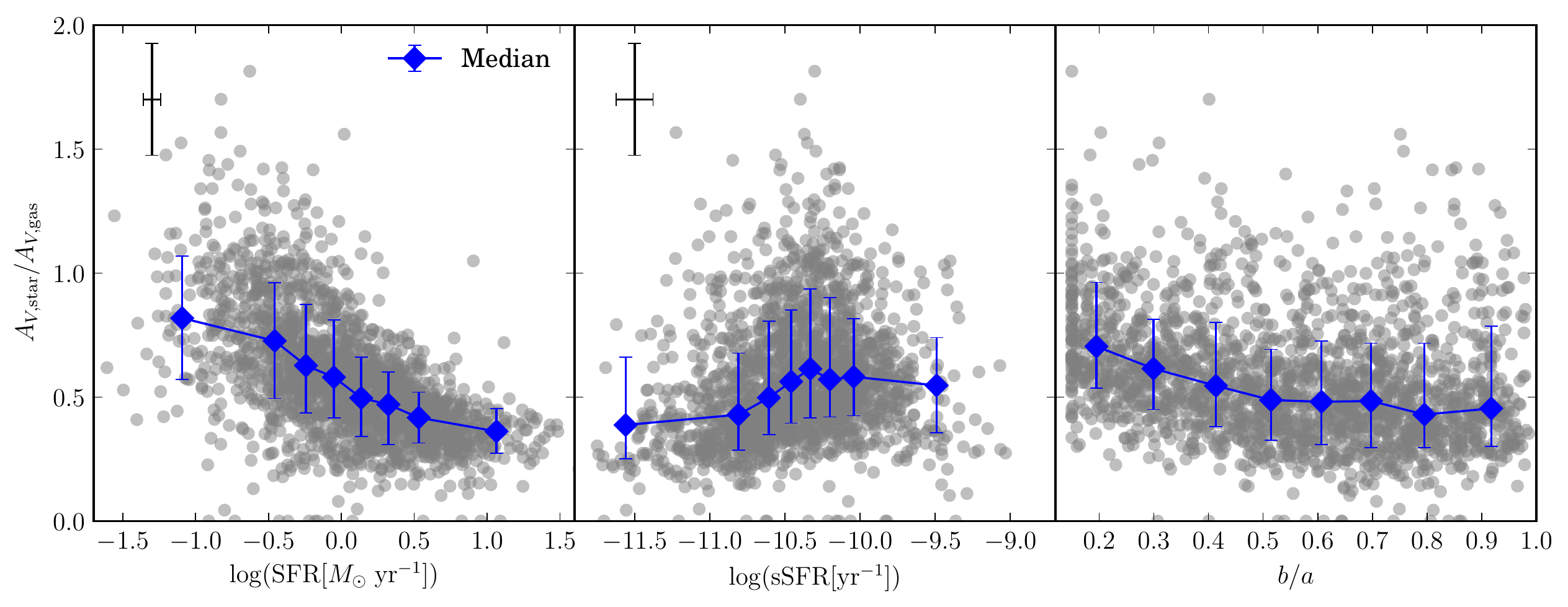}
    \caption{Global \rav\ ratio as a function of SFR (left), sSFR (middle), and $b/a$ (right). The blue diamonds with error bars indicate the medians and corresponding $1\sigma$ scatters (i.e., 16\%--84\% range) of the binned distributions for all galaxies. These bins are adjusted to equalize the number of galaxies in each bin for each set of bins. The typical uncertainties are given in the upper left corner, except for the $b/a$ subplot.
    \label{fig:sfr_ssfr_b2a_rav}}
    \end{figure*}

The SFR shows a moderate correlation with the global \rav, while sSFR and $b/a$ (i.e., inclination) exhibit very weak correlations with the global \rav. These correlations are shown in Figure \ref{fig:sfr_ssfr_b2a_rav}, together with a median curve and the corresponding scatters around the medians of the binned distributions for each case. The \rav\ ratio decreases with increasing SFR, while the scatters become smaller toward the high-SFR end. Although the median curves of \rav\ in sSFR and inclination are fairly flat, slight elevations can be identified when $\log(\mathrm{sSFR[yr^{-1}]})>-10.5$ and toward smaller $b/a$ (i.e., larger inclination). The global \avstar\ has a moderate anticorrelation with $b/a$, while the global \avgas\ shows a very weak anticorrelation with $b/a$. A similar result was reported by \cite{Yip2010}, in which the stellar attenuation is found to increase with inclination, while the \ha/\hb\ ratio remains nearly unchanged with $b/a$. Such behaviors are consistent with the picture of the two-component dust model (e.g., \citealt{Charlot2000,Wild2011}). In this model, \avstar\ arises from the diffuse ISM that distributes more homogeneously in the stellar disk (e.g., \citealt{Mosenkov2019}), so that galaxies with larger inclinations would have larger \avstar\ due to the increasing dust column density. The \avgas\ describes the combination of the dust effect from the dense birth clouds, which are more likely spherical, and the diffuse ISM but is dominated by the former (e.g., \citealt{Wild2011}). Therefore, the inclination only has a small effect on \avgas, and the global \rav\ would show a weak correlation with the inclination.

\cite{Koyama2019} presented a detailed study of the relation between the global \rav\footnote{\cite{Koyama2019} used $E(B-V)_{\mathrm{star}}/E(B-V)_{\mathrm{gas}}$ to indicate the ratio between stellar and nebular attenuation. In the following discussion, we use \rav\ instead of $E(B-V)_{\mathrm{star}}/E(B-V)_{\mathrm{gas}}$ when quoting their results due to the fact that $A_{V,\mathrm{star}}/A_{V,\mathrm{gas}}\propto E(B-V)_{\mathrm{star}}/E(B-V)_{\mathrm{gas}}$ and a constant $R_V$ are assumed for all galaxies when either stellar or nebular attenuation is derived in \cite{Koyama2019}.} and $M_*$, SFR, and sSFR using local SFG samples. They found that \rav\ decreases with increasing $M_*$, decreases at $-1\lesssim\log(\mathrm{SFR}[M_{\odot}~\mathrm{yr}^{-1}])\lesssim0.5$, and then increases at $0.5\lesssim\log(\mathrm{SFR}[M_{\odot}~\mathrm{yr}^{-1}])\lesssim1.5$ and toward the high-sSFR end for their SDSS/{\it Galaxy Evolution Explorer (GALEX)}/{\it Wide-field Infrared Survey Explorer (WISE)} sample. Within the parameter ranges explored by our sample, the overall trends of these three correlations are in good agreement with ours. By stacking the SDSS spectra of local SFGs, \cite{Zahid2017} reported that the trend between the global \rav\ and $M_*$ increases first at $M_*\lesssim10^{9.5}~M_{\odot}$ and then decreases toward the high-$M_*$ end with an increasing slope, whereas the same relation from both this work and \cite{Koyama2019} can be well described by a monotonic decreasing function with a gradual flattening at the high-$M_*$ end. However, due to the different data treatment between our work (or \citealt{Koyama2019}) and \cite{Zahid2017} (i.e., individual calculation \vs\ stacking), it is difficult to find out the reasons for these conflicts at present. Such different observations need to be understood in future studies.

As for the global dust attenuations, Table \ref{tab:cc} shows that \avstar\ only correlates with $b/a$, while \avgas\ moderately correlates with $M_*$, SFR, and gas-phase metallicity; weakly correlates with $b/a$; and does not correlate with sSFR. Similar to the local case, the correlations between \rav\ and the other physical properties discussed above are mainly driven by the correlations between \avgas\ and these properties. However, we also note that, compared to \avgas, \rav\ has slightly stronger correlations (in terms of either $r$ or $\rho$) with $M_*$, SFR, and $b/a$ and a comparable correlation with gas-phase metallicity, indicating the nonnegligible role of \avstar\ in these correlations.

\subsection{Correlations along Scaling Relations}
\label{subsec:scaling_relations}

All of the physical properties discussed above can be linked to $M_*$ except the inclination (i.e., $b/a$) via two well-studied scaling relations of SFGs. On the one hand, the star-forming main sequence (SFMS; e.g., \citealt{Speagle2014}) connects the SFR and sSFR to $M_*$. On the other hand, the metal enrichment in the ISM is closely related to the growth of galaxies and can be characterized by the well-known mass--metallicity relation (MZR; e.g., \citealt{Lequeux1979,Tremonti2004}).

    \begin{figure*}[ht]
    \centering
    \includegraphics[width=0.9\textwidth]{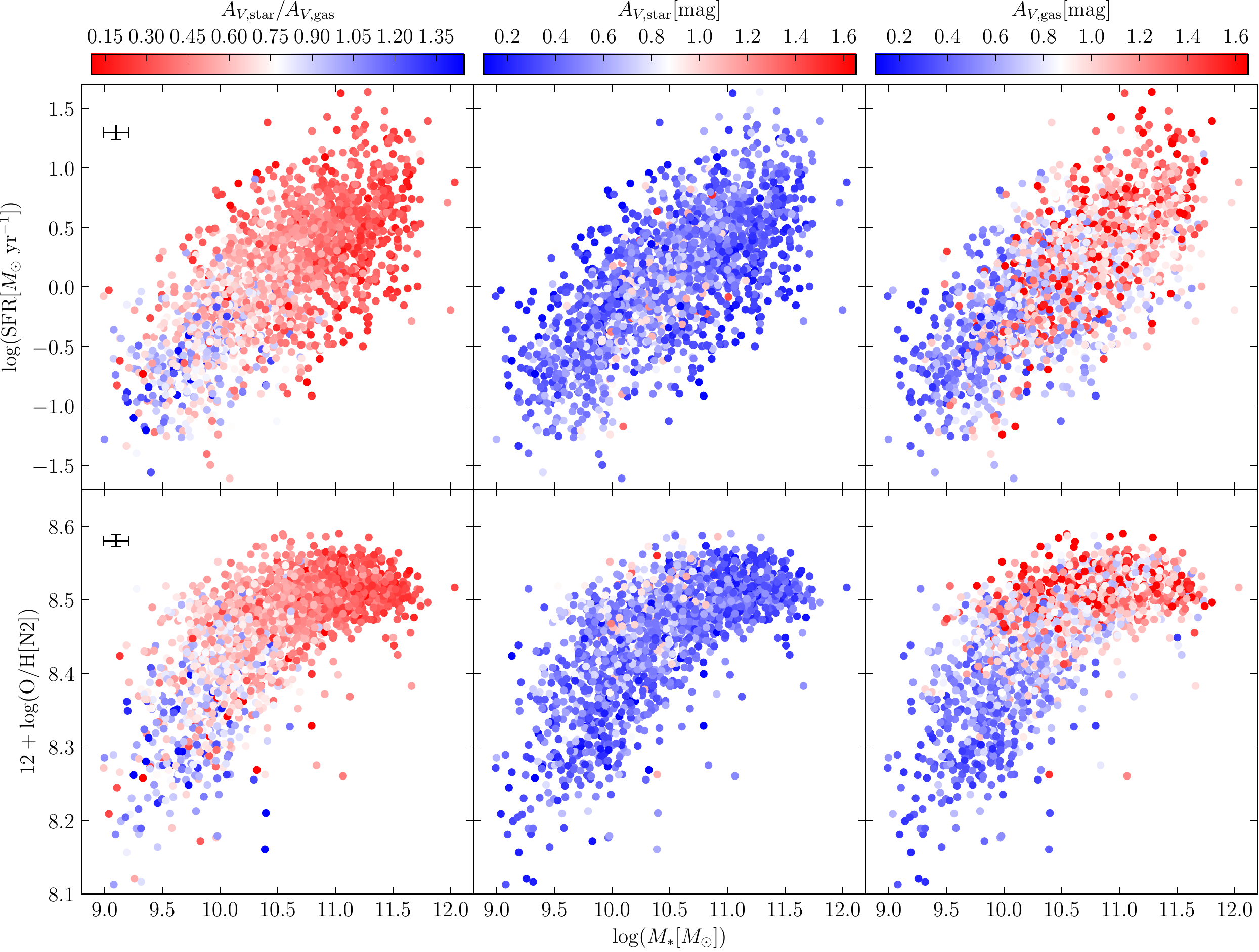}
    \caption{The SFMS (top row) and MZR (bottom row) of our selected galaxies, color-coded by their global \rav\ ratio (left), \avstar\ (middle), and \avgas\ (right). The typical uncertainties are shown in the upper left corner of the left panels.
    \label{fig:ms_mzr}}
    \end{figure*}

We present the SFMS and MZR of our sample in Figure \ref{fig:ms_mzr}, color-coding by their global \rav\ ratios and dust attenuations. Both plots exhibit a gradual trend of decreasing \rav\ along the sequence toward the high-$M_*$ end. The trend of \rav\ in our SFMS resembles the one using the \ha-based SFR ($\mathrm{SFR_{H\alpha}}$) in \cite{Koyama2019}; however, their fiducial total SFR is derived from a combination of UV and IR observations ($\mathrm{SFR_{UV+FIR}}$) for which the global \rav\ presents an evident trend across the SFMS. Their SDSS/\textit{GALEX}/\textit{WISE} sample suggests that galaxies with higher SFR tend to have a larger \rav\ value at fixed $M_*$. By comparing the $\mathrm{SFR_{H\alpha}}$ and $\mathrm{SFR_{UV+IR}}$, they further found that galaxies with $\mathrm{SFR_{H\alpha}}<\mathrm{SFR_{UV+IR}}$ tend to have substantially larger \rav. However, such a trend is inevitable because both the stellar and nebular attenuation are directly linked to the corresponding SFR in their calculation. The $E(B-V)_{\mathrm{star}}$ in \cite{Koyama2019} is derived from the $\mathrm{SFR_{UV+IR}}$ in the form of
\begin{equation}
    \begin{split}
    E(B-V)_{\mathrm{star}}&\propto\log(\mathrm{SFR_{UV+IR}/SFR_{UV}})\\
    &=\log(1+\mathrm{SFR_{IR}/SFR_{UV}}),
    \end{split}
\end{equation}
in which $\mathrm{SFR_{IR}}$ and $\mathrm{SFR_{UV}}$ are SFRs obtained from IR and UV data, respectively, while the $E(B-V)_{\mathrm{gas}}$ can be connected to $\mathrm{SFR_{H\alpha}}$ via
\begin{equation}
    \begin{split}
    A_{\mathrm{H\alpha}}&\propto E(B-V)_{\mathrm{gas}}\\
    &\propto\log(1+\mathrm{SFR_{H\alpha,atten}/SFR_{H\alpha,obs}}),
    \end{split}
\end{equation}
in which $\mathrm{SFR_{H\alpha,atten}}$ and $\mathrm{SFR_{H\alpha,obs}}$ are SFRs calculated from attenuated and observed \ha\ fluxes, respectively. Then we can obtain
\begin{equation}
    \frac{A_{V,\mathrm{star}}}{A_{V,\mathrm{gas}}}\propto\frac{\log(1+\mathrm{SFR_{IR}/SFR_{UV}})}{\log(1+\mathrm{SFR_{H\alpha,atten}/SFR_{H\alpha,obs}})}.
\end{equation}
Although \cite{Koyama2019} claimed that the physical reason for the difference between $\mathrm{SFR_{H\alpha}}$ and $\mathrm{SFR_{UV+IR}}$ is unclear, we can expect that either the overestimation of $\mathrm{SFR_{IR}}$, e.g., due to the large PSF of the {\it WISE} {\it W}4 band \citep{Wright2010} and contamination from their neighbors, or the underestimation of $\mathrm{SFR_{H\alpha,atten}}$, e.g., due to saturation in emission line attenuation \citep{Calabro2018}, would lead to an elevation on \rav\ if the unattenuated parts are thought to be more reliable. Therefore, we suggest that an SFR-independent $E(B-V)_{\mathrm{star}}$ (in calculation) and an IR-clear sample are required to solve this puzzle.

As suggested in Section \ref{subsec:observational_evidences}, \avstar\ exhibits a relatively uniform distribution in either SFMS or MZR, while \avgas\ dominates the observed trends along the two scaling relations for \rav. These results imply that the properties of dust in the diffuse ISM do not significantly alter as galaxies grow, while the dust in \hii\ regions and/or the surrounding dense clouds is more sensitive to the galaxy evolution.

Since \avgas\ can be connected to dust mass surface density and also gas mass surface density (e.g., \citealt{Boquien2013,Brinchmann2013,Kreckel2013,Kreckel2016}), one would speculate that the observed \rav--SFR relation might stem from the more basic Kennicutt--Schmidt star formation law \citep{Schmidt1959,Kennicutt1998a} and leads to the \rav--$M_*$ and \rav--\oh\ relations via the SFMS and MZR, respectively. We believe that the Kennicutt--Schmidt law can partly account for the \rav--SFR relation; however, it cannot fully explain the observed trends along the two scaling relations due to the following reasons. First, Table \ref{tab:cc} shows that both $M_*$ and gas-phase metallicity more strongly correlate with \avgas\ compared to SFR. If the \avgas--SFR relation is more fundamental and connects \avgas\ with $M_*$ and gas-phase metallicity via the SFMS and MZR, respectively, we would expect a stronger correlation with SFR compared to $M_*$ and gas-phase metallicity. Second, comparing with \avgas, \rav\ displays stronger correlations with $M_*$ and SFR and a comparable correlation with gas-phase metallicity, implying that although \avstar\ exhibits very weak correlations with these physical properties, its role is still nonnegligible. Third, our partial correlation analysis discussed below suggests that when SFR is controlled, $M_*$ still moderately correlates with \rav, and these three parameters seem to play comparable roles in the correlation.

\begin{deluxetable*}{cCCCCCCC}[htb]
\tablecaption{Partial Correlations between \rav\ and Some Physical Properties\label{tab:pcor}}
\tablecolumns{3}
\tablewidth{0pt}
\tablehead{
\colhead{Variables\tablenotemark{a}} & \colhead{$M_*$} & \colhead{$M_*$} & \colhead{SFR} & \colhead{\oh} & \colhead{$M_*$} & \colhead{SFR} & \colhead{\oh} \\
\colhead{Controlled\tablenotemark{b}} & \colhead{SFR} & \colhead{\oh} & \colhead{$M_*$} & \colhead{$M_*$} & \colhead{SFR and \oh} & \colhead{$M_*$ and \oh} & \colhead{$M_*$ and SFR}
}
\startdata
$r$ & -0.42 & -0.43 & -0.25 & -0.30 & -0.16 & -0.31 & -0.35 \\
$\rho$ & -0.50 & -0.52 & -0.24 & -0.23 & -0.26 & -0.29 & -0.29 \\
\enddata
\tablenotetext{a}{For each column, partial correlation coefficients are calculated between the physical property in this row and \rav, while the physical properties shown in the second row are controlled. Here $M_*$ and SFR are in log scale, and \oh\ is the N2-based one. Pearson correlation coefficients $r$ and Spearman rank correlation coefficients $\rho$ are derived.}
\tablenotetext{b}{Physics properties that are controlled in partial correlation analysis.}
\end{deluxetable*}

\cite{Koyama2019} speculated that the decrease of \rav\ with increasing SFR at the low-SFR side originates from the relation between $M_*$ and \rav\ because of the SFMS. Similarly, the correlation between \rav\ and gas-phase metallicity revealed by Figure \ref{fig:oh_rav} also might result from the $M_*$--\rav\ relation due to the tight MZR. To test these conjectures, a partial correlation analysis is performed utilizing the R ppcor package\footnote{\url{https://cran.r-project.org/web/packages/ppcor/}} \citep{Kim2015}. We calculate the Pearson correlation coefficients $r$ and Spearman rank correlation coefficients $\rho$ to see how \rav\ correlates with $M_*$, SFR, and \oh[N2], respectively, when the other one or two physical properties are controlled.

The results are presented in Table \ref{tab:pcor} in which the first four columns show the cases of controlling only one parameter and the last three columns are the cases of controlling two parameters simultaneously. When only one parameter is controlled, $M_*$ gives the strongest correlation with \rav, regardless of the controlled parameters and calculation methods. Comparing with the coefficients listed in Table \ref{tab:cc}, it is evident that all three parameters can partly explain the corresponding correlations, while $M_*$ seems to have a dominant effect. In the cases of controlling two physical properties at the same time, the effect of $M_*$ becomes much smaller compared to SFR and metallicity when considering linear correlation. However, if nonlinear correlation (i.e., $\rho$) is also taken into account, the three parameters have comparable roles in the correlations with \rav. Given the bridge role of $M_*$ in connecting SFR and metallicity, we argue that the apparently dominant effect of $M_*$ when controlling only one parameter has a significant contribution from metallicity (SFR) via the MZR (SFMS). Therefore, we conclude that $M_*$, SFR, and gas-phase metallicity play comparable roles in the correlations with these physical properties, and either of them can fully account for the trends unveiled by Figure \ref{fig:ms_mzr}.

\subsection{A Possible Picture Based on Dust Growth}
\label{subsec:possible_picture}

Both Table \ref{tab:cc} and Figure \ref{fig:ms_mzr} suggest that the correlations between \rav\ and other galactic properties (i.e, $M_*$, SFR, and gas-phase metallicity), as well as the trends along the scaling relations, are mainly driven by the global \avgas. Integrating along the line of sight of Equation (\ref{eq:a_lambda}) results in (e.g., \citealt{Galliano2018})
\begin{equation}
    A_{\lambda} = 1.086\times\Sigma_{\mathrm{dust}} = 1.086\times\Sigma_{\mathrm{gas}} \cdot \mathrm{D/G},
\end{equation}
where $\Sigma_{\mathrm{dust}}$ and $\Sigma_{\mathrm{gas}}$ are the mass surface densities of dust and gas, respectively. Obviously, an increasing \avgas\ with $M_*$ implies more dust in more massive galaxies, which is equivalent to more gas and/or a larger D/G ratio. The increasing dust mass and total gas mass with $M_*$ in SFGs were reported by many previous works (e.g., \citealt{DeVis2017a,Casasola2019}). Considering the factor related to dust, here we only focus on the possible role of the D/G ratio in the observed correlations, attempting to provide an alternative origin of these correlations in the view of dust evolution.

In the analysis of the local \rav, we suggest that in addition to the geometry of stars/dust, the local physical conditions of ionized gas also play an important role in determining the \rav\ value. The D/G ratio is used to connect the local gas-phase metallicity and ionization parameter to dust attenuation. On a galactic scale, the behaviors of the global D/G ratio are found to be controlled by some dust processes accompanied by galaxy growth \citep{Asano2013,DeVis2017,DeVis2019,Aoyama2018,Galliano2018a}.

As we mentioned in Section \ref{subsubec:exp_phyconds}, metallicity is the main physical property that drives the D/G ratio \citep{Engelbracht2008,Asano2013,Remy-Ruyer2014}. This metallicity dependence can be described by the following picture \citep{Galliano2018}: when galaxies are metal-poor, dust grains are mainly produced by stellar ejecta, and the D/G is proportional to metallicity with a small dust-to-metal ratio; at intermediate metallicities, metal accretion onto dust grains becomes efficient, and the D/G increases steeply; and when galaxies become more metal-rich, dust growth is limited by metallicity again, but with a larger dust-to-metal ratio, due to the depletion of metals in the ISM\footnote{We note that the range of our N2-based \oh\ is already in the most metal-rich (ISM growth) regime of the \cite{Galliano2018} picture. However, the absolute value of gas-phase metallicity derived from different calibrations varies up to 0.7 dex \citep{Kewley2008}. On the other hand, both our $M_*$ and \oh\ ranges cover the predicted regimes in which dust growth by accretion goes from being efficient to being saturated in the hydrodynamic simulation of \cite{Aoyama2018}. Therefore, here we focus on the picture evolved with the relative value of metallicity, rather than the absolute value.}.

Such a picture is well reproduced by hydrodynamic simulation. As displayed in \cite{Aoyama2018}, for intermediate-mass galaxies ($10^{8.5}~M_{\odot}\lesssim M_*\lesssim 10^{10}~M_{\odot}$ in their simulation), dust growth by accretion becomes active and the D/G--Z relation becomes steeper. A large scatter of the relation in this region emerges, suggesting that the D/G is strongly impacted by local internal conditions and star formation histories (SFHs) in this mass range \citep{Asano2013,Remy-Ruyer2014}. For the most massive galaxies ($M_*\gtrsim 10^{10}~M_{\odot}$), dust growth by accretion is limited by metallicity and saturated. Here the MZR is already reproduced in their simulation and naturally connects the D/G to $M_*$. In addition, Table \ref{tab:cc} shows that the global \avgas\ has slightly weaker (but opposite sign) correlations with $M_*$ and SFR compared to those of \rav\ and comparable correlations with oxygen abundances, irrespective of metallicity diagnostics. However, the global \avstar\ shows almost no correlations with all of the properties listed in Table \ref{tab:cc} except the $b/a$ ratio. Combining the predictions from simulations with the observed evidence, we speculate that, similar to the local case, the elevated global D/G ratio in more massive SFGs leads to the increase of the overall \avgas\ and thus a smaller \rav.

Furthermore, the $M_*$-dependent scatter shown in Figure \ref{fig:m_rav} can be explained as the scatter of D/G, which is found to be large for $10^{8.5}~M_{\odot}\lesssim M_*\lesssim 10^{10}~M_{\odot}$ and much smaller for $M_*\gtrsim 10^{10}~M_{\odot}$ in the simulation of \cite{Aoyama2018}. Recent IFU studies also suggest that the global stellar mass has a nonnegligible role in regulating the local metallicity \citep{Gao2018}, especially for low-mass galaxies, probably due to their shallower gravitational potential wells for which it is easier to lose their metal by galactic outflows \citep{Chisholm2018}. Because the global \rav\ used in this work is derived from the mode of the local ones, the $M_*$ effect on local metallicity should also impact the global \rav. Hence, besides the local internal processes and SFHs, $M_*$ itself may be another factor that drives the $M_*$-dependent scatter in the $M_*$--\rav\ relation.

Therefore, the evolution of dust growth in birth clouds accompanied by galaxy growth might be one plausible origin for our observed behaviors of the global \rav.

\section{Summary}
\label{sec:summary}

In this work, we construct a sample of \hii\ regions using the IFU data of the MaNGA survey released from the SDSS DR15 and investigate how the \rav\ ratio varies with other physical properties on subgalactic and galactic scales.

On a subgalactic scale, we find that the \avgas\ indeed correlates with the \avstar, while the correlation becomes much stronger for spaxels with higher \sigha. The local \rav\ is found to have moderate correlations with DIG indicators (i.e., the N2S2 and S2 indices, \sigha) and tracers of gas-phase metallicity and ionization parameter. Among these parameters, the N2S2 index shows the strongest correlation with the local \rav\ ratio. The \rav\ ratio of DIG-dominated spaxels tends to be systematically larger than that of classic \hii-dominated spaxels. Local physical conditions of ionized gas show significant effects on the dust attenuation. Metal-poor spaxels with high ionization parameters tend to suffer less nebular attenuation (i.e., smaller \avgas) and thus larger \rav\ values for either \hii- or DIG-dominated subsamples. We argue that the systematic difference in \rav\ between DIG- and \hii-dominated spaxels can be explained by the stars/dust geometry based on the two-component dust model \citep{Charlot2000}, while the dependences on metallicity and ionization parameters can be attributed to the change of dust properties in different local environments. We further suggest that the metallicity- and ionization parameter--dependent dust-to-gas mass ratio could be one possible parameter to connect the physical conditions with dust attenuation. The median maps of dust attenuation (and the ratio) on the BPT diagrams form a low-\avgas\ and high-\rav\ sequence that can be resolved into DIG-dominated and metal-poor spaxels. The local analyses reveal that both geometry between stars and gas/dust and physical conditions are important in determining the local \rav\ values.

The local \rav\ within each galaxy converges to one constant at the high-\sigha\ end, which is also the mode of the distribution of \rav\ and is taken as the global \rav.

On a galactic scale, this global \rav\ ratio indeed varies from galaxy to galaxy and correlates with physical properties such as stellar mass, SFR, and metallicity. It is found to show a strong correlation with stellar mass, moderate correlations with SFR and metallicity, and weak correlations with $b/a$ and sSFR. The SFGs that have larger $M_*$ and higher SFR and are more metal-rich tend to have smaller \rav\ ratios. A gradual trend of decreasing \rav\ toward the high-$M_*$ end along the SFMS and MZR is found. Partial correlation analysis demonstrates that $M_*$, SFR, and gas-phase metallicity have comparable effects in the observed correlations, and no single property can fully account for these correlations. We suggest that the metallicity-dependent dust-to-gas ratio, together with the dust growth process accompanied by galaxy growth, might partly explain these observations, although we cannot rule out any contribution from the geometry effect.

Overall, our results highlight the importance of local physical conditions in dust processes, as well as the level of dust attenuation. High spatial resolution IR observations are required to give a quantitative conclusion. In addition, the most uncertain parameter in this study is the local \avstar, for which more constraints from IR observation are required. Fortunately, the Mid-Infrared Instrument of the {\it James Webb Space Telescope} is able to provide mid-IR (MIR) images of nearby galaxies from 5 to 28.5 $\mu$m with unprecedented spatial resolution \citep{Rieke2015}. Such high-quality MIR observations, in combination with the optical IFU data and the well-calibrated monochromatic estimators of total IR luminosity (e.g., \citealt{Boquien2010,Elbaz2010,Lin2016}), will allow us to put more constraints on the derivation of \avstar\ and thus help us to better understand the relation between \rav\ and other parameters on a subkiloparsec scale.


\acknowledgments

The authors would like to thank Daniela Calzetti for very helpful suggestions and discussions. This work is supported by the National Key R\&D Program of China (2015CB857004, 2017YFA0402600) and the National Natural Science Foundation of China (NSFC; Nos. 11320101002, 11421303, and 11433005). Z.L. gratefully acknowledges support from the China Scholarship Council (No. 201806340211).

This paper makes use of the MaNGA-Pipe3D data products. We thank the IA-UNAM MaNGA team for creating this catalog and Conacyt Project CB-285080 for supporting them.

Funding for the Sloan Digital Sky Survey IV has been provided by the Alfred P. Sloan Foundation, the U.S. Department of Energy Office of Science, and the Participating Institutions. The SDSS-IV acknowledges support and resources from the Center for High-Performance Computing at the University of Utah. The SDSS website is www.sdss.org.The SDSS-IV is managed by the Astrophysical Research Consortium for the Participating Institutions of the SDSS Collaboration, including the Brazilian Participation Group, the Carnegie Institution for Science, Carnegie Mellon University, the Chilean Participation Group, the French Participation Group, Harvard-Smithsonian Center for Astrophysics, Instituto de Astrof\'isica de Canarias, The Johns Hopkins University, Kavli Institute for the Physics and Mathematics of the Universe (IPMU)/University of Tokyo, the Korean Participation Group, Lawrence Berkeley National Laboratory, 
Leibniz Institut f\"ur Astrophysik Potsdam (AIP),  
Max-Planck-Institut f\"ur Astronomie (MPIA Heidelberg), 
Max-Planck-Institut f\"ur Astrophysik (MPA Garching), 
Max-Planck-Institut f\"ur Extraterrestrische Physik (MPE), 
National Astronomical Observatories of China, New Mexico State University, 
New York University, the University of Notre Dame, 
Observat\'ario Nacional/MCTI, The Ohio State University, 
Pennsylvania State University, Shanghai Astronomical Observatory, 
the United Kingdom Participation Group,
Universidad Nacional Aut\'onoma de M\'exico, the University of Arizona, 
the University of Colorado Boulder, the University of Oxford, the University of Portsmouth, 
the University of Utah, the University of Virginia, the University of Washington, the University of Wisconsin, 
Vanderbilt University, and Yale University.

%



\software{Astropy \citep{AstropyCollaboration2013,AstropyCollaboration2018}, IPython \citep{Perez2007}, Matplotlib \citep{Hunter2007}, Numpy \citep{Oliphant2006}, MPFIT \citep{Markwardt2009}.}



\appendix
\section{Effects of S/N criteria on Emission Lines}
\label{appen:snr_cut}

As mentioned in Section \ref{subsec:sample}, we require our sample to have an S/N of the Balmer line (\ha\ and \hb; S/N$_{\mathrm{Bal}}$) greater than 5 and an S/N of the other main emission lines (\oiii$\lambda$5007, \nii$\lambda$6584, and \sii$\lambda$6717; S/N$_{\mathrm{eml}}$) greater than 3 to ensure reliable estimates of \avgas\ and emission line indices. Here we examine that whether these S/N criteria affect our results.

    \begin{figure*}[htb]
    \centering
    \includegraphics[width=\textwidth]{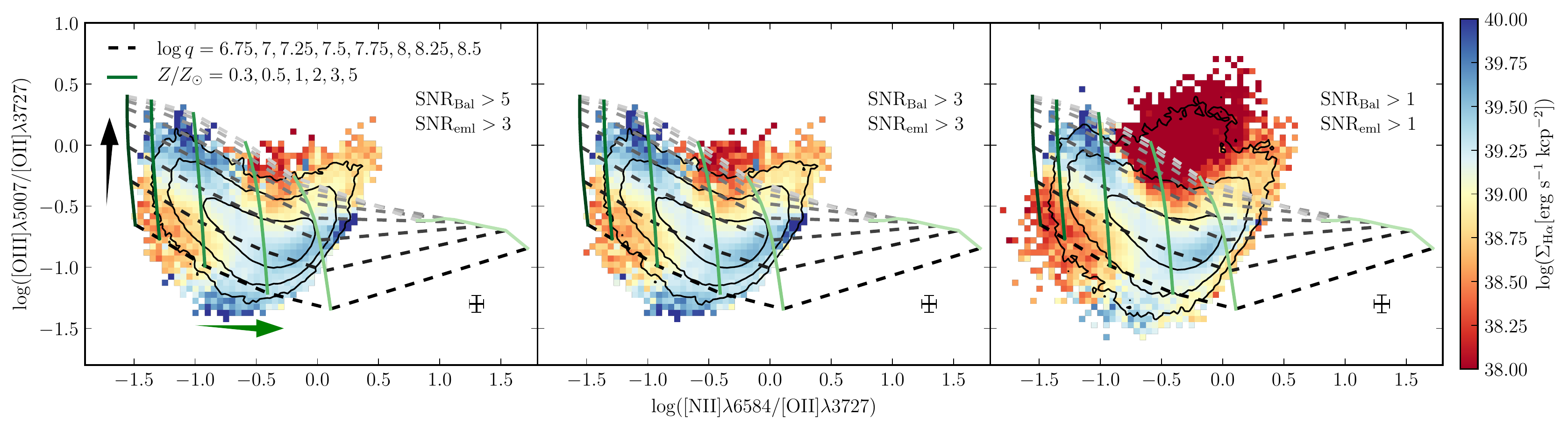}
    \caption{Median maps of \sigha\ on the N2O2 \vs\ O32 diagram for spaxels with S/N$_{\mathrm{Bal}}>5$ and S/N$_{\mathrm{eml}}>3$ (left), S/N$_{\mathrm{Bal}}>3$ and S/N$_{\mathrm{eml}}>3$ (middle), and S/N$_{\mathrm{Bal}}>1$ and S/N$_{\mathrm{eml}}>1$ (right). The symbols are the same as in Figure \ref{fig:zq_ha}.
    \label{fig:zq_ha_compsnr}}
    \end{figure*}

\begin{figure}[htb]
    \centering
    \includegraphics[width=0.5\textwidth]{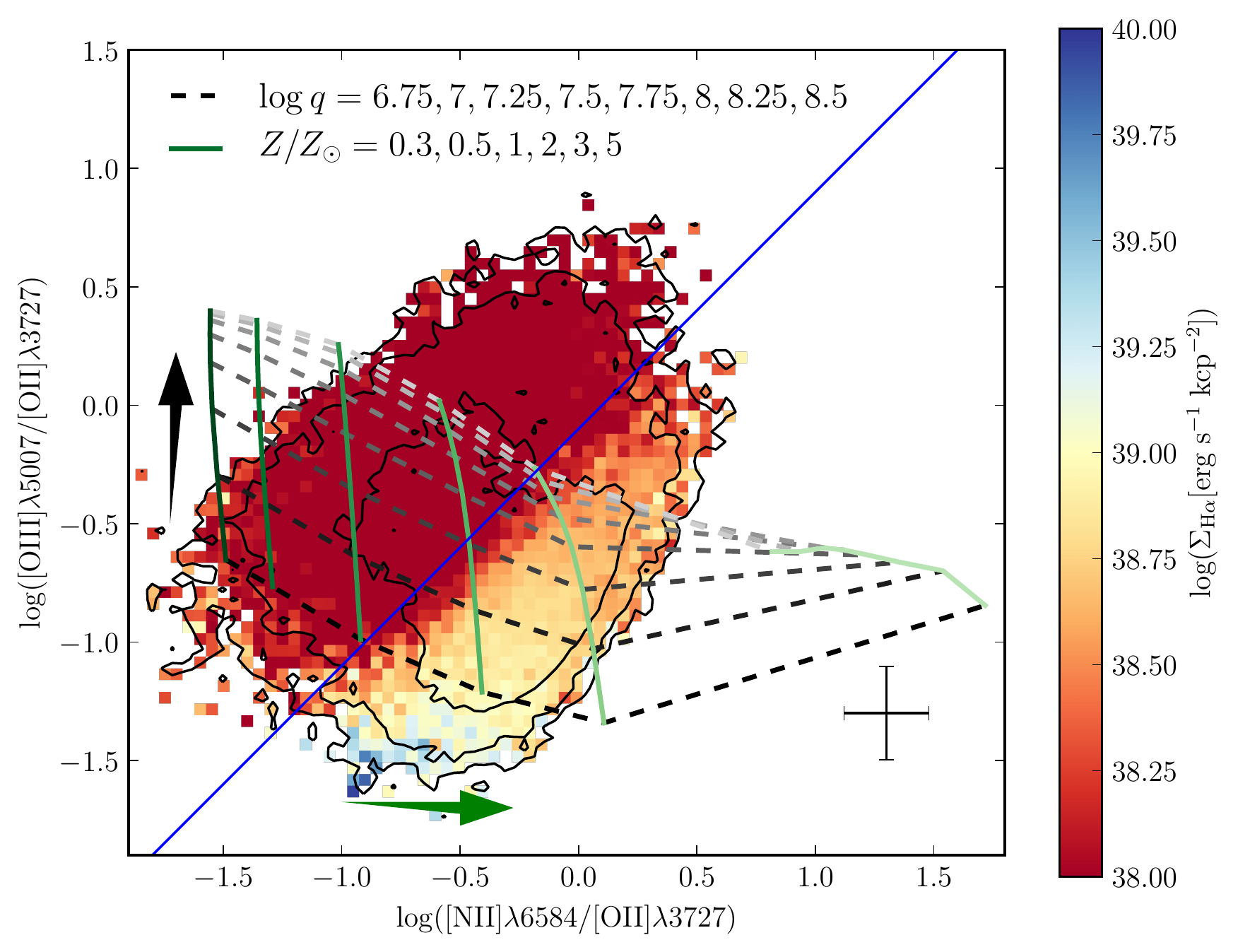}
    \caption{Median map of \sigha\ on the N2O2 \vs\ O32 diagram for spaxels included by S/N$_{\mathrm{Bal}}>1$ and S/N$_{\mathrm{eml}}>1$ but excluded by S/N$_{\mathrm{Bal}}>3$ and S/N$_{\mathrm{eml}}>3$. The blue line indicates a visually selected boundary of $\log(\oiii\lambda5007/\oii\lambda3727)=\log(\nii\lambda6584/\oii\lambda3727)-0.1$ that is used to divided the spaxels into low-\sigha\ (upper) and high-\sigha\ (lower) branches. Other symbols are the same as in Figure \ref{fig:zq_ha}.
    \label{fig:z_q_ha_snr1_3}}
    \end{figure}

In Figure \ref{fig:zq_ha_compsnr}, we compare the \sigha\ maps on the N2O2 \vs\ O32 diagram for spaxels selected from three S/N cuts: S/N$_{\mathrm{Bal}}>5$ and S/N$_{\mathrm{eml}}>3$, S/N$_{\mathrm{Bal}}>3$ and S/N$_{\mathrm{eml}}>3$, and S/N$_{\mathrm{Bal}}>1$ and S/N$_{\mathrm{eml}}>1$, where the first one is the fiducial cut adopted by this work. Clearly, the selected sample is nearly unaffected by relaxing the S/N$_{\mathrm{Bal}}$ limit from 5 to 3 when the S/N$_{\mathrm{eml}}$ cut is fixed to 3. However, setting both the S/N$_{\mathrm{Bal}}$ and S/N$_{\mathrm{eml}}$ limits to 1 introduces a low-\sigha\ regime within which spaxels should be classified as DIG-dominated regions. This additional regime also cannot be covered by the model grid of \hii\ regions from \cite{Dopita2013}. To understand these low-\sigha\ spaxels, we further select spaxels included by S/N$_{\mathrm{Bal}}>1$ and S/N$_{\mathrm{eml}}>1$ but excluded by S/N$_{\mathrm{Bal}}>3$ and S/N$_{\mathrm{eml}}>3$, and plot their \sigha\ map in Figure \ref{fig:z_q_ha_snr1_3}.

Intriguingly, the selected subsample exhibits a significant trend of increasing \sigha\ from the high-\oiii$\lambda$5007 and low-\nii$\lambda6584$ corner to the low-\oiii$\lambda$5007 and high-\nii$\lambda6584$ corner. The bimodal 68\% contour indicates that this subsample might be a combination of two populations with different average \sigha\ and even different excitation mechanisms. We visually define a demarcation expressed as
\begin{equation}
    \log\left(\frac{\oiii\lambda5007}{\oii\lambda3727}\right)=\log\left(\frac{\nii\lambda6584}{\oii\lambda3727}\right) - 0.1
\end{equation}
to cross two saddle points of the 68\% contour and separate the subsample into low-\sigha\ (upper) and high-\sigha\ (lower) branches. The median \sigha\ and the corresponding $1\sigma$ ranges are $37.78_{-0.32}^{+0.42}$ and $38.71_{-0.56}^{+0.41}$ for the upper and lower branches, respectively.

    \begin{figure*}[htb]
    \centering
    \includegraphics[width=0.8\textwidth]{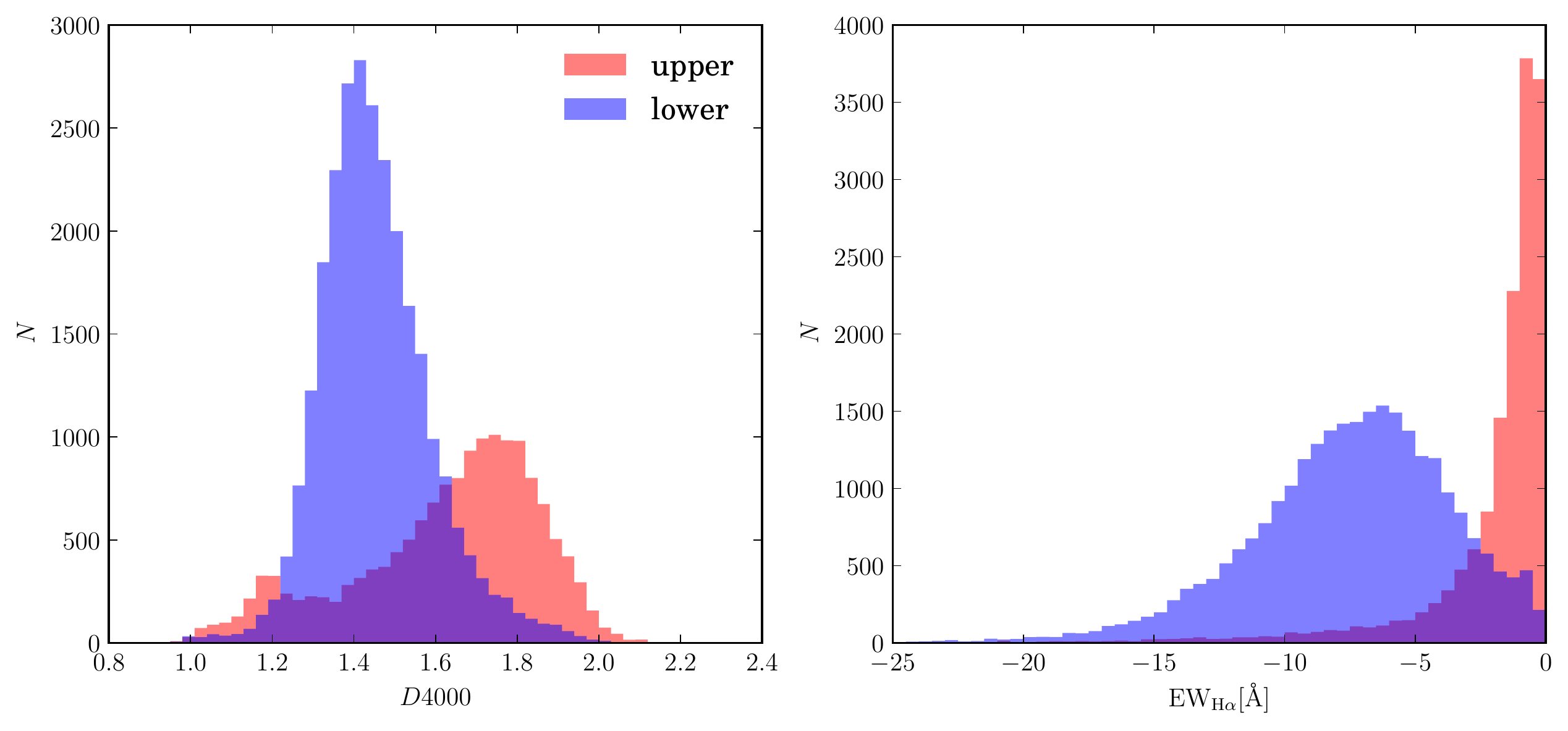}
    \caption{Histogram of $D4000$ (left) and $\mathrm{EW_{H\alpha}}$ (right) for spaxels at the upper (red) and lower (blue) branches defined in Figure \ref{fig:z_q_ha_snr1_3}.
    \label{fig:hist_snr1_3}}
    \end{figure*}

    \begin{figure*}[htb]
    \centering
    \includegraphics[width=\textwidth]{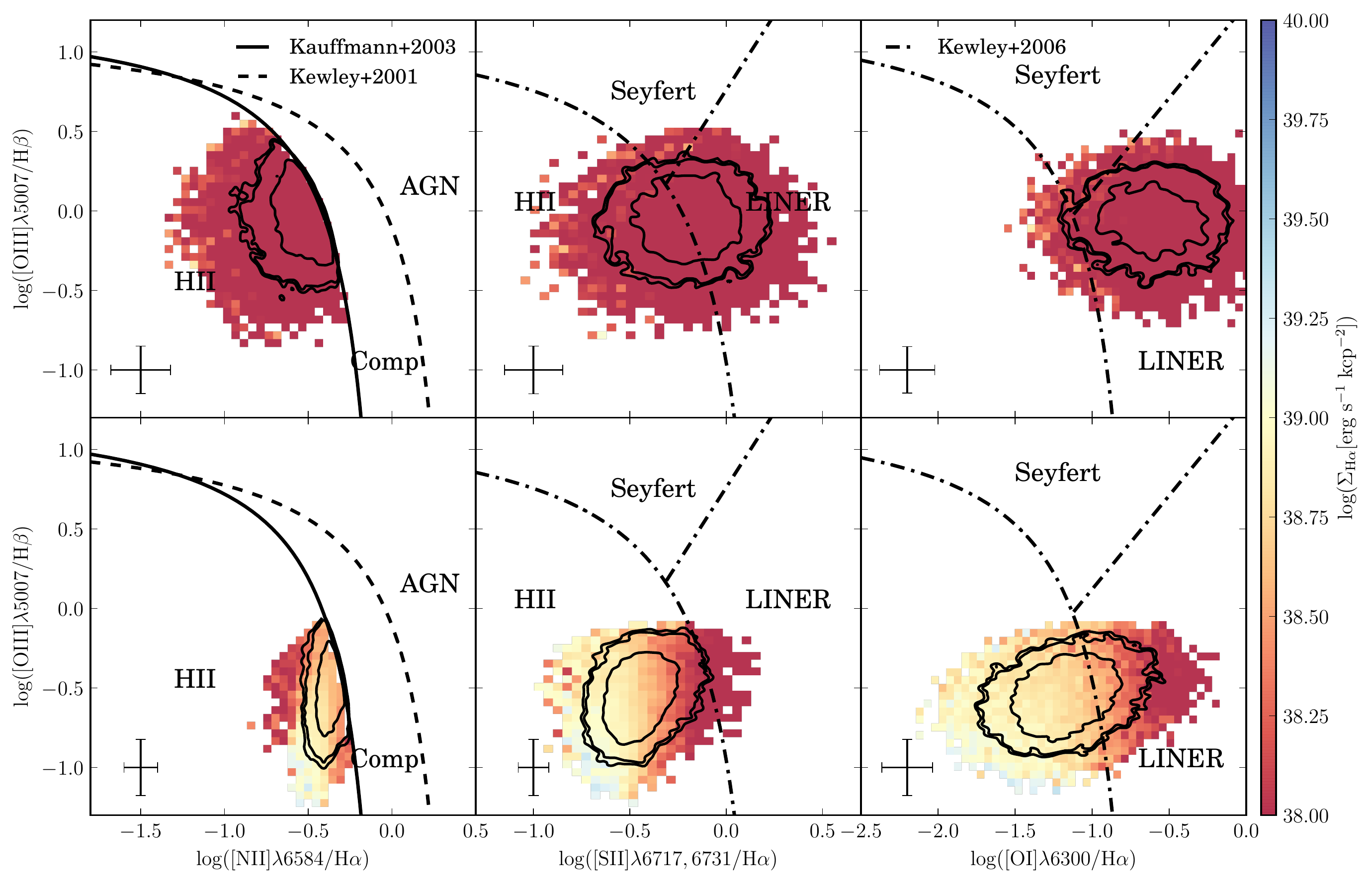}
    \caption{Median maps of \sigha\ on the BPT diagrams for spaxels at the upper (top) and lower (bottom) branches defined in Figure \ref{fig:z_q_ha_snr1_3}. Boundaries from \cite{Kauffmann2003a}, \cite{Kewley2001}, and \cite{Kewley2006} are denoted by solid, dashed, and dashed--dotted black curves, respectively. The contours encompass 68\%, 95\%, and 99\% of the spaxels on each diagram. The typical uncertainties are shown in the lower left corner.
    \label{fig:bpt_snr1_3}}
    \end{figure*}

To reveal the possible origins of the upper and lower branches, we present their histograms of $D4000$ and $\mathrm{EW_{H\alpha}}$ in Figure \ref{fig:hist_snr1_3} and the \sigha\ maps on the BPT diagrams in Figure \ref{fig:bpt_snr1_3}. The two branches have distinct distributions for both $D4000$ and $\mathrm{EW_{H\alpha}}$. The medians and 68\% ranges of $D4000$ are $1.68_{-0.30}^{+0.16}$ and $1.44_{-0.11}^{+0.14}$ for the upper and lower branches, respectively, while the same parameters are $-1.05_{-2.28}^{+0.66}$ and $-7.16_{-3.95}^{+3.33}$ for $\mathrm{EW_{H\alpha}}$. Obviously, the upper branch is dominated by an old stellar population for which star formation is almost quenched, whereas the lower one suggests a relatively young population with a moderate level of star formation. For the BPT diagrams, nearly all of the spaxels at the upper branch locate in the LINER regime on the \oi--\oiii\ diagram, and half of them lie beyond the \hii\ regime on the \sii--\oiii\ diagram. For the lower branch, nearly all spaxels are within the \hii\ regime in the \sii--\oiii\ plane, and only small fraction of spaxels with lower \sigha\ move toward the LINER regime of the \oi--\oiii\ plane.

The distinct distributions of the above properties of stellar population and ionized gas indicate the different physical origins of the two branches. The upper branch is comprised of an old stellar population with emission lines arising from LI(N)ER-like regions or the so-called hDIG in \cite{Lacerda2018}, which is heated by hot, low-mass, evolved stars. Spaxels at the lower branch are more likely faint \hii\ regions. Comparison between Figures \ref{fig:zq_ha_compsnr} and \ref{fig:z_q_ha_snr1_3} shows that our adopted S/N criteria indeed reduce the covered N2O2--O32 parameter space of non-star-forming origins, while the star-forming regime (i.e., the lower branch of Figure \ref{fig:z_q_ha_snr1_3} or the main star-forming cloud revealed in Figure \ref{fig:zq_ha}) is almost unaffected. Given the aim of studying the dust attenuation ratio for classical \hii\ regions, we concluded that the fiducial S/N cuts (i.e., S/N$_{\mathrm{Bal}}>5$ \& S/N$_{\mathrm{eml}}>3$) are suitable, and relaxing the criteria to smaller S/N values cannot change our results.




\bibliography{ms}

\begin{thebibliography}{}
\providecommand\natexlab[1]{#1}
\providecommand\JournalTitle[1]{#1}

\bibitem[{Aguado {et~al.}(2019)Aguado, Ahumada, Almeida, Anderson, Andrews,
  Anguiano, Aquino~Ort{\'{\i}}z, Arag{\'o}n-Salamanca, Argudo-Fern{\'a}ndez,
  Aubert, Avila-Reese, Badenes, Barboza~Rembold, Barger, Barrera-Ballesteros,
  Bates, Bautista, Beaton, Beers, Belfiore, Bernardi, Bershady, Beutler, Bird,
  Bizyaev, Blanc, Blanton, Blomqvist, Bolton, Boquien, Borissova, Bovy, Brandt,
  Brinkmann, Brownstein, Bundy, Burgasser, Byler, Cano~Diaz, Cappellari,
  Carrera, Cervantes~Sodi, Chen, Cherinka, Doohyun~Choi, Chung, Coffey,
  Comerford, Comparat, Covey, da~Silva~Ilha, da~Costa, Dai, Damke, Darling,
  Davies, Dawson, de~Sainte~Agathe, Deconto~Machado, Del~Moro, De~Lee,
  Diamond-Stanic, Dom{\'{\i}}nguez~S{\'a}nchez, Donor, Drory, du~Mas~des
  Bourboux, Duckworth, Dwelly, Ebelke, Emsellem, Escoffier,
  Fern{\'a}ndez-Trincado, Feuillet, Fischer, Fleming, Fraser-McKelvie,
  Freischlad, Frinchaboy, Fu, Galbany, Garcia-Dias,
  Garc{\'{\i}}a-Hern{\'a}ndez, Garma~Oehmichen, Geimba~Maia, Gil-Mar{\'{\i}}n,
  Grabowski, Gu, Guo, Ha, Harrington, Hasselquist, Hayes, Hearty,
  Hernandez~Toledo, Hicks, Hogg, Holley-Bockelmann, Holtzman, Hsieh, Hunt,
  Hwang, Ibarra-Medel, Jimenez~Angel, Johnson, Jones, J{\"o}nsson, Kinemuchi,
  Kollmeier, Krawczyk, Kreckel, Kruk, Lacerna, Lan, Lane, Law, Lee, Li, Lian,
  Lin, Lin, Lintott, Long, Longa-Pe{\~n}a, Mackereth, de~la Macorra, Majewski,
  Malanushenko, Manchado, Maraston, Mariappan, Marinelli, Marques-Chaves,
  Masseron, Masters, McDermid, Medina~Pe{\~n}a, Meneses-Goytia, Merloni,
  Merrifield, Meszaros, Minniti, Minsley, Muna, Myers, Nair, Correa~do
  Nascimento, Newman, Nitschelm, Olmstead, Oravetz, Oravetz, Ortega~Minakata,
  Pace, Padilla, Palicio, Pan, Pan, Parikh, Parker, Peirani, Penny, Percival,
  Perez-Fournon, Peterken, Pinsonneault, Prakash, Raddick, Raichoor, Riffel,
  Riffel, Rix, Robin, Roman-Lopes, Rose, Ross, Rossi, Rowlands, Rubin,
  S{\'a}nchez, S{\'a}nchez-Gallego, Sayres, Schaefer, Schiavon, Schimoia,
  Schlafly, Schlegel, Schneider, Schultheis, Seo, Shamsi, Shao, Shen, Shetty,
  Simonian, Smethurst, Sobeck, Souter, Spindler, Stark, Stassun, Steinmetz,
  Storchi-Bergmann, Stringfellow, Su{\'a}rez, Sun, Taghizadeh-Popp, Talbot,
  Tayar, Thakar, Thomas, Tissera, Tojeiro, Troup, Unda-Sanzana, Valenzuela,
  Vargas-Maga{\~n}a, V{\'a}zquez-Mata, Wake, Weaver, Weijmans, Westfall, Wild,
  Wilson, Woods, Yan, Yang, Zamora, Zasowski, Zhang, Zheng, Zheng, Zhu, Zinn,
  \& Zou}]{Aguado2019}
Aguado, D.~S., Ahumada, R., Almeida, A., {et~al.} 2019,
  \href{http://dx.doi.org/10.3847/1538-4365/aaf651}{\JournalTitle{\apjs}, 240,
  23}

\bibitem[{Alatalo {et~al.}(2016)Alatalo, Cales, Rich, Appleton, Kewley, Lacy,
  Lanz, Medling, \& Nyland}]{Alatalo2016}
Alatalo, K., Cales, S.~L., Rich, J.~A., {et~al.} 2016,
  \href{http://dx.doi.org/10.3847/0067-0049/224/2/38}{\JournalTitle{\apjs},
  224, 38}

\bibitem[{Aoyama {et~al.}(2018)Aoyama, Hou, Hirashita, Nagamine, \&
  Shimizu}]{Aoyama2018}
Aoyama, S., Hou, K.-C., Hirashita, H., Nagamine, K., \& Shimizu, I. 2018,
  \href{http://dx.doi.org/10.1093/mnras/sty1431}{\JournalTitle{\mnras}, 478,
  4905}

\bibitem[{Aoyama {et~al.}(2017)Aoyama, Hou, Shimizu, Hirashita, Todoroki, Choi,
  \& Nagamine}]{Aoyama2017}
Aoyama, S., Hou, K.-C., Shimizu, I., {et~al.} 2017,
  \href{http://dx.doi.org/10.1093/mnras/stw3061}{\JournalTitle{\mnras}, 466,
  105}

\bibitem[{Asano {et~al.}(2013)Asano, Takeuchi, Hirashita, \& Inoue}]{Asano2013}
Asano, R.~S., Takeuchi, T.~T., Hirashita, H., \& Inoue, A.~K. 2013,
  \href{http://dx.doi.org/10.5047/eps.2012.04.014}{\JournalTitle{Earth,
  Planets, and Space}, 65, 213}

\bibitem[{{Astropy Collaboration} {et~al.}(2013){Astropy Collaboration},
  Robitaille, Tollerud, Greenfield, Droettboom, Bray, Aldcroft, Davis,
  Ginsburg, Price-Whelan, Kerzendorf, Conley, Crighton, Barbary, Muna,
  Ferguson, Grollier, Parikh, Nair, Unther, Deil, Woillez, Conseil, Kramer,
  Turner, Singer, Fox, Weaver, Zabalza, Edwards, Azalee~Bostroem, Burke, Casey,
  Crawford, Dencheva, Ely, Jenness, Labrie, Lim, Pierfederici, Pontzen, Ptak,
  Refsdal, Servillat, \& Streicher}]{AstropyCollaboration2013}
{Astropy Collaboration}, Robitaille, T.~P., Tollerud, E.~J., {et~al.} 2013,
  \href{http://dx.doi.org/10.1051/0004-6361/201322068}{\JournalTitle{\aap},
  558, A33}

\bibitem[{{Astropy Collaboration} {et~al.}(2018){Astropy Collaboration},
  Price-Whelan, Sip{\H o}cz, G{\"u}nther, Lim, Crawford, Conseil, Shupe, Craig,
  Dencheva, Ginsburg, VanderPlas, Bradley, P{\'e}rez-Su{\'a}rez, de~Val-Borro,
  Aldcroft, Cruz, Robitaille, Tollerud, Ardelean, Babej, Bach, Bachetti,
  Bakanov, Bamford, Barentsen, Barmby, Baumbach, Berry, Biscani, Boquien,
  Bostroem, Bouma, Brammer, Bray, Breytenbach, Buddelmeijer, Burke, Calderone,
  Cano~Rodr{\'{\i}}guez, Cara, Cardoso, Cheedella, Copin, Corrales, Crichton,
  D'Avella, Deil, Depagne, Dietrich, Donath, Droettboom, Earl, Erben, Fabbro,
  Ferreira, Finethy, Fox, Garrison, Gibbons, Goldstein, Gommers, Greco,
  Greenfield, Groener, Grollier, Hagen, Hirst, Homeier, Horton, Hosseinzadeh,
  Hu, Hunkeler, Ivezi{\'c}, Jain, Jenness, Kanarek, Kendrew, Kern, Kerzendorf,
  Khvalko, King, Kirkby, Kulkarni, Kumar, Lee, Lenz, Littlefair, Ma, Macleod,
  Mastropietro, McCully, Montagnac, Morris, Mueller, Mumford, Muna, Murphy,
  Nelson, Nguyen, Ninan, N{\"o}the, Ogaz, Oh, Parejko, Parley, Pascual, Patil,
  Patil, Plunkett, Prochaska, Rastogi, Reddy~Janga, Sabater, Sakurikar,
  Seifert, Sherbert, Sherwood-Taylor, Shih, Sick, Silbiger, Singanamalla,
  Singer, Sladen, Sooley, Sornarajah, Streicher, Teuben, Thomas, Tremblay,
  Turner, Terr{\'o}n, van Kerkwijk, de~la Vega, Watkins, Weaver, Whitmore,
  Woillez, Zabalza, \& Contributors}]{AstropyCollaboration2018}
{Astropy Collaboration}, Price-Whelan, A.~M., Sip{\H o}cz, B.~M., {et~al.}
  2018, \href{http://dx.doi.org/10.3847/1538-3881/aabc4f}{\JournalTitle{\aj},
  156, 123}

\bibitem[{Baldwin {et~al.}(1981)Baldwin, Phillips, \& Terlevich}]{Baldwin1981}
Baldwin, J.~A., Phillips, M.~M., \& Terlevich, R. 1981,
  \href{http://dx.doi.org/10.1086/130766}{\JournalTitle{\pasp}, 93, 5}

\bibitem[{{Battisti} {et~al.}(2016){Battisti}, {Calzetti}, \&
  {Chary}}]{Battisti2016}
{Battisti}, A.~J., {Calzetti}, D., \& {Chary}, R.-R. 2016,
  \href{http://dx.doi.org/10.3847/0004-637X/818/1/13}{\JournalTitle{\apj}, 818,
  13}

\bibitem[{Battisti {et~al.}(2017)Battisti, Calzetti, \& Chary}]{Battisti2017a}
Battisti, A.~J., Calzetti, D., \& Chary, R.-R. 2017,
  \href{http://dx.doi.org/10.3847/1538-4357/aa9a43}{\JournalTitle{\apj}, 851,
  90}

\bibitem[{Blanton \& Roweis(2007)}]{Blanton2007}
Blanton, M.~R., \& Roweis, S. 2007,
  \href{http://dx.doi.org/10.1086/510127}{\JournalTitle{\aj}, 133, 734}

\bibitem[{Blanton {et~al.}(2017)Blanton, Bershady, Abolfathi, Albareti,
  Allende~Prieto, Almeida, Alonso-Garc{\'{\i}}a, Anders, Anderson, Andrews,
  Aquino-Ort{\'{\i}}z, Arag{\'o}n-Salamanca, Argudo-Fern{\'a}ndez, Armengaud,
  Aubourg, Avila-Reese, Badenes, Bailey, Barger, Barrera-Ballesteros, Bartosz,
  Bates, Baumgarten, Bautista, Beaton, Beers, Belfiore, Bender, Berlind,
  Bernardi, Beutler, Bird, Bizyaev, Blanc, Blomqvist, Bolton, Boquien,
  Borissova, van~den Bosch, Bovy, Brandt, Brinkmann, Brownstein, Bundy,
  Burgasser, Burtin, Busca, Cappellari, Delgado~Carigi, Carlberg,
  Carnero~Rosell, Carrera, Chanover, Cherinka, Cheung, G{\'o}mez Maqueo~Chew,
  Chiappini, Doohyun~Choi, Chojnowski, Chuang, Chung, Cirolini, Clerc, Cohen,
  Comparat, da~Costa, Cousinou, Covey, Crane, Croft, Cruz-Gonzalez,
  Garrido~Cuadra, Cunha, Damke, Darling, Davies, Dawson, de~la Macorra,
  Dell'Agli, De~Lee, Delubac, Di~Mille, Diamond-Stanic, Cano-D{\'{\i}}az,
  Donor, Downes, Drory, du~Mas~des Bourboux, Duckworth, Dwelly, Dyer, Ebelke,
  Eigenbrot, Eisenstein, Emsellem, Eracleous, Escoffier, Evans, Fan,
  Fern{\'a}ndez-Alvar, Fernandez-Trincado, Feuillet, Finoguenov, Fleming,
  Font-Ribera, Fredrickson, Freischlad, Frinchaboy, Fuentes, Galbany,
  Garcia-Dias, Garc{\'{\i}}a-Hern{\'a}ndez, Gaulme, Geisler, Gelfand,
  Gil-Mar{\'{\i}}n, Gillespie, Goddard, Gonzalez-Perez, Grabowski, Green,
  Grier, Gunn, Guo, Guy, Hagen, Hahn, Hall, Harding, Hasselquist, Hawley,
  Hearty, Gonzalez~Hern{\'a}ndez, Ho, Hogg, Holley-Bockelmann, Holtzman,
  Holzer, Huehnerhoff, Hutchinson, Hwang, Ibarra-Medel, da~Silva~Ilha, Ivans,
  Ivory, Jackson, Jensen, Johnson, Jones, J{\"o}nsson, Jullo, Kamble,
  Kinemuchi, Kirkby, Kitaura, Klaene, Knapp, Kneib, Kollmeier, Lacerna, Lane,
  Lang, Law, Lazarz, Lee, Le~Goff, Liang, Li, Li, Lian, Lima, Lin, Lin,
  Bertran~de Lis, Liu, de~Icaza~Lizaola, Long, Lucatello, Lundgren, MacDonald,
  Deconto~Machado, MacLeod, Mahadevan, Geimba~Maia, Maiolino, Majewski,
  Malanushenko, Malanushenko, Manchado, Mao, Maraston, Marques-Chaves,
  Masseron, Masters, McBride, McDermid, McGrath, McGreer, Medina~Pe{\~n}a,
  Melendez, Merloni, Merrifield, Meszaros, Meza, Minchev, Minniti, Miyaji,
  More, Mulchaey, M{\"u}ller-S{\'a}nchez, Muna, Munoz, Myers, Nair, Nandra,
  Correa~do Nascimento, Negrete, Ness, Newman, Nichol, Nidever, Nitschelm,
  Ntelis, O'Connell, Oelkers, Oravetz, Oravetz, Pace, Padilla,
  Palanque-Delabrouille, Alonso~Palicio, Pan, Parejko, Parikh, P{\^a}ris, Park,
  Patten, Peirani, Pellejero-Ibanez, Penny, Percival, Perez-Fournon, Petitjean,
  Pieri, Pinsonneault, Pisani, Poleski, Prada, Prakash, Queiroz, Raddick,
  Raichoor, Barboza~Rembold, Richstein, Riffel, Riffel, Rix, Robin, Rockosi,
  Rodr{\'{\i}}guez-Torres, Roman-Lopes, Rom{\'a}n-Z{\'u}{\~n}iga, Rosado, Ross,
  Rossi, Ruan, Ruggeri, Rykoff, Salazar-Albornoz, Salvato, S{\'a}nchez, Aguado,
  S{\'a}nchez-Gallego, Santana, Santiago, Sayres, Schiavon, da~Silva~Schimoia,
  Schlafly, Schlegel, Schneider, Schultheis, Schuster, Schwope, Seo, Shao,
  Shen, Shetrone, Shull, Simon, Skinner, Skrutskie, Slosar, Smith, Sobeck,
  Sobreira, Somers, Souto, Stark, Stassun, Stauffer, Steinmetz,
  Storchi-Bergmann, Streblyanska, Stringfellow, Su{\'a}rez, Sun, Suzuki,
  Szigeti, Taghizadeh-Popp, Tang, Tao, Tayar, Tembe, Teske, Thakar, Thomas,
  Thompson, Tinker, Tissera, Tojeiro, Hernandez~Toledo, de~la Torre, Tremonti,
  Troup, Valenzuela, Martinez~Valpuesta, Vargas-Gonz{\'a}lez,
  Vargas-Maga{\~n}a, Vazquez, Villanova, Vivek, Vogt, Wake, Walterbos, Wang,
  Weaver, Weijmans, Weinberg, Westfall, Whelan, Wild, Wilson, Wood-Vasey,
  Wylezalek, Xiao, Yan, Yang, Ybarra, Y{\`e}che, Zakamska, Zamora, Zarrouk,
  Zasowski, Zhang, Zhao, Zheng, Zheng, Zhou, Zhou, Zhu, Zoccali, \&
  Zou}]{Blanton2017}
Blanton, M.~R., Bershady, M.~A., Abolfathi, B., {et~al.} 2017,
  \href{http://dx.doi.org/10.3847/1538-3881/aa7567}{\JournalTitle{\aj}, 154,
  28}

\bibitem[{Boquien {et~al.}(2010)Boquien, Bendo, Calzetti, Dale, Engelbracht,
  Kennicutt, Lee, van Zee, \& Moustakas}]{Boquien2010}
Boquien, M., Bendo, G., Calzetti, D., {et~al.} 2010,
  \href{http://dx.doi.org/10.1088/0004-637X/713/1/626}{\JournalTitle{\apj},
  713, 626}

\bibitem[{Boquien {et~al.}(2013)Boquien, Boselli, Buat, Baes, Bendo, Boissier,
  Ciesla, Cooray, Cortese, Eales, Koda, Lebouteiller, De~Looze, Smith,
  Spinoglio, \& Wilson}]{Boquien2013}
Boquien, M., Boselli, A., Buat, V., {et~al.} 2013,
  \href{http://dx.doi.org/10.1051/0004-6361/201220768}{\JournalTitle{\aap},
  554, A14}

\bibitem[{Boquien {et~al.}(2015)Boquien, Calzetti, Aalto, Boselli, Braine,
  Buat, Combes, Israel, Kramer, Lord, Rela{\~n}o, Rosolowsky, Stacey,
  Tabatabaei, van~der Tak, van~der Werf, Verley, \& Xilouris}]{Boquien2015}
Boquien, M., Calzetti, D., Aalto, S., {et~al.} 2015,
  \href{http://dx.doi.org/10.1051/0004-6361/201423518}{\JournalTitle{\aap},
  578, A8}

\bibitem[{Brinchmann {et~al.}(2013)Brinchmann, Charlot, Kauffmann, Heckman,
  White, \& Tremonti}]{Brinchmann2013}
Brinchmann, J., Charlot, S., Kauffmann, G., {et~al.} 2013,
  \href{http://dx.doi.org/10.1093/mnras/stt551}{\JournalTitle{\mnras}, 432,
  2112}

\bibitem[{Bruzual~A.(1983)}]{BruzualA.1983}
Bruzual~A., G. 1983,
  \href{http://dx.doi.org/10.1086/161352}{\JournalTitle{\apj}, 273, 105}

\bibitem[{Bundy {et~al.}(2015)Bundy, Bershady, Law, Yan, Drory, MacDonald,
  Wake, Cherinka, S{\'a}nchez-Gallego, Weijmans, Thomas, Tremonti, Masters,
  Coccato, Diamond-Stanic, Arag{\'o}n-Salamanca, Avila-Reese, Badenes,
  Falc{\'o}n-Barroso, Belfiore, Bizyaev, Blanc, Bland-Hawthorn, Blanton,
  Brownstein, Byler, Cappellari, Conroy, Dutton, Emsellem, Etherington,
  Frinchaboy, Fu, Gunn, Harding, Johnston, Kauffmann, Kinemuchi, Klaene,
  Knapen, Leauthaud, Li, Lin, Maiolino, Malanushenko, Malanushenko, Mao,
  Maraston, McDermid, Merrifield, Nichol, Oravetz, Pan, Parejko, Sanchez,
  Schlegel, Simmons, Steele, Steinmetz, Thanjavur, Thompson, Tinker, van~den
  Bosch, Westfall, Wilkinson, Wright, Xiao, \& Zhang}]{Bundy2015}
Bundy, K., Bershady, M.~A., Law, D.~R., {et~al.} 2015,
  \href{http://dx.doi.org/10.1088/0004-637X/798/1/7}{\JournalTitle{\apj}, 798,
  7}

\bibitem[{Calabr{\`o} {et~al.}(2018)Calabr{\`o}, Daddi, Cassata, Onodera,
  Gobat, Puglisi, Jin, Liu, Amor{\'{\i}}n, Arimoto, Boquien, Carraro, Elbaz,
  Ibar, Juneau, Mannucci, M{\'e}ndez~Hern{\'a}nez, Oliva, Rodighiero,
  Valentino, \& Zanella}]{Calabro2018}
Calabr{\`o}, A., Daddi, E., Cassata, P., {et~al.} 2018,
  \href{http://dx.doi.org/10.3847/2041-8213/aad33e}{\JournalTitle{\apjl}, 862,
  L22}

\bibitem[{Calzetti(1997)}]{Calzetti1997}
Calzetti, D. 1997, \href{http://dx.doi.org/10.1063/1.53764}{in American
  Institute of Physics Conference Series, Vol. 408, The Ultraviolet Universe at
  Low and High Redshift, ed. W.~H. {Waller}}, 403

\bibitem[{Calzetti(2001)}]{Calzetti2001}
Calzetti, D. 2001,
  \href{http://dx.doi.org/10.1086/324269}{\JournalTitle{\pasp}, 113, 1449}

\bibitem[{{Calzetti} {et~al.}(2000){Calzetti}, {Armus}, {Bohlin}, {Kinney},
  {Koornneef}, \& {Storchi-Bergmann}}]{Calzetti2000}
{Calzetti}, D., {Armus}, L., {Bohlin}, R.~C., {et~al.} 2000,
  \href{http://dx.doi.org/10.1086/308692}{\JournalTitle{\apj}, 533, 682}

\bibitem[{{Calzetti} {et~al.}(1994){Calzetti}, {Kinney}, \&
  {Storchi-Bergmann}}]{Calzetti1994}
{Calzetti}, D., {Kinney}, A.~L., \& {Storchi-Bergmann}, T. 1994,
  \href{http://dx.doi.org/10.1086/174346}{\JournalTitle{\apj}, 429, 582}

\bibitem[{Cano-D{\'{\i}}az {et~al.}(2016)Cano-D{\'{\i}}az, S{\'a}nchez,
  Zibetti, Ascasibar, Bland-Hawthorn, Ziegler, Gonz{\'a}lez~Delgado, Walcher,
  Garc{\'{\i}}a-Benito, Mast, Mendoza-P{\'e}rez, Falc{\'o}n-Barroso, Galbany,
  Husemann, Kehrig, Marino, S{\'a}nchez-Bl{\'a}zquez, L{\'o}pez-Cob{\'a},
  L{\'o}pez-S{\'a}nchez, \& Vilchez}]{Cano-Diaz2016}
Cano-D{\'{\i}}az, M., S{\'a}nchez, S.~F., Zibetti, S., {et~al.} 2016,
  \href{http://dx.doi.org/10.3847/2041-8205/821/2/L26}{\JournalTitle{\apjl},
  821, L26}

\bibitem[{{Cardelli} {et~al.}(1989){Cardelli}, {Clayton}, \&
  {Mathis}}]{Cardelli1989}
{Cardelli}, J.~A., {Clayton}, G.~C., \& {Mathis}, J.~S. 1989,
  \href{http://dx.doi.org/10.1086/167900}{\JournalTitle{\apj}, 345, 245}

\bibitem[{Casasola {et~al.}(2019)Casasola, Bianchi, Vis, Magrini, Corbelli,
  Clark, Fritz, Nersesian, Viaene, Baes, Cassara', Davies, Looze, Dobbels,
  Galametz, Galliano, Jones, Madden, Mosenkov, Trcka, \&
  Xilouris}]{Casasola2019}
Casasola, V., Bianchi, S., Vis, P.~D., {et~al.} 2019,
  \href{http://arxiv.org/abs/1911.09187v1}{{\sffamily 1911.09187v1}}

\bibitem[{{Charlot} \& {Fall}(2000)}]{Charlot2000}
{Charlot}, S., \& {Fall}, S.~M. 2000,
  \href{http://dx.doi.org/10.1086/309250}{\JournalTitle{\apj}, 539, 718}

\bibitem[{Chevallard {et~al.}(2013)Chevallard, Charlot, Wandelt, \&
  Wild}]{Chevallard2013}
Chevallard, J., Charlot, S., Wandelt, B., \& Wild, V. 2013,
  \href{http://dx.doi.org/10.1093/mnras/stt523}{\JournalTitle{\mnras}, 432,
  2061}

\bibitem[{Chiang {et~al.}(2018)Chiang, Sandstrom, Chastenet, Johnson, Leroy, \&
  Utomo}]{Chiang2018}
Chiang, I.-D., Sandstrom, K.~M., Chastenet, J., {et~al.} 2018,
  \href{http://dx.doi.org/10.3847/1538-4357/aadc5f}{\JournalTitle{\apj}, 865,
  117}

\bibitem[{Chisholm {et~al.}(2018)Chisholm, Tremonti, \&
  Leitherer}]{Chisholm2018}
Chisholm, J., Tremonti, C., \& Leitherer, C. 2018,
  \href{http://dx.doi.org/10.1093/mnras/sty2380}{\JournalTitle{\mnras}, 481,
  1690}

\bibitem[{Cid~Fernandes {et~al.}(2013)Cid~Fernandes, P{\'e}rez,
  Garc{\'{\i}}a~Benito, Gonz{\'a}lez~Delgado, de~Amorim, S{\'a}nchez, Husemann,
  Falc{\'o}n~Barroso, S{\'a}nchez-Bl{\'a}zquez, Walcher, \&
  Mast}]{CidFernandes2013}
Cid~Fernandes, R., P{\'e}rez, E., Garc{\'{\i}}a~Benito, R., {et~al.} 2013,
  \href{http://dx.doi.org/10.1051/0004-6361/201220616}{\JournalTitle{\aap},
  557, A86}

\bibitem[{Croom {et~al.}(2012)Croom, Lawrence, Bland-Hawthorn, Bryant, Fogarty,
  Richards, Goodwin, Farrell, Miziarski, Heald, Jones, Lee, Colless, Brough,
  Hopkins, Bauer, Birchall, Ellis, Horton, Leon-Saval, Lewis,
  L{\'o}pez-S{\'a}nchez, Min, Trinh, \& Trowland}]{Croom2012}
Croom, S.~M., Lawrence, J.~S., Bland-Hawthorn, J., {et~al.} 2012,
  \href{http://dx.doi.org/10.1111/j.1365-2966.2011.20365.x}{\JournalTitle{\mnras},
  421, 872}

\bibitem[{De~Vis {et~al.}(2017{\natexlab{a}})De~Vis, Dunne, Maddox, Gomez,
  Clark, Bauer, Viaene, Schofield, Baes, Baker, Bourne, Driver, Dye, Eales,
  Furlanetto, Ivison, Robotham, Rowlands, Smith, Smith, Valiante, \&
  Wright}]{DeVis2017a}
De~Vis, P., Dunne, L., Maddox, S., {et~al.} 2017{\natexlab{a}},
  \href{http://dx.doi.org/10.1093/mnras/stw2501}{\JournalTitle{\mnras}, 464,
  4680}

\bibitem[{De~Vis {et~al.}(2017{\natexlab{b}})De~Vis, Gomez, Schofield, Maddox,
  Dunne, Baes, Cigan, Clark, Gomez, Lara-L{\'o}pez, \& Owers}]{DeVis2017}
De~Vis, P., Gomez, H.~L., Schofield, S.~P., {et~al.} 2017{\natexlab{b}},
  \href{http://dx.doi.org/10.1093/mnras/stx981}{\JournalTitle{\mnras}, 471,
  1743}

\bibitem[{De~Vis {et~al.}(2019)De~Vis, Jones, Viaene, Casasola, Clark, Baes,
  Bianchi, Cassara, Davies, De~Looze, Galametz, Galliano, Lianou, Madden,
  Manilla-Robles, Mosenkov, Nersesian, Roychowdhury, Xilouris, \&
  Ysard}]{DeVis2019}
De~Vis, P., Jones, A., Viaene, S., {et~al.} 2019,
  \href{http://dx.doi.org/10.1051/0004-6361/201834444}{\JournalTitle{\aap},
  623, A5}

\bibitem[{Dole {et~al.}(2006)Dole, Lagache, Puget, Caputi, Fern{\'a}ndez-Conde,
  Le~Floc'h, Papovich, P{\'e}rez-Gonz{\'a}lez, Rieke, \& Blaylock}]{Dole2006}
Dole, H., Lagache, G., Puget, J.-L., {et~al.} 2006,
  \href{http://dx.doi.org/10.1051/0004-6361:20054446}{\JournalTitle{\aap}, 451,
  417}

\bibitem[{Dopita(1997)}]{Dopita1997}
Dopita, M.~A. 1997,
  \href{http://dx.doi.org/10.1086/310804}{\JournalTitle{\apjl}, 485, L41}

\bibitem[{Dopita {et~al.}(2000)Dopita, Kewley, Heisler, \&
  Sutherland}]{Dopita2000}
Dopita, M.~A., Kewley, L.~J., Heisler, C.~A., \& Sutherland, R.~S. 2000,
  \href{http://dx.doi.org/10.1086/309538}{\JournalTitle{\apj}, 542, 224}

\bibitem[{Dopita {et~al.}(2013)Dopita, Sutherland, Nicholls, Kewley, \&
  Vogt}]{Dopita2013}
Dopita, M.~A., Sutherland, R.~S., Nicholls, D.~C., Kewley, L.~J., \& Vogt, F.
  P.~A. 2013,
  \href{http://dx.doi.org/10.1088/0067-0049/208/1/10}{\JournalTitle{\apjs},
  208, 10}

\bibitem[{Draine(2009)}]{Draine2009}
Draine, B.~T. 2009, \href{http://adsabs.harvard.edu/abs/2009ASPC..414..453D}{in
  Astronomical Society of the Pacific Conference Series, Vol. 414, Cosmic Dust
  - Near and Far, ed. T.~{Henning}, E.~{Gr{\"u}n}, \& J.~{Steinacker}}, 453

\bibitem[{Dwek(1998)}]{Dwek1998}
Dwek, E. 1998, \href{http://dx.doi.org/10.1086/305829}{\JournalTitle{\apj},
  501, 643}

\bibitem[{Elbaz {et~al.}(2010)Elbaz, Hwang, Magnelli, Daddi, Aussel, Altieri,
  Amblard, Andreani, Arumugam, Auld, Babbedge, Berta, Blain, Bock, Bongiovanni,
  Boselli, Buat, Burgarella, Castro-Rodriguez, Cava, Cepa, Chanial, Chary,
  Cimatti, Clements, Conley, Conversi, Cooray, Dickinson, Dominguez, Dowell,
  Dunlop, Dwek, Eales, Farrah, F{\"o}rster~Schreiber, Fox, Franceschini, Gear,
  Genzel, Glenn, Griffin, Gruppioni, Halpern, Hatziminaoglou, Ibar, Isaak,
  Ivison, Lagache, Le~Borgne, Le~Floc'h, Levenson, Lu, Lutz, Madden, Maffei,
  Magdis, Mainetti, Maiolino, Marchetti, Mortier, Nguyen, Nordon, O'Halloran,
  Okumura, Oliver, Omont, Page, Panuzzo, Papageorgiou, Pearson, Perez~Fournon,
  P{\'e}rez~Garc{\'{\i}}a, Poglitsch, Pohlen, Popesso, Pozzi, Rawlings,
  Rigopoulou, Riguccini, Rizzo, Rodighiero, Roseboom, Rowan-Robinson,
  Saintonge, Sanchez~Portal, Santini, Sauvage, Schulz, Scott, Seymour, Shao,
  Shupe, Smith, Stevens, Sturm, Symeonidis, Tacconi, Trichas, Tugwell, Vaccari,
  Valtchanov, Vieira, Vigroux, Wang, Ward, Wright, Xu, \& Zemcov}]{Elbaz2010}
Elbaz, D., Hwang, H.~S., Magnelli, B., {et~al.} 2010,
  \href{http://dx.doi.org/10.1051/0004-6361/201014687}{\JournalTitle{\aap},
  518, L29}

\bibitem[{Engelbracht {et~al.}(2008)Engelbracht, Rieke, Gordon, Smith, Werner,
  Moustakas, Willmer, \& Vanzi}]{Engelbracht2008}
Engelbracht, C.~W., Rieke, G.~H., Gordon, K.~D., {et~al.} 2008,
  \href{http://dx.doi.org/10.1086/529513}{\JournalTitle{\apj}, 678, 804}

\bibitem[{Flores-Fajardo {et~al.}(2011)Flores-Fajardo, Morisset, Stasi{\'n}ska,
  \& Binette}]{Flores-Fajardo2011}
Flores-Fajardo, N., Morisset, C., Stasi{\'n}ska, G., \& Binette, L. 2011,
  \href{http://dx.doi.org/10.1111/j.1365-2966.2011.18848.x}{\JournalTitle{\mnras},
  415, 2182}

\bibitem[{Galliano(2018)}]{Galliano2018}
Galliano, F. 2018,
  \href{http://dx.doi.org/10.1093/mnras/sty189}{\JournalTitle{\mnras}, 476,
  1445}

\bibitem[{Galliano {et~al.}(2018)Galliano, Galametz, \& Jones}]{Galliano2018a}
Galliano, F., Galametz, M., \& Jones, A.~P. 2018,
  \href{http://dx.doi.org/10.1146/annurev-astro-081817-051900}{\JournalTitle{\araa},
  56, 673}

\bibitem[{Galliano {et~al.}(2011)Galliano, Hony, Bernard, Bot, Madden,
  Roman-Duval, Galametz, Li, Meixner, Engelbracht, Lebouteiller, Misselt,
  Montiel, Panuzzo, Reach, \& Skibba}]{Galliano2011}
Galliano, F., Hony, S., Bernard, J.-P., {et~al.} 2011,
  \href{http://dx.doi.org/10.1051/0004-6361/201117952}{\JournalTitle{\aap},
  536, A88}

\bibitem[{Gao {et~al.}(2018)Gao, Wang, Kong, Lin, Liu, Liu, Liu, Hu,
  Berhane~Teklu, Chen, \& Zhao}]{Gao2018}
Gao, Y., Wang, E., Kong, X., {et~al.} 2018,
  \href{http://dx.doi.org/10.3847/1538-4357/aae9f1}{\JournalTitle{\apj}, 868,
  89}

\bibitem[{Garn {et~al.}(2010)Garn, Sobral, Best, Geach, Smail, Cirasuolo,
  Dalton, Dunlop, McLure, \& Farrah}]{Garn2010}
Garn, T., Sobral, D., Best, P.~N., {et~al.} 2010,
  \href{http://dx.doi.org/10.1111/j.1365-2966.2009.16042.x}{\JournalTitle{\mnras},
  402, 2017}

\bibitem[{Gjergo {et~al.}(2018)Gjergo, Granato, Murante, Ragone-Figueroa,
  Tornatore, \& Borgani}]{Gjergo2018}
Gjergo, E., Granato, G.~L., Murante, G., {et~al.} 2018,
  \href{http://dx.doi.org/10.1093/mnras/sty1564}{\JournalTitle{\mnras}, 479,
  2588}

\bibitem[{Grasha {et~al.}(2019)Grasha, Calzetti, Adamo, Kennicutt, Elmegreen,
  Messa, Dale, Fedorenko, Mahadevan, Grebel, {et~al.}}]{Grasha2019}
Grasha, K., Calzetti, D., Adamo, A., {et~al.} 2019,
  \href{http://dx.doi.org/10.1093/mnras/sty3424}{\JournalTitle{Monthly Notices
  of the Royal Astronomical Society}, 483, 4707}

\bibitem[{Haffner {et~al.}(2009)Haffner, Dettmar, Beckman, Wood, Slavin,
  Giammanco, Madsen, Zurita, \& Reynolds}]{Haffner2009}
Haffner, L.~M., Dettmar, R.-J., Beckman, J.~E., {et~al.} 2009,
  \href{http://dx.doi.org/10.1103/RevModPhys.81.969}{\JournalTitle{Reviews of
  Modern Physics}, 81, 969}

\bibitem[{Hirashita \& Aoyama(2019)}]{Hirashita2019}
Hirashita, H., \& Aoyama, S. 2019,
  \href{http://dx.doi.org/10.1093/mnras/sty2838}{\JournalTitle{\mnras}, 482,
  2555}

\bibitem[{Hirashita \& Voshchinnikov(2014)}]{Hirashita2014}
Hirashita, H., \& Voshchinnikov, N.~V. 2014,
  \href{http://dx.doi.org/10.1093/mnras/stt1997}{\JournalTitle{\mnras}, 437,
  1636}

\bibitem[{Hollyhead {et~al.}(2015)Hollyhead, Bastian, Adamo, Silva-Villa, Dale,
  Ryon, \& Gazak}]{Hollyhead2015}
Hollyhead, K., Bastian, N., Adamo, A., {et~al.} 2015,
  \href{http://dx.doi.org/10.1093/mnras/stv331}{\JournalTitle{Monthly Notices
  of the Royal Astronomical Society}, 449, 1106}

\bibitem[{Hoopes \& Walterbos(2003)}]{Hoopes2003}
Hoopes, C.~G., \& Walterbos, R. A.~M. 2003,
  \href{http://dx.doi.org/10.1086/367954}{\JournalTitle{\apj}, 586, 902}

\bibitem[{Hsieh {et~al.}(2017)Hsieh, Lin, Lin, Pan, Hsu, S{\'a}nchez,
  Cano-D{\'{\i}}az, Zhang, Yan, Barrera-Ballesteros, Boquien, Riffel,
  Brownstein, Cruz-Gonz{\'a}lez, Hagen, Ibarra, Pan, Bizyaev, Oravetz, \&
  Simmons}]{Hsieh2017}
Hsieh, B.~C., Lin, L., Lin, J.~H., {et~al.} 2017,
  \href{http://dx.doi.org/10.3847/2041-8213/aa9d80}{\JournalTitle{\apjl}, 851,
  L24}

\bibitem[{Hunt \& Hirashita(2009)}]{Hunt2009}
Hunt, L.~K., \& Hirashita, H. 2009,
  \href{http://dx.doi.org/10.1051/0004-6361/200912020}{\JournalTitle{\aap},
  507, 1327}

\bibitem[{Hunter(2007)}]{Hunter2007}
Hunter, J.~D. 2007,
  \href{http://dx.doi.org/10.1109/MCSE.2007.55}{\JournalTitle{Computing in
  Science and Engineering}, 9, 90}

\bibitem[{Kaplan {et~al.}(2016)Kaplan, Jogee, Kewley, Blanc, Weinzirl, Song,
  Drory, Luo, \& van~den Bosch}]{Kaplan2016}
Kaplan, K.~F., Jogee, S., Kewley, L., {et~al.} 2016,
  \href{http://dx.doi.org/10.1093/mnras/stw1422}{\JournalTitle{\mnras}, 462,
  1642}

\bibitem[{Kauffmann {et~al.}(2003)Kauffmann, Heckman, White, Charlot, Tremonti,
  Brinchmann, Bruzual, Peng, Seibert, Bernardi, Blanton, Brinkmann, Castander,
  Cs{\'a}bai, Fukugita, Ivezic, Munn, Nichol, Padmanabhan, Thakar, Weinberg, \&
  York}]{Kauffmann2003}
Kauffmann, G., Heckman, T.~M., White, S. D.~M., {et~al.} 2003,
  \href{http://dx.doi.org/10.1046/j.1365-8711.2003.06291.x}{\JournalTitle{\mnras},
  341, 33}

\bibitem[{{Kauffmann} {et~al.}(2003){Kauffmann}, {Heckman}, {Tremonti},
  {Brinchmann}, {Charlot}, {White}, {Ridgway}, {Brinkmann}, {Fukugita}, {Hall},
  {Ivezi{\'c}}, {Richards}, \& {Schneider}}]{Kauffmann2003a}
{Kauffmann}, G., {Heckman}, T.~M., {Tremonti}, C., {et~al.} 2003,
  \href{http://dx.doi.org/10.1111/j.1365-2966.2003.07154.x}{\JournalTitle{\mnras},
  346, 1055}

\bibitem[{Kennicutt(1998{\natexlab{a}})}]{Kennicutt1998a}
Kennicutt, Jr., R.~C. 1998{\natexlab{a}},
  \href{http://dx.doi.org/10.1086/305588}{\JournalTitle{\apj}, 498, 541}

\bibitem[{Kennicutt(1998{\natexlab{b}})}]{Kennicutt1998}
---. 1998{\natexlab{b}},
  \href{http://dx.doi.org/10.1146/annurev.astro.36.1.189}{\JournalTitle{\araa},
  36, 189}

\bibitem[{Kewley {et~al.}(2001)Kewley, Dopita, Sutherland, Heisler, \&
  Trevena}]{Kewley2001}
Kewley, L.~J., Dopita, M.~A., Sutherland, R.~S., Heisler, C.~A., \& Trevena, J.
  2001, \href{http://dx.doi.org/10.1086/321545}{\JournalTitle{\apj}, 556, 121}

\bibitem[{Kewley \& Ellison(2008)}]{Kewley2008}
Kewley, L.~J., \& Ellison, S.~L. 2008,
  \href{http://dx.doi.org/10.1086/587500}{\JournalTitle{\apj}, 681, 1183}

\bibitem[{Kewley {et~al.}(2006)Kewley, Groves, Kauffmann, \&
  Heckman}]{Kewley2006}
Kewley, L.~J., Groves, B., Kauffmann, G., \& Heckman, T. 2006,
  \href{http://dx.doi.org/10.1111/j.1365-2966.2006.10859.x}{\JournalTitle{\mnras},
  372, 961}

\bibitem[{Kim(2015)}]{Kim2015}
Kim, S. 2015, \JournalTitle{Communications for statistical applications and
  methods}, 22, 665

\bibitem[{{Kong} {et~al.}(2004){Kong}, {Charlot}, {Brinchmann}, \&
  {Fall}}]{Kong2004}
{Kong}, X., {Charlot}, S., {Brinchmann}, J., \& {Fall}, S.~M. 2004,
  \href{http://dx.doi.org/10.1111/j.1365-2966.2004.07556.x}{\JournalTitle{\mnras},
  349, 769}

\bibitem[{Koyama {et~al.}(2019)Koyama, Shimakawa, Yamamura, Kodama, \&
  Hayashi}]{Koyama2019}
Koyama, Y., Shimakawa, R., Yamamura, I., Kodama, T., \& Hayashi, M. 2019,
  \href{http://dx.doi.org/10.1093/pasj/psy113}{\JournalTitle{\pasj}, 71, 8}

\bibitem[{Koyama {et~al.}(2015)Koyama, Kodama, Hayashi, Shimakawa, Yamamura,
  Egusa, Oi, Tanaka, Tadaki, Takita, \& Makiuti}]{Koyama2015}
Koyama, Y., Kodama, T., Hayashi, M., {et~al.} 2015,
  \href{http://dx.doi.org/10.1093/mnras/stv1599}{\JournalTitle{\mnras}, 453,
  879}

\bibitem[{Kreckel {et~al.}(2016)Kreckel, Blanc, Schinnerer, Groves, Adamo,
  Hughes, \& Meidt}]{Kreckel2016}
Kreckel, K., Blanc, G.~A., Schinnerer, E., {et~al.} 2016,
  \href{http://dx.doi.org/10.3847/0004-637X/827/2/103}{\JournalTitle{\apj},
  827, 103}

\bibitem[{Kreckel {et~al.}(2013)Kreckel, Groves, Schinnerer, Johnson, Aniano,
  Calzetti, Croxall, Draine, Gordon, Crocker, Dale, Hunt, Kennicutt, Meidt,
  Smith, \& Tabatabaei}]{Kreckel2013}
Kreckel, K., Groves, B., Schinnerer, E., {et~al.} 2013,
  \href{http://dx.doi.org/10.1088/0004-637X/771/1/62}{\JournalTitle{\apj}, 771,
  62}

\bibitem[{Lacerda {et~al.}(2018)Lacerda, Cid~Fernandes, Couto, Stasi{\'n}ska,
  Garc{\'{\i}}a-Benito, Vale~Asari, P{\'e}rez, Gonz{\'a}lez~Delgado,
  S{\'a}nchez, \& de~Amorim}]{Lacerda2018}
Lacerda, E. A.~D., Cid~Fernandes, R., Couto, G.~S., {et~al.} 2018,
  \href{http://dx.doi.org/10.1093/mnras/stx3022}{\JournalTitle{\mnras}, 474,
  3727}

\bibitem[{Law {et~al.}(2016)Law, Cherinka, Yan, Andrews, Bershady, Bizyaev,
  Blanc, Blanton, Bolton, Brownstein, Bundy, Chen, Drory, D'Souza, Fu, Jones,
  Kauffmann, MacDonald, Masters, Newman, Parejko, S{\'a}nchez-Gallego,
  S{\'a}nchez, Schlegel, Thomas, Wake, Weijmans, Westfall, \& Zhang}]{Law2016}
Law, D.~R., Cherinka, B., Yan, R., {et~al.} 2016,
  \href{http://dx.doi.org/10.3847/0004-6256/152/4/83}{\JournalTitle{\aj}, 152,
  83}

\bibitem[{Lawton {et~al.}(2010)Lawton, Gordon, Babler, Block, Bolatto, Bracker,
  Carlson, Engelbracht, Hora, Indebetouw, Madden, Meade, Meixner, Misselt, Oey,
  Oliveira, Robitaille, Sewilo, Shiao, Vijh, \& Whitney}]{Lawton2010}
Lawton, B., Gordon, K.~D., Babler, B., {et~al.} 2010,
  \href{http://dx.doi.org/10.1088/0004-637X/716/1/453}{\JournalTitle{\apj},
  716, 453}

\bibitem[{Lequeux {et~al.}(1979)Lequeux, Peimbert, Rayo, Serrano, \&
  Torres-Peimbert}]{Lequeux1979}
Lequeux, J., Peimbert, M., Rayo, J.~F., Serrano, A., \& Torres-Peimbert, S.
  1979,
  \href{http://adsabs.harvard.edu/abs/1979A%26A....80..155L}{\JournalTitle{\aap},
  80, 155}

\bibitem[{Lin {et~al.}(2016)Lin, Fang, \& Kong}]{Lin2016}
Lin, Z., Fang, G., \& Kong, X. 2016,
  \href{http://dx.doi.org/10.3847/0004-6256/152/6/191}{\JournalTitle{\aj}, 152,
  191}

\bibitem[{Lin {et~al.}(2017)Lin, Hu, Kong, Gao, Zou, Wang, Cheng, Fang, Lin, \&
  Wang}]{LinZ2017}
Lin, Z., Hu, N., Kong, X., {et~al.} 2017,
  \href{http://dx.doi.org/10.3847/1538-4357/aa6f14}{\JournalTitle{\apj}, 842,
  97}

\bibitem[{Liu {et~al.}(2018)Liu, Wang, Lin, Gao, Liu, Berhane~Teklu, \&
  Kong}]{Liu2018}
Liu, Q., Wang, E., Lin, Z., {et~al.} 2018,
  \href{http://dx.doi.org/10.3847/1538-4357/aab3d5}{\JournalTitle{\apj}, 857,
  17}

\bibitem[{Ly {et~al.}(2014)Ly, Malkan, Nagao, Kashikawa, Shimasaku, \&
  Hayashi}]{Ly2014}
Ly, C., Malkan, M.~A., Nagao, T., {et~al.} 2014,
  \href{http://dx.doi.org/10.1088/0004-637X/780/2/122}{\JournalTitle{\apj},
  780, 122}

\bibitem[{Madsen {et~al.}(2006)Madsen, Reynolds, \& Haffner}]{Madsen2006}
Madsen, G.~J., Reynolds, R.~J., \& Haffner, L.~M. 2006,
  \href{http://dx.doi.org/10.1086/508441}{\JournalTitle{\apj}, 652, 401}

\bibitem[{Mao {et~al.}(2018)Mao, Lin, \& Kong}]{Mao2018}
Mao, Y.-W., Lin, L., \& Kong, X. 2018,
  \href{http://dx.doi.org/10.3847/1538-4357/aaa29e}{\JournalTitle{\apj}, 853,
  151}

\bibitem[{Marino {et~al.}(2013)Marino, Rosales-Ortega, S{\'a}nchez, Gil~de Paz,
  V{\'{\i}}lchez, Miralles-Caballero, Kehrig, P{\'e}rez-Montero, Stanishev,
  Iglesias-P{\'a}ramo, D{\'{\i}}az, Castillo-Morales, Kennicutt,
  L{\'o}pez-S{\'a}nchez, Galbany, Garc{\'{\i}}a-Benito, Mast, Mendez-Abreu,
  Monreal-Ibero, Husemann, Walcher, Garc{\'{\i}}a-Lorenzo, Masegosa, Del
  Olmo~Orozco, Mour{\~a}o, Ziegler, Moll{\'a}, Papaderos,
  S{\'a}nchez-Bl{\'a}zquez, Gonz{\'a}lez~Delgado, Falc{\'o}n-Barroso, Roth,
  van~de Ven, \& Team}]{Marino2013}
Marino, R.~A., Rosales-Ortega, F.~F., S{\'a}nchez, S.~F., {et~al.} 2013,
  \href{http://dx.doi.org/10.1051/0004-6361/201321956}{\JournalTitle{\aap},
  559, A114}

\bibitem[{Markwardt(2009)}]{Markwardt2009}
Markwardt, C.~B. 2009,
  \href{http://adsabs.harvard.edu/abs/2009ASPC..411..251M}{in Astronomical
  Society of the Pacific Conference Series, Vol. 411, Astronomical Data
  Analysis Software and Systems XVIII, ed. D.~A. {Bohlender}, D.~{Durand}, \&
  P.~{Dowler}}, 251

\bibitem[{Meurer {et~al.}(1999)Meurer, Heckman, \& Calzetti}]{Meurer1999}
Meurer, G.~R., Heckman, T.~M., \& Calzetti, D. 1999,
  \href{http://dx.doi.org/10.1086/307523}{\JournalTitle{\apj}, 521, 64}

\bibitem[{Mosenkov {et~al.}(2019)Mosenkov, Baes, Bianchi, Casasola,
  Cassar{\`a}, Clark, Davies, De~Looze, De~Vis, Fritz, Galametz, Galliano,
  Jones, Lianou, Madden, Nersesian, Smith, Tr{\v c}ka, Verstocken, Viaene,
  Vika, \& Xilouris}]{Mosenkov2019}
Mosenkov, A.~V., Baes, M., Bianchi, S., {et~al.} 2019,
  \href{http://dx.doi.org/10.1051/0004-6361/201833932}{\JournalTitle{\aap},
  622, A132}

\bibitem[{Narayanan {et~al.}(2018{\natexlab{a}})Narayanan, Conroy, Dav{\'e},
  Johnson, \& Popping}]{Narayanan2018}
Narayanan, D., Conroy, C., Dav{\'e}, R., Johnson, B.~D., \& Popping, G.
  2018{\natexlab{a}},
  \href{http://dx.doi.org/10.3847/1538-4357/aaed25}{\JournalTitle{\apj}, 869,
  70}

\bibitem[{Narayanan {et~al.}(2018{\natexlab{b}})Narayanan, Dav{\'e}, Johnson,
  Thompson, Conroy, \& Geach}]{Narayanan2018a}
Narayanan, D., Dav{\'e}, R., Johnson, B.~D., {et~al.} 2018{\natexlab{b}},
  \href{http://dx.doi.org/10.1093/mnras/stx2860}{\JournalTitle{\mnras}, 474,
  1718}

\bibitem[{Nicholls {et~al.}(2012)Nicholls, Dopita, \&
  Sutherland}]{Nicholls2012}
Nicholls, D.~C., Dopita, M.~A., \& Sutherland, R.~S. 2012,
  \href{http://dx.doi.org/10.1088/0004-637X/752/2/148}{\JournalTitle{The
  Astrophysical Journal}, 752, 148}

\bibitem[{Oliphant(2006)}]{Oliphant2006}
Oliphant, T.~E. 2006, A Guide to NumPy, Vol.~1 (Trelgol Publishing USA)

\bibitem[{Pannella {et~al.}(2015)Pannella, Elbaz, Daddi, Dickinson, Hwang,
  Schreiber, Strazzullo, Aussel, Bethermin, Buat, Charmandaris, Cibinel,
  Juneau, Ivison, Le~Borgne, Le~Floc'h, Leiton, Lin, Magdis, Morrison,
  Mullaney, Onodera, Renzini, Salim, Sargent, Scott, Shu, \&
  Wang}]{Pannella2015}
Pannella, M., Elbaz, D., Daddi, E., {et~al.} 2015,
  \href{http://dx.doi.org/10.1088/0004-637X/807/2/141}{\JournalTitle{\apj},
  807, 141}

\bibitem[{P{\'e}rez \& Granger(2007)}]{Perez2007}
P{\'e}rez, F., \& Granger, B.~E. 2007,
  \href{http://dx.doi.org/10.1109/MCSE.2007.53}{\JournalTitle{Computing in
  Science \& Engineering}, 9, 21}

\bibitem[{Pettini \& Pagel(2004)}]{Pettini2004}
Pettini, M., \& Pagel, B. E.~J. 2004,
  \href{http://dx.doi.org/10.1111/j.1365-2966.2004.07591.x}{\JournalTitle{\mnras},
  348, L59}

\bibitem[{Popping {et~al.}(2017)Popping, Puglisi, \& Norman}]{Popping2017}
Popping, G., Puglisi, A., \& Norman, C.~A. 2017,
  \href{http://dx.doi.org/10.1093/mnras/stx2202}{\JournalTitle{\mnras}, 472,
  2315}

\bibitem[{{Price} {et~al.}(2014){Price}, {Kriek}, {Brammer}, {Conroy},
  {F{\"o}rster Schreiber}, {Franx}, {Fumagalli}, {Lundgren}, {Momcheva},
  {Nelson}, {Skelton}, {van Dokkum}, {Whitaker}, \& {Wuyts}}]{Price2014}
{Price}, S.~H., {Kriek}, M., {Brammer}, G.~B., {et~al.} 2014,
  \href{http://dx.doi.org/10.1088/0004-637X/788/1/86}{\JournalTitle{\apj}, 788,
  86}

\bibitem[{Puglisi {et~al.}(2016)Puglisi, Rodighiero, Franceschini, Talia,
  Cimatti, Baronchelli, Daddi, Renzini, Schawinski, Mancini, Silverman,
  Gruppioni, Lutz, Berta, \& Oliver}]{Puglisi2016}
Puglisi, A., Rodighiero, G., Franceschini, A., {et~al.} 2016,
  \href{http://dx.doi.org/10.1051/0004-6361/201526782}{\JournalTitle{\aap},
  586, A83}

\bibitem[{{Qin} {et~al.}(2019){Qin}, {Zheng}, {Wuyts}, {Pan}, \&
  {Ren}}]{Qin2019a}
{Qin}, J., {Zheng}, X.~Z., {Wuyts}, S., {Pan}, Z., \& {Ren}, J. 2019,
  \href{http://dx.doi.org/10.3847/1538-4357/ab4a04}{\JournalTitle{\apj}, 886,
  28}

\bibitem[{{Reddy} {et~al.}(2015){Reddy}, {Kriek}, {Shapley}, {Freeman},
  {Siana}, {Coil}, {Mobasher}, {Price}, {Sanders}, \& {Shivaei}}]{Reddy2015}
{Reddy}, N.~A., {Kriek}, M., {Shapley}, A.~E., {et~al.} 2015,
  \href{http://dx.doi.org/10.1088/0004-637X/806/2/259}{\JournalTitle{\apj},
  806, 259}

\bibitem[{Rela{\~n}o {et~al.}(2018)Rela{\~n}o, De~Looze, Kennicutt, Lisenfeld,
  Dariush, Verley, Braine, Tabatabaei, Kramer, Boquien, Xilouris, \&
  Gratier}]{Relano2018}
Rela{\~n}o, M., De~Looze, I., Kennicutt, R.~C., {et~al.} 2018,
  \href{http://dx.doi.org/10.1051/0004-6361/201732347}{\JournalTitle{\aap},
  613, A43}

\bibitem[{R{\'e}my-Ruyer {et~al.}(2014)R{\'e}my-Ruyer, Madden, Galliano,
  Galametz, Takeuchi, Asano, Zhukovska, Lebouteiller, Cormier, Jones, Bocchio,
  Baes, Bendo, Boquien, Boselli, DeLooze, Doublier-Pritchard, Hughes,
  Karczewski, \& Spinoglio}]{Remy-Ruyer2014}
R{\'e}my-Ruyer, A., Madden, S.~C., Galliano, F., {et~al.} 2014,
  \href{http://dx.doi.org/10.1051/0004-6361/201322803}{\JournalTitle{\aap},
  563, A31}

\bibitem[{Rieke {et~al.}(2015)Rieke, Wright, B{\"o}ker, Bouwman, Colina,
  Glasse, Gordon, Greene, G{\"u}del, Henning, Justtanont, Lagage, Meixner,
  N{\o}rgaard-Nielsen, Ray, Ressler, van Dishoeck, \& Waelkens}]{Rieke2015}
Rieke, G.~H., Wright, G.~S., B{\"o}ker, T., {et~al.} 2015,
  \href{http://dx.doi.org/10.1086/682252}{\JournalTitle{\pasp}, 127, 584}

\bibitem[{Salmon {et~al.}(2016)Salmon, Papovich, Long, Willner, Finkelstein,
  Ferguson, Dickinson, Duncan, Faber, Hathi, Koekemoer, Kurczynski, Newman,
  Pacifici, P{\'e}rez-Gonz{\'a}lez, \& Pforr}]{Salmon2016}
Salmon, B., Papovich, C., Long, J., {et~al.} 2016,
  \href{http://dx.doi.org/10.3847/0004-637X/827/1/20}{\JournalTitle{\apj}, 827,
  20}

\bibitem[{Salpeter(1955)}]{Salpeter1955}
Salpeter, E.~E. 1955,
  \href{http://dx.doi.org/10.1086/145971}{\JournalTitle{\apj}, 121, 161}

\bibitem[{S{\'a}nchez {et~al.}(2012)S{\'a}nchez, Kennicutt, Gil~de Paz, van~de
  Ven, V{\'{\i}}lchez, Wisotzki, Walcher, Mast, Aguerri, Albiol-P{\'e}rez,
  Alonso-Herrero, Alves, Bakos, Bart{\'a}kov{\'a}, Bland-Hawthorn, Boselli,
  Bomans, Castillo-Morales, Cortijo-Ferrero, de~Lorenzo-C{\'a}ceres, Del~Olmo,
  Dettmar, D{\'{\i}}az, Ellis, Falc{\'o}n-Barroso, Flores, Gallazzi,
  Garc{\'{\i}}a-Lorenzo, Gonz{\'a}lez~Delgado, Gruel, Haines, Hao, Husemann,
  Igl{\'e}sias-P{\'a}ramo, Jahnke, Johnson, Jungwiert, Kalinova, Kehrig, Kupko,
  L{\'o}pez-S{\'a}nchez, Lyubenova, Marino, M{\'a}rmol-Queralt{\'o},
  M{\'a}rquez, Masegosa, Meidt, Mendez-Abreu, Monreal-Ibero, Montijo,
  Mour{\~a}o, Palacios-Navarro, Papaderos, Pasquali, Peletier, P{\'e}rez,
  P{\'e}rez, Quirrenbach, Rela{\~n}o, Rosales-Ortega, Roth, Ruiz-Lara,
  S{\'a}nchez-Bl{\'a}zquez, Sengupta, Singh, Stanishev, Trager, Vazdekis,
  Viironen, Wild, Zibetti, \& Ziegler}]{Sanchez2012}
S{\'a}nchez, S.~F., Kennicutt, R.~C., Gil~de Paz, A., {et~al.} 2012,
  \href{http://dx.doi.org/10.1051/0004-6361/201117353}{\JournalTitle{\aap},
  538, A8}

\bibitem[{S{\'a}nchez {et~al.}(2013)S{\'a}nchez, Rosales-Ortega, Jungwiert,
  Iglesias-P{\'a}ramo, V{\'{\i}}lchez, Marino, Walcher, Husemann, Mast,
  Monreal-Ibero, Cid~Fernandes, P{\'e}rez, Gonz{\'a}lez~Delgado,
  Garc{\'{\i}}a-Benito, Galbany, van~de Ven, Jahnke, Flores, Bland-Hawthorn,
  L{\'o}pez-S{\'a}nchez, Stanishev, Miralles-Caballero, D{\'{\i}}az,
  S{\'a}nchez-Blazquez, Moll{\'a}, Gallazzi, Papaderos, Gomes, Gruel,
  P{\'e}rez, Ruiz-Lara, Florido, de~Lorenzo-C{\'a}ceres, Mendez-Abreu, Kehrig,
  Roth, Ziegler, Alves, Wisotzki, Kupko, Quirrenbach, Bomans, \&
  Collaboration}]{Sanchez2013}
S{\'a}nchez, S.~F., Rosales-Ortega, F.~F., Jungwiert, B., {et~al.} 2013,
  \href{http://dx.doi.org/10.1051/0004-6361/201220669}{\JournalTitle{\aap},
  554, A58}

\bibitem[{S{\'a}nchez {et~al.}(2016{\natexlab{a}})S{\'a}nchez, P{\'e}rez,
  S{\'a}nchez-Bl{\'a}zquez, Gonz{\'a}lez, Ros{\'a}lez-Ortega,
  Cano-D{\'{\i}}~az, L{\'o}pez-Cob{\'a}, Marino, Gil~de Paz, Moll{\'a},
  L{\'o}pez-S{\'a}nchez, Ascasibar, \& Barrera-Ballesteros}]{Sanchez2016a}
S{\'a}nchez, S.~F., P{\'e}rez, E., S{\'a}nchez-Bl{\'a}zquez, P., {et~al.}
  2016{\natexlab{a}},
  \href{http://adsabs.harvard.edu/abs/2016RMxAA..52...21S}{\JournalTitle{\rmxaa},
  52, 21}

\bibitem[{S{\'a}nchez {et~al.}(2016{\natexlab{b}})S{\'a}nchez, P{\'e}rez,
  S{\'a}nchez-Bl{\'a}zquez, Garc{\'{\i}}a-Benito, Ibarra-Mede, Gonz{\'a}lez,
  Rosales-Ortega, S{\'a}nchez-Menguiano, Ascasibar, Bitsakis, Law,
  Cano-D{\'{\i}}az, L{\'o}pez-Cob{\'a}, Marino, Gil~de Paz,
  L{\'o}pez-S{\'a}nchez, Barrera-Ballesteros, Galbany, Mast, Abril-Melgarejo,
  \& Roman-Lopes}]{Sanchez2016}
---. 2016{\natexlab{b}},
  \href{http://adsabs.harvard.edu/abs/2016RMxAA..52..171S}{\JournalTitle{\rmxaa},
  52, 171}

\bibitem[{S{\'a}nchez {et~al.}(2017)S{\'a}nchez, Barrera-Ballesteros,
  S{\'a}nchez-Menguiano, Walcher, Marino, Galbany, Bland-Hawthorn,
  Cano-D{\'{\i}}az, Garc{\'{\i}}a-Benito, L{\'o}pez-Cob{\'a}, Zibetti, Vilchez,
  Igl{\'e}sias-P{\'a}ramo, Kehrig, L{\'o}pez~S{\'a}nchez, Duarte~Puertas, \&
  Ziegler}]{Sanchez2017}
S{\'a}nchez, S.~F., Barrera-Ballesteros, J.~K., S{\'a}nchez-Menguiano, L.,
  {et~al.} 2017,
  \href{http://dx.doi.org/10.1093/mnras/stx808}{\JournalTitle{\mnras}, 469,
  2121}

\bibitem[{S{\'a}nchez {et~al.}(2018)S{\'a}nchez, Avila-Reese, Hernandez-Toledo,
  Cortes-Su{\'a}rez, Rodr{\'{\i}}guez-Puebla, Ibarra-Medel, Cano-D{\'{\i}}az,
  Barrera-Ballesteros, Negrete, Calette, de~Lorenzo-C{\'a}ceres,
  Ortega-Minakata, Aquino, Valenzuela, Clemente, Storchi-Bergmann, Riffel,
  Schimoia, Riffel, Rembold, Brownstein, Pan, Yates, Mallmann, \&
  Bitsakis}]{Sanchez2018}
S{\'a}nchez, S.~F., Avila-Reese, V., Hernandez-Toledo, H., {et~al.} 2018,
  \href{http://adsabs.harvard.edu/abs/2018RMxAA..54..217S}{\JournalTitle{\rmxaa},
  54, 217}

\bibitem[{Schmidt(1959)}]{Schmidt1959}
Schmidt, M. 1959, \href{http://dx.doi.org/10.1086/146614}{\JournalTitle{\apj},
  129, 243}

\bibitem[{Speagle {et~al.}(2014)Speagle, Steinhardt, Capak, \&
  Silverman}]{Speagle2014}
Speagle, J.~S., Steinhardt, C.~L., Capak, P.~L., \& Silverman, J.~D. 2014,
  \href{http://dx.doi.org/10.1088/0067-0049/214/2/15}{\JournalTitle{\apjs},
  214, 15}

\bibitem[{Storey \& Hummer(1995)}]{Storey1995}
Storey, P.~J., \& Hummer, D.~G. 1995,
  \href{http://dx.doi.org/10.1093/mnras/272.1.41}{\JournalTitle{\mnras}, 272,
  41}

\bibitem[{Theios {et~al.}(2019)Theios, Steidel, Strom, Rudie, Trainor, \&
  Reddy}]{Theios2019}
Theios, R.~L., Steidel, C.~C., Strom, A.~L., {et~al.} 2019,
  \href{http://dx.doi.org/10.3847/1538-4357/aaf386}{\JournalTitle{\apj}, 871,
  128}

\bibitem[{Tremonti {et~al.}(2004)Tremonti, Heckman, Kauffmann, Brinchmann,
  Charlot, White, Seibert, Peng, Schlegel, Uomoto, Fukugita, \&
  Brinkmann}]{Tremonti2004}
Tremonti, C.~A., Heckman, T.~M., Kauffmann, G., {et~al.} 2004,
  \href{http://dx.doi.org/10.1086/423264}{\JournalTitle{\apj}, 613, 898}

\bibitem[{Viaene {et~al.}(2017)Viaene, Sarzi, Baes, Fritz, \&
  Puerari}]{Viaene2017}
Viaene, S., Sarzi, M., Baes, M., Fritz, J., \& Puerari, I. 2017,
  \href{http://dx.doi.org/10.1093/mnras/stx1781}{\JournalTitle{\mnras}, 472,
  1286}

\bibitem[{Viero {et~al.}(2013)Viero, Moncelsi, Quadri, Arumugam, Assef,
  B{\'e}thermin, Bock, Bridge, Casey, Conley, Cooray, Farrah, Glenn, Heinis,
  Ibar, Ikarashi, Ivison, Kohno, Marsden, Oliver, Roseboom, Schulz, Scott,
  Serra, Vaccari, Vieira, Wang, Wardlow, Wilson, Yun, \& Zemcov}]{Viero2013}
Viero, M.~P., Moncelsi, L., Quadri, R.~F., {et~al.} 2013,
  \href{http://dx.doi.org/10.1088/0004-637X/779/1/32}{\JournalTitle{\apj}, 779,
  32}

\bibitem[{Whitmore {et~al.}(2014)Whitmore, Brogan, Chandar, Evans, Hibbard,
  Johnson, Leroy, Privon, Remijan, \& Sheth}]{Whitmore2014}
Whitmore, B.~C., Brogan, C., Chandar, R., {et~al.} 2014,
  \href{http://dx.doi.org/10.1088/0004-637X/795/2/156}{\JournalTitle{The
  Astrophysical Journal}, 795, 156}

\bibitem[{Wild {et~al.}(2011)Wild, Charlot, Brinchmann, Heckman, Vince,
  Pacifici, \& Chevallard}]{Wild2011}
Wild, V., Charlot, S., Brinchmann, J., {et~al.} 2011,
  \href{http://dx.doi.org/10.1111/j.1365-2966.2011.19367.x}{\JournalTitle{\mnras},
  417, 1760}

\bibitem[{Wright {et~al.}(2010)Wright, Eisenhardt, Mainzer, Ressler, Cutri,
  Jarrett, Kirkpatrick, Padgett, McMillan, Skrutskie, Stanford, Cohen, Walker,
  Mather, Leisawitz, Gautier, McLean, Benford, Lonsdale, Blain, Mendez, Irace,
  Duval, Liu, Royer, Heinrichsen, Howard, Shannon, Kendall, Walsh, Larsen,
  Cardon, Schick, Schwalm, Abid, Fabinsky, Naes, \& Tsai}]{Wright2010}
Wright, E.~L., Eisenhardt, P. R.~M., Mainzer, A.~K., {et~al.} 2010,
  \href{http://dx.doi.org/10.1088/0004-6256/140/6/1868}{\JournalTitle{\aj},
  140, 1868}

\bibitem[{Wuyts {et~al.}(2011)Wuyts, F{\"o}rster~Schreiber, Lutz, Nordon,
  Berta, Altieri, Andreani, Aussel, Bongiovanni, Cepa, Cimatti, Daddi, Elbaz,
  Genzel, Koekemoer, Magnelli, Maiolino, McGrath, P{\'e}rez~Garc{\'{\i}}a,
  Poglitsch, Popesso, Pozzi, Sanchez-Portal, Sturm, Tacconi, \&
  Valtchanov}]{Wuyts2011}
Wuyts, S., F{\"o}rster~Schreiber, N.~M., Lutz, D., {et~al.} 2011,
  \href{http://dx.doi.org/10.1088/0004-637X/738/1/106}{\JournalTitle{\apj},
  738, 106}

\bibitem[{Wuyts {et~al.}(2013)Wuyts, F{\"o}rster~Schreiber, Nelson, van Dokkum,
  Brammer, Chang, Faber, Ferguson, Franx, Fumagalli, Genzel, Grogin, Kocevski,
  Koekemoer, Lundgren, Lutz, McGrath, Momcheva, Rosario, Skelton, Tacconi,
  van~der Wel, \& Whitaker}]{Wuyts2013}
Wuyts, S., F{\"o}rster~Schreiber, N.~M., Nelson, E.~J., {et~al.} 2013,
  \href{http://dx.doi.org/10.1088/0004-637X/779/2/135}{\JournalTitle{\apj},
  779, 135}

\bibitem[{Yan {et~al.}(2016)Yan, Bundy, Law, Bershady, Andrews, Cherinka,
  Diamond-Stanic, Drory, MacDonald, S{\'a}nchez-Gallego, Thomas, Wake,
  Weijmans, Westfall, Zhang, Arag{\'o}n-Salamanca, Belfiore, Bizyaev, Blanc,
  Blanton, Brownstein, Cappellari, D'Souza, Emsellem, Fu, Gaulme, Graham,
  Goddard, Gunn, Harding, Jones, Kinemuchi, Li, Li, Maiolino, Mao, Maraston,
  Masters, Merrifield, Oravetz, Pan, Parejko, Sanchez, Schlegel, Simmons,
  Thanjavur, Tinker, Tremonti, van~den Bosch, \& Zheng}]{Yan2016}
Yan, R., Bundy, K., Law, D.~R., {et~al.} 2016,
  \href{http://dx.doi.org/10.3847/0004-6256/152/6/197}{\JournalTitle{\aj}, 152,
  197}

\bibitem[{Ye {et~al.}(2016)Ye, Zou, Lin, Lian, Hu, \& Kong}]{Ye2016}
Ye, C., Zou, H., Lin, L., {et~al.} 2016,
  \href{http://dx.doi.org/10.3847/0004-637X/826/2/209}{\JournalTitle{\apj},
  826, 209}

\bibitem[{Yip {et~al.}(2010)Yip, Szalay, Wyse, Dobos, Budav{\'a}ri, \&
  Csabai}]{Yip2010}
Yip, C.-W., Szalay, A.~S., Wyse, R. F.~G., {et~al.} 2010,
  \href{http://dx.doi.org/10.1088/0004-637X/709/2/780}{\JournalTitle{\apj},
  709, 780}

\bibitem[{Zahid {et~al.}(2017)Zahid, Kudritzki, Conroy, Andrews, \&
  Ho}]{Zahid2017}
Zahid, H.~J., Kudritzki, R.-P., Conroy, C., Andrews, B., \& Ho, I.-T. 2017,
  \href{http://dx.doi.org/10.3847/1538-4357/aa88ae}{\JournalTitle{\apj}, 847,
  18}

\bibitem[{Zhang {et~al.}(2017)Zhang, Yan, Bundy, Bershady, Haffner, Walterbos,
  Maiolino, Tremonti, Thomas, Drory, Jones, Belfiore, S{\'a}nchez,
  Diamond-Stanic, Bizyaev, Nitschelm, Andrews, Brinkmann, Brownstein, Cheung,
  Li, Law, Roman~Lopes, Oravetz, Pan, Storchi~Bergmann, \& Simmons}]{Zhang2017}
Zhang, K., Yan, R., Bundy, K., {et~al.} 2017,
  \href{http://dx.doi.org/10.1093/mnras/stw3308}{\JournalTitle{\mnras}, 466,
  3217}

\end{thebibliography}



\end{CJK*}
\end{document}